\documentclass[11pt]{article}
\pdfoutput=1

\usepackage{geometry,graphicx,amsmath,amssymb,amsfonts,mathdots,mathrsfs,caption,subcaption, setspace, wrapfig}
\usepackage[usenames,dvipsnames,svgnames,table]{xcolor}
\usepackage{graphicx}
\usepackage{jheppub}
\usepackage{epsfig}
\usepackage{longtable}
\usepackage{braket}
\usepackage{comment}

\graphicspath{{./}{./Figs/}}

\newcommand{\met}{\ensuremath{{\not\mathrel{E}}_T}}
\newcommand{\n}{{\chi}^0_1}
\newcommand{\nn}{{\chi}^0_2}
\newcommand{\nnn}{{\chi}^0_3}
\newcommand{\nnnn}{{\chi}^0_4}
\newcommand{\x}{{\chi}_1^\pm}
\newcommand{\xx}{{\chi}_2^\pm}

\newcommand{\xii}{{\chi}_i^\pm}
\newcommand{\xj}{{\chi}_j^\pm}
\newcommand{\nii}{{\chi}_i^0}
\newcommand{\nj}{{\chi}_j^0}

\newcommand{\tev}{\text{TeV}}
\newcommand{\gev}{\text{GeV}}
\newcommand{\mev}{\text{MeV}}
\newcommand{\beq}{\begin{eqnarray}}
\newcommand{\eeq}{\end{eqnarray}}
\newcommand{\bea}{\begin{eqnarray}}
\newcommand{\eea}{\end{eqnarray}}

\begin{document}


\title{Electroweakino Constraints from LHC Data}
\author[]{Travis A. W. Martin,}
\author[]{David E. Morrissey}
\affiliation[]{TRIUMF, 4004 Wesbrook Mall, Vancouver, Canada V6T 2A3}

\emailAdd{tmartin@triumf.ca}
\emailAdd{dmorri@triumf.ca}

\abstract{
We investigate the sensitivity of existing LHC 
searches to the charginos and neutralinos of the MSSM 
when all the other superpartners are decoupled.  
In this limit, the underlying parameter space reduces to 
a simple four-dimensional set $\{M_1,\,M_2,\,\mu,\,\tan\beta\}$.  
We examine the constraints placed on this parameter space by
a broad range of LHC searches taking into account the full set
of relevant production and decay channels.  We find that the exclusions
implied by these searches exceed existing limits from LEP only
for smaller values of the Bino mass $M_1 \lesssim 150\,\gev$.  
Our results have implications for MSSM dark matter
and electroweak baryogenesis.
}

\maketitle

\section{Introduction}

  Supersymmetry~(SUSY) is a well-motivated possibility for new physics,
and is one of the main discovery targets of the 
Large Hadron Collider~(LHC).  A broad range of SUSY searches have been
performed by the ATLAS and CMS collaborations with up to 5~fb$^{-1}$ 
of data at $7\,\tev$ and 20~fb$^{-1}$ of data at $8\,\tev$.  
Despite this great effort, no conclusive signals beyond the predictions 
of the Standard Model~(SM) have been observed so far.

  The absence of new signals puts constraints on the masses of the 
SM superpartners.   The strongest bounds apply to the light-flavour
squarks and the gluino, and can be as large as $m_{\tilde{q}/\tilde{g}}
\gtrsim 1500\,\gev$~\cite{Chatrchyan:2014lfa,Aad:2014wea}.
Limits on stops and sbottoms, which must not be too heavy if they are
to protect the naturalness of the weak scale~\cite{Barbieri:1987fn}, 
range between $m_{\tilde{t}/\tilde{b}} \gtrsim 200\!-\!700\,\gev$
depending on how they decay~\cite{Aad:2013ija,TheATLAScollaboration:2013aia,Aad:2014qaa,Chatrchyan:2013xna,CMS:2013cfa}.  
In contrast, superpartners that are uncharged under QCD can 
often be much lighter while remaining consistent with the current data.

  In the present work we study the implications of existing ATLAS and CMS
searches on the charginos and neutralinos of the minimal supersymmetric
standard model~(MSSM).  These states, which we will refer to collectively
as \emph{electroweakinos}, are mixtures of the superpartners of the electroweak
vector and Higgs bosons, and they take the form of four Majorana-fermion 
neutralinos $\chi_i^0$ ($i=1\!-\!4$ with $|m_i| \leq |m_{i+1}|$) 
and two Dirac-fermion charginos $\chi_i^{\pm}$ ($i=1,2$ with $|m_1|\leq |m_2|$).
All other superpartners, namely the sfermions and the gluino,
are assumed to be heavy enough that they can be neglected.  
Such a spectrum is motivated by the non-observation of squarks or the gluino, 
and can occur in theories of natural~\cite{Hall:1990ac,Essig:2011qg,Brust:2011tb,Papucci:2011wy,Baer:2012cf},
or (mini-)split~\cite{Wells:2004di,ArkaniHamed:2004fb,Giudice:2004tc,ArkaniHamed:2004yi,Hall:2011jd,Ibe:2011aa,Ibe:2012hu,Arvanitaki:2012ps,ArkaniHamed:2012gw} supersymmetry.  
This leads to a relatively simple parameter space of four variables: 
$\left\{M_1,\,M_2,\,\mu,\,\tan\beta\right\}$.

  A number of dedicated searches for electroweakinos have been performed
by the LHC collaborations~\cite{1596278,ATLAS:2013rla,TheATLAScollaboration:2013hha}.  These searches focus primarily on final states with multiple leptons, 
and their results have been interpreted mainly within the context 
of simplified models~\cite{Alwall:2008ag,Alves:2011wf}.
Our work extends these results in three important ways.
First, while simplified models are very useful in modelling key features
of the production and decay processes, they do not capture the full dynamics
of the MSSM.  For example, multiple production channels can contribute 
importantly to the signal, and individual states can have many 
significant decay paths~\cite{Drees:2013wra,Kraml:2013mwa,
Papucci:2014rja,Barnard:2014joa}.
Second, we investigate the sensitivity to electroweakinos 
of a much broader range of searches than were considered by the ATLAS and 
CMS collaborations in this context.  And third, we translate the search results 
into exclusions on $\left\{M_1,M_2,\mu,\tan\beta\right\}$, 
which has only recently been attempted in a limited way by the collaborations.  
This is useful for comparing with indirect limits on the electroweakinos, 
such as from flavour mixing and CP violation~\cite{Bharucha:2013epa}, 
precision electroweak tests~\cite{Martin:2004id}, 
Higgs production and decay rates~\cite{Casas:2013pta,Batell:2013bka}, 
and cosmological processes 
like electroweak baryogenesis~\cite{Cirigliano:2009yd,Kozaczuk:2011vr} and
dark matter production~\cite{ArkaniHamed:2006mb,Giudice:2010wb,Cheung:2012qy}.  

  The implications of LHC searches on electroweakinos have also been
the subject of many recent theoretical studies.  These analyses
often concentrate on specific collider 
topologies~\cite{Cabrera:2012gf,Baer:2013yha,Delannoy:2013ata} 
or kinematic regimes~\cite{Gori:2013ala,Han:2013kza,Han:2013usa,Schwaller:2013baa,Baer:2014cua,Han:2014kaa,Anandakrishnan:2014exa,Bramante:2014dza,Han:2014xoa}, 
or are focussed on specific dark-matter-motivated 
scenarios~\cite{Baer:2008ih,Belanger:2012jn,Liu:2013gba,Calibbi:2013poa,
Pierce:2013rda,Low:2014aa,Boehm:2013qva,Cirelli:2014dsa}.
Relative to these studies, we attempt to cover the MSSM parameter space
more broadly, and without imposing any restrictions motivated by cosmology.
At the same time, our analysis is more focussed on the electroweakinos
than the detailed MSSM parameter scans considered in 
Refs.~\cite{CahillRowley:2012cb,Cahill-Rowley:2013yla}.

  The outline of this paper is as follows.  After this introduction, 
we describe the parameter ranges to be studied, their relationship 
to the spectrum, and their effect on production and decay processes 
in Section~\ref{sec:model}.  In Section~\ref{sec:search} we describe 
in detail the methods used to reinterpret the LHC results.
Next, in Section~\ref{sec:constraint} we give our results in terms
of exclusions on the underlying electroweakino parameters.  Finally,
Section~\ref{sec:conc} is reserved for our conclusions.
Formulas for the relevant masses and couplings are collected in
Appendices~\ref{sec:appa} and \ref{sec:appb}.

\section{Masses and Mixings, Production and Decay\label{sec:model}}

  In this work we study the electroweakinos of the MSSM in the 
limit where all other superpartners (and the additional Higgs bosons)
are much heavier.  To be concrete, we set the values of the sfermion and
gluino soft mass parameters to $2000\,\gev$ together with $m_A = 1500\,\gev$,
which effectively decouples these states from the LHC searches to be studied.
This leaves a four-dimensional parameter space for the electroweakinos
consisting of $\left\{M_1,\,M_2,\,\mu,\,\tan\beta\right\}$. 
We take as input the running values of these parameters
at the scale $M_S = 2000\,\gev$ and fix $\tan\beta=10$.\footnote{The running
of these parameters below this scale is mild, and we find nearly 
identical results using the same input values defined instead
at $M_S=300\,\gev$.}
Explicit tree-level expressions for the electroweakino masses and couplings
in terms of these parameters are collected in Appendices~\ref{sec:appa} 
and \ref{sec:appb}.
  
  In most of the regions of interest, the diagonal mass-matrix elements, 
set by $M_1$ and $M_2$ for the gauginos and $\mu$ for the Higgsinos, 
are significantly larger than the off-diagonal elements, which 
are proportional to $m_Z$.  As a result, the mass eigenstates 
tend to be closely aligned with the underlying gauge eigenstates unless
there is a degeneracy among the diagonal terms.  We will therefore
speak frequently of Bino-like, Wino-like, and Higgsino-like mass eigenstates.
The two Higgsino-like neutralino states coincide with the linear combinations
\beq
\widetilde{H}_{\pm}^0 = \frac{1}{\sqrt{2}}
\left(\widetilde{H}_u^0\pm\widetilde{H}_d^0\right) \ .
\eeq
Away from degeneracies, the mixing of the Higgsinos $\widetilde{H}_{\pm}^0$
with the gaugino $\widetilde{\lambda}_a$ ($a=1,2$) 
is proportional to $m_Z/|\mu \pm M_a|$, 
where $M_a$ is the relevant gaugino mass. 
Note that when $\mu$ and $M_a$ have the same sign, the $\widetilde{H}_-^0$
state mixes more strongly with the gaugino than 
the $\widetilde{H}_+^0$~\cite{ArkaniHamed:2006mb}.
Mixing between different gaugino-like states requires two
small mixing factors and is further suppressed.
We will apply these considerations below to explain the relative 
production and decay rates of the physical chargino and neutralino states.  
  
  The relatively small mixing away from degeneracies also motivates
us to focus on a specific value of $\tan\beta=10$.  This parameter only 
enters into the properties of the electroweakinos through the off-diagonal 
elements of the mixing matrices (and in the direct couplings to the Higgs boson), 
as described in Appendices~\ref{sec:appa} and \ref{sec:appb}. 
Thus, we expect qualitatively similar results for production rates
and decay fractions throughout the range
 $2 \leq \tan\beta \leq 50$ except where the splitting 
between two of $M_1$, $M_2$, or $\mu$ becomes small.

\subsection{LHC Production Rates}

\begin{figure}[ttt]
   \begin{center}
	\includegraphics[width=0.6\textwidth]{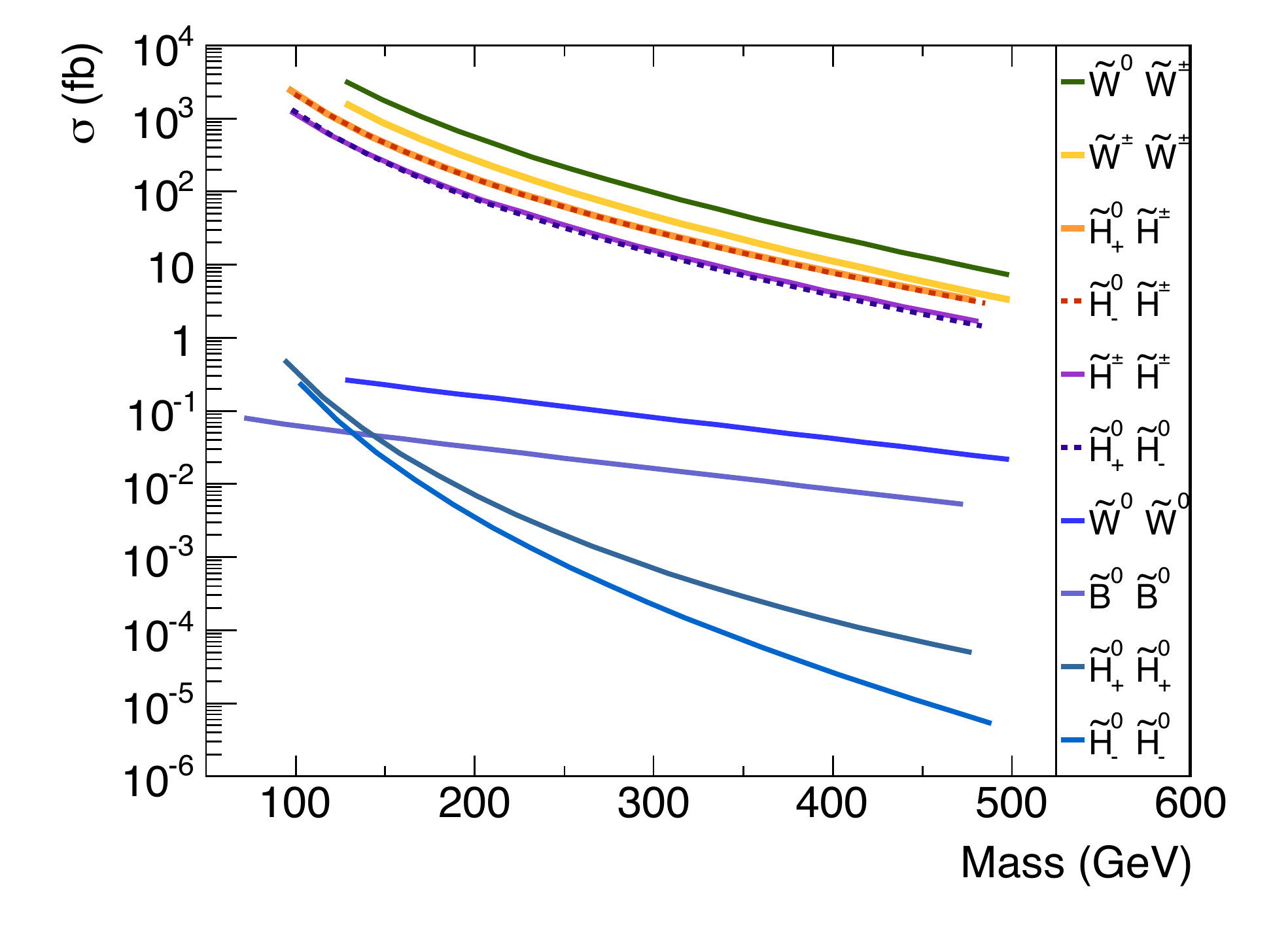}
   \end{center}
   \vspace{-4mm}
   \caption{Production cross sections of the electroweakinos 
at the LHC with $\sqrt{s}=8\,\tev$ in the limit of nearly pure
gauge eigenstates.  One of $M_1$, $M_2$, or $\mu$ is varied
independently with the other values set to $1000\,\tev$
and all other MSSM parameters set to $2000\,\gev$.
The ``Mass'' label refers to the average
mass of the two states being produced. Higgsino states are expressed 
in terms of $\widetilde{H}_\pm^0 = (\widetilde{H}_u^0 \pm \widetilde{H}_d^0)/\sqrt(2)$.
}
\label{fig:pure}
\end{figure}

   In Fig.~\ref{fig:pure} we show the LHC8 production cross-sections 
of Bino-like, Wino-like, and Higgsino-like states for $\tan\beta=10$
with all other parameters taken to be much larger:
$m = 2000\,\gev$ for the sfermions and gluino and $m = 1000\,\gev$
for the other electroweakino parameters.
The physical masses in this limit are given approximately
by $M_1,\,M_2$, or $\mu$.  
The neutral Higgsino-like states are labelled in order of 
increasing mass and correspond to the linear combinations 
$\widetilde{H}_1^0 \sim \widetilde{H}_-^0$ and
$\widetilde{H}_2^0 \sim \widetilde{H}_+^0$ defined above.
All rates shown in the figure are computed
at leading order~(LO) with MadGraph~5~\cite{Alwall:2011uj} 
and cross-checked in Prospino2.1~\cite{Beenakker:1996ed,Beenakker:1999xh}.

  These production cross sections are dominated by processes
with intermediate electroweak vector bosons.  Both the Winos
and Higgsinos couple to vector bosons through their gauge-covariant
derivatives.   This leads to unsuppressed couplings for
$\chi_i^0\chi_j^{\pm}W^{\mp}$ and $\chi_i^{\pm}\chi_j^{\mp}Z^0$
when both states are pure Wino or pure Higgsino.  
In contrast, the neutralino couplings $\chi_i^0\chi_j^0Z^0$ 
involve only the Higgsino states, and are also suppressed when
the $i=j$ state is Higgsino-like.  The absence of a direct Wino coupling 
to the $Z^0$ arises because $\widetilde{W}^0$ has $t^3=Y=0$.
Thus, the production rates of $\widetilde{W}^0\widetilde{W}^0$
and $\widetilde{B}^0\widetilde{B}^0$ are suppressed since
both processes require two Higgsino mixings in the amplitude.
The very small $\widetilde{H}^0_{i}\widetilde{H}^0_i$ rates
are due to a cancellation in the pure Higgsino limit reflecting
the fact that the corresponding mass eigenstates approach Dirac states
with only a vector coupling to the $Z^0$ as $m_Z/\mu\to 0$.
Production through a $W^{\pm}$ is generally larger than via
the neutral vector bosons.

\begin{figure}[ttt]
   \begin{center}
	\includegraphics[width=0.34\textwidth]{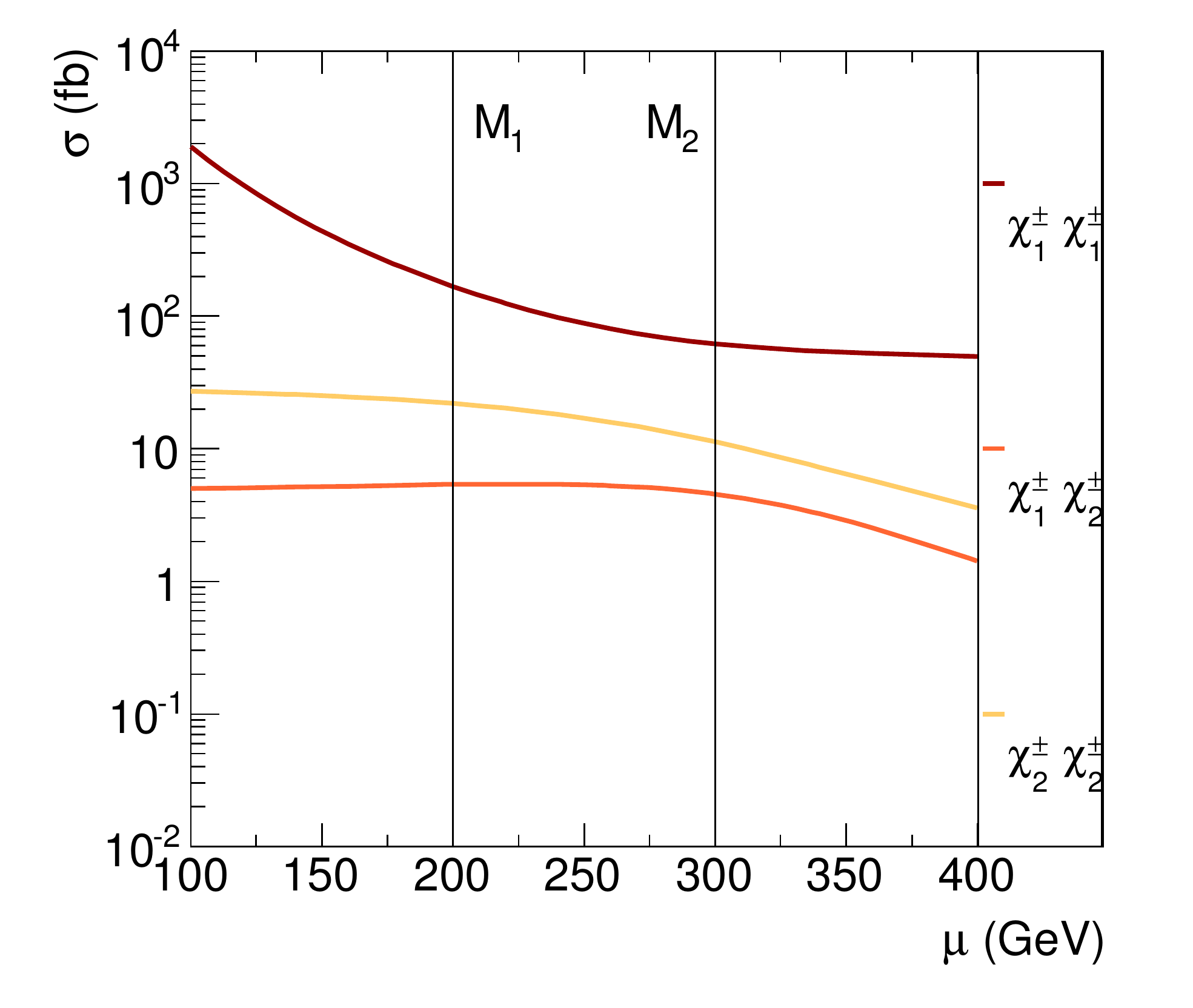}
\hspace{-0.5cm}
	\includegraphics[width=0.34\textwidth]{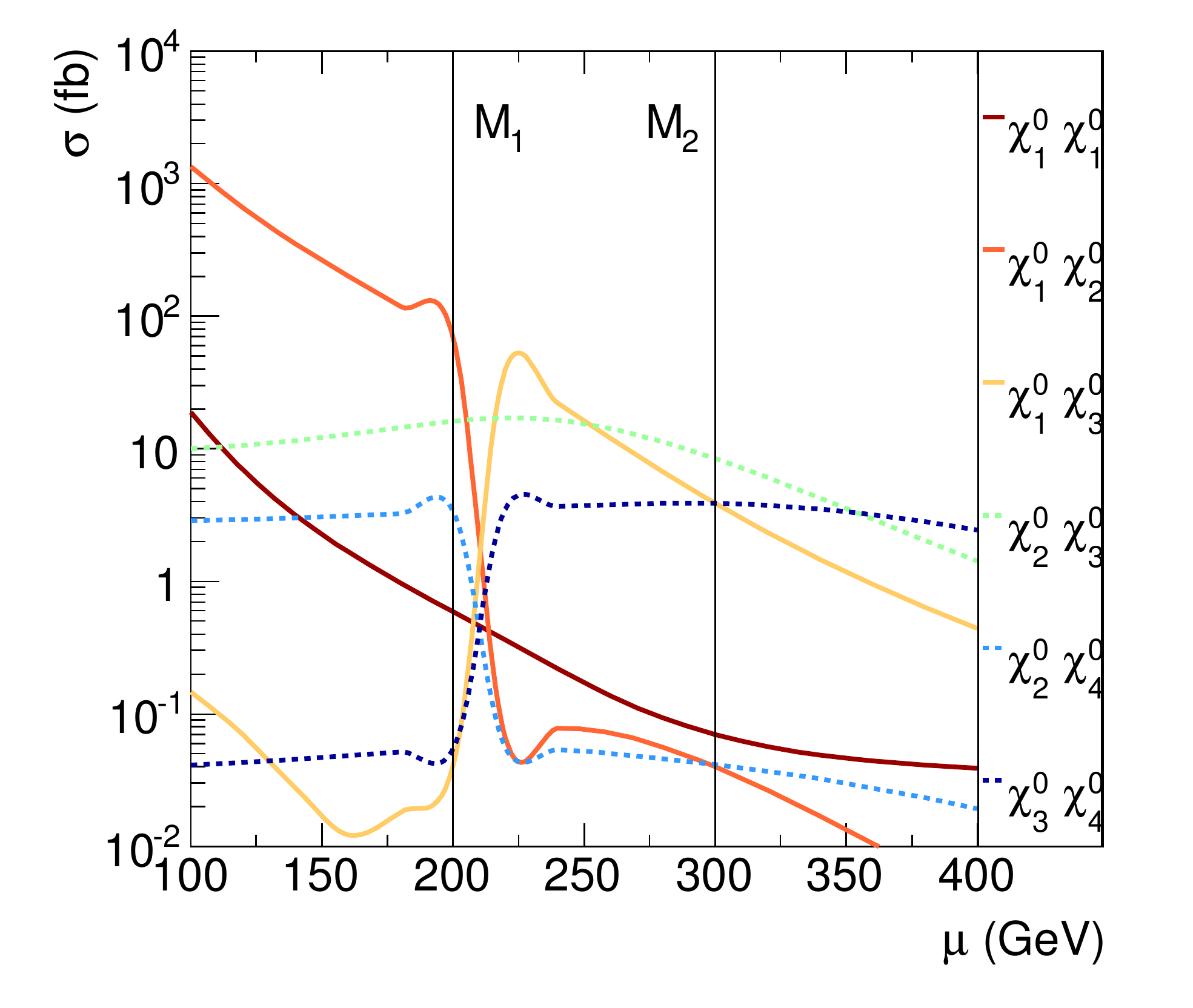}
\hspace{-0.5cm}
	\includegraphics[width=0.34\textwidth]{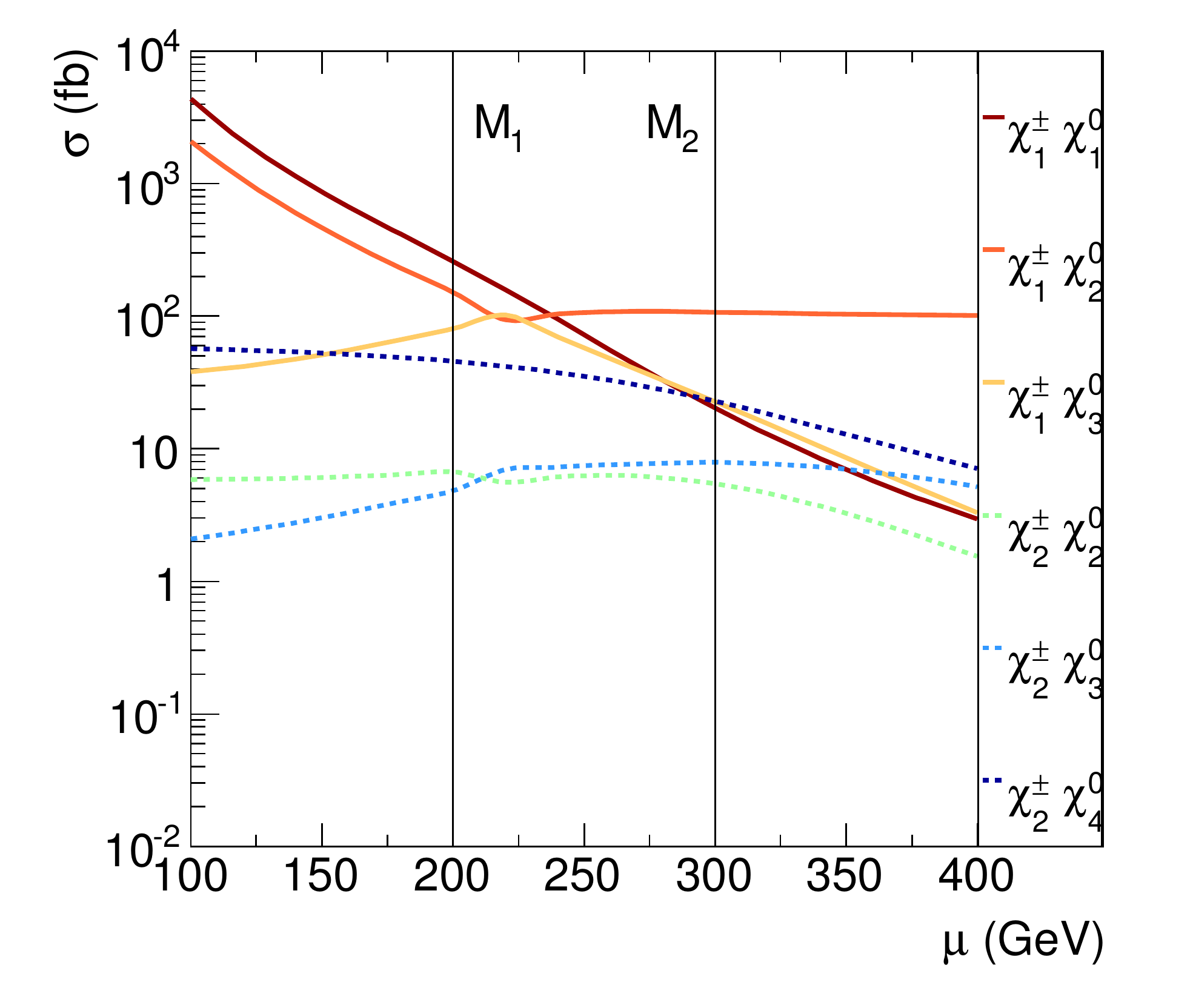}
   \end{center}
   \vspace{-4mm}
   \caption{Leading electroweakino production cross sections as 
a function of $\mu$ for $M_1=200\,\gev$ and $M_2 = 300\,\gev$.  
The leftmost panel shows the chargino-chargino rates, 
the middle panel shows the dominant neutralino-neutralino rates, 
and the rightmost panel shows the largest chargino-neutralino rates.
}
\label{fig:domxsec0}
\end{figure}

\begin{figure}[ttt]
   \begin{center}
	\includegraphics[width=0.34\textwidth]{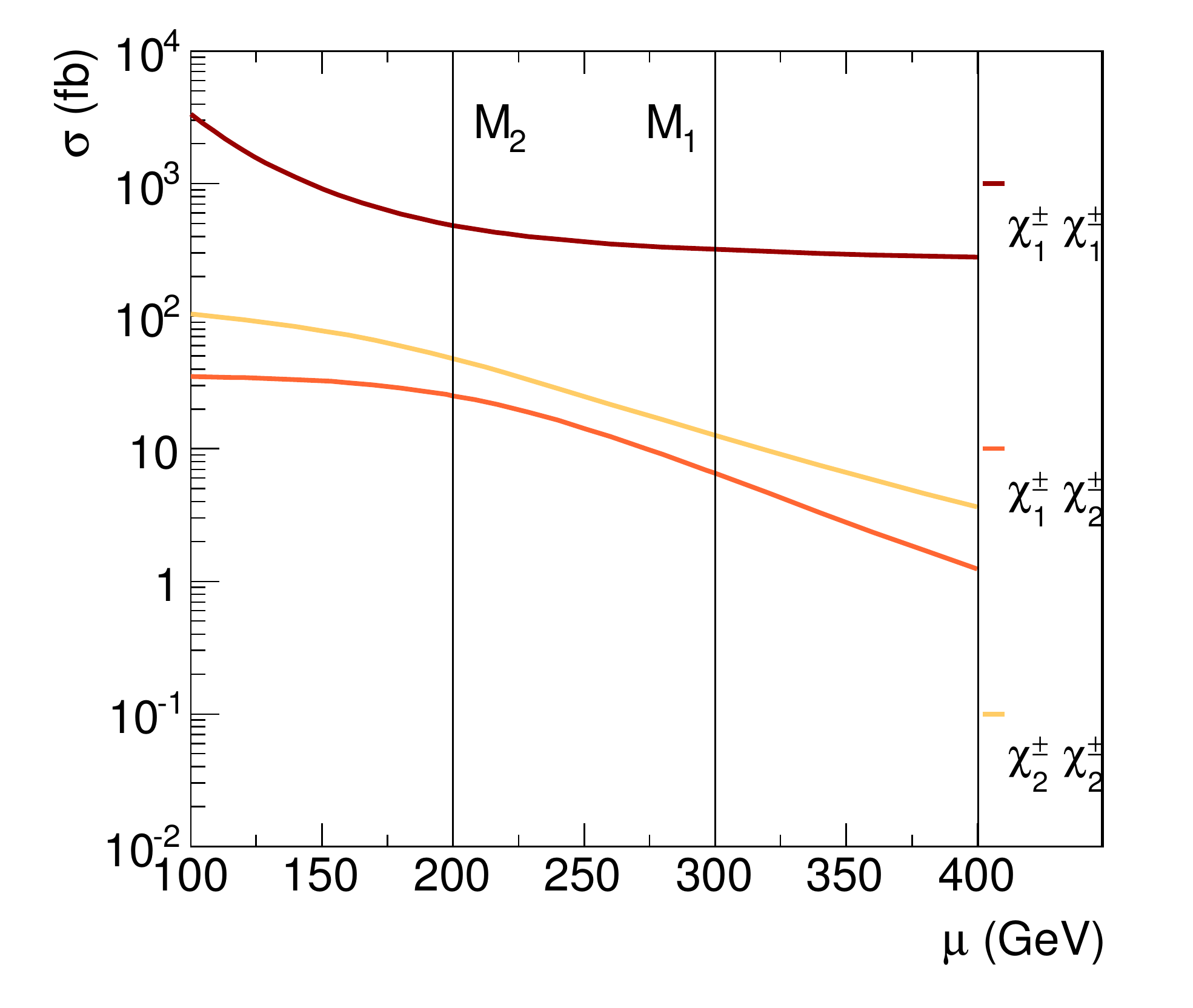}
        \hspace{-0.5cm}
	\includegraphics[width=0.34\textwidth]{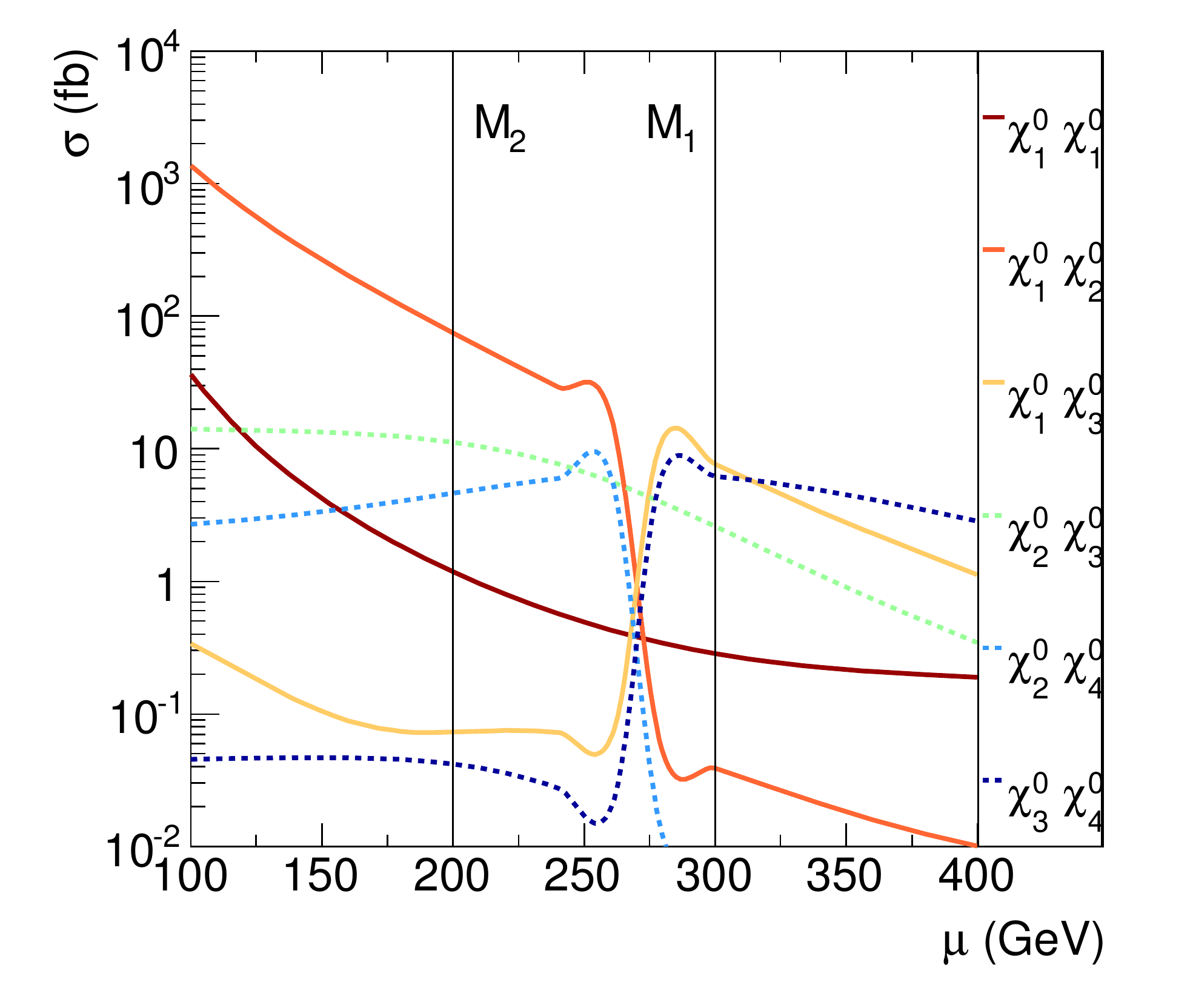}
        \hspace{-0.5cm}
	\includegraphics[width=0.34\textwidth]{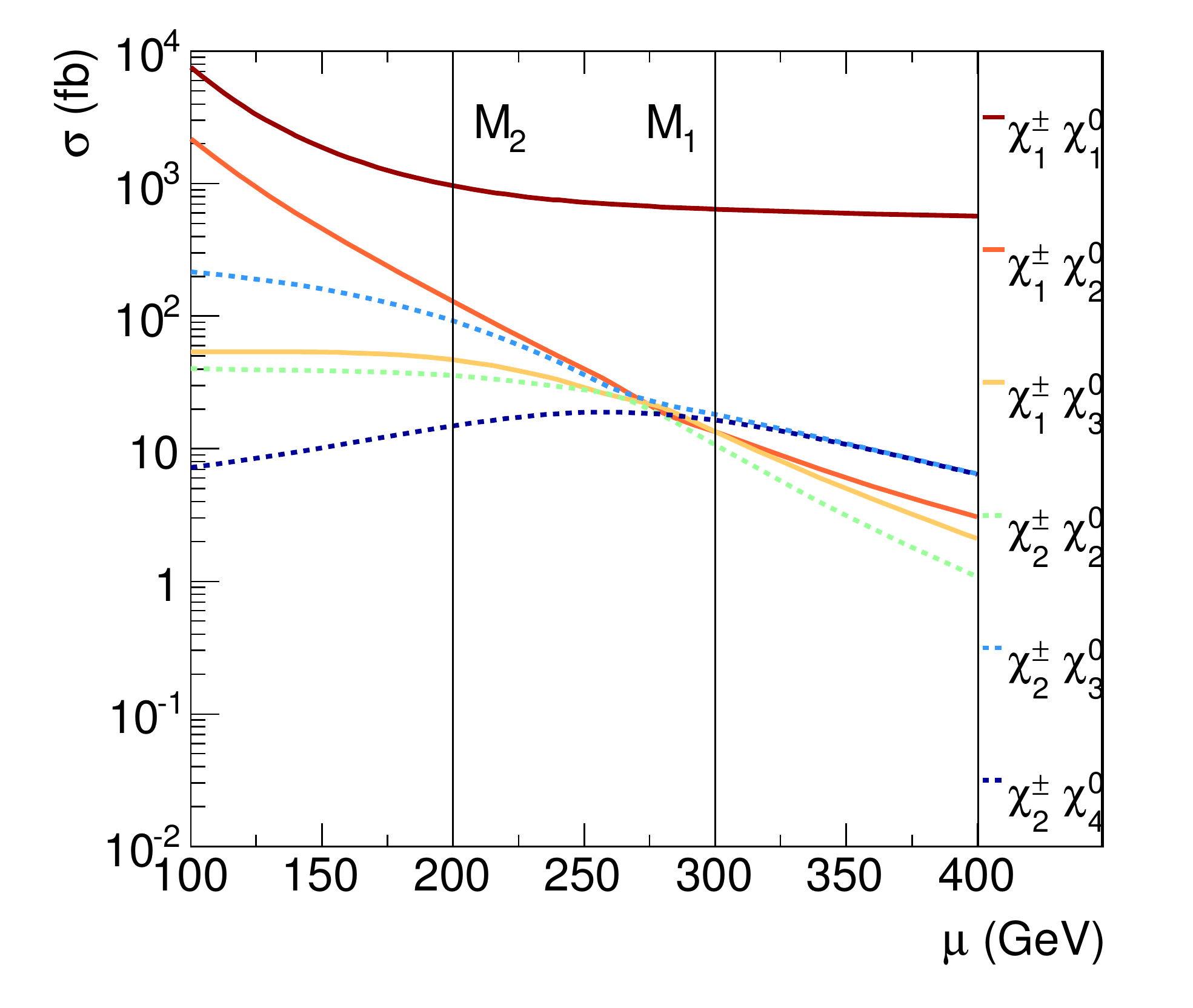}
   \end{center}
   \vspace{-4mm}
   \caption{Leading electroweakino production cross sections as 
a function of $\mu$ for $M_1=200\,\gev$ and $M_2 = 300\,\gev$.
The leftmost panel shows the chargino-chargino rates, 
the middle panel shows the dominant neutralino-neutralino rates, 
and the rightmost panel shows the largest chargino-neutralino rates.
}
\label{fig:domxsec1}
\end{figure}

  In Fig.~\ref{fig:domxsec0} we show 
the dominant chargino and neutralino production cross sections 
as a function of the $\mu$ parameter for $\tan\beta=10$ and fixed 
$(M_1,M_2)=(200,300)\,\gev$.  Similar plots for  
$(M_1,M_2)=(300,200)\,\gev$ are shown in Fig.~\ref{fig:domxsec1}.
These figures can be understood in terms of the gauge-eigenstate content
of the corresponding mass eigenstates within the six possible hierarchies
of $M_1$, $M_2$, and $\mu$.


  Of the processes shown in Figs.~\ref{fig:domxsec0} and \ref{fig:domxsec1}, 
the chargino pair production rates in the leftmost panels are
the easiest to understand.  Here, the $\x$ state is Higgsino-like
for $\mu < M_2$ and evolves smoothly into a Wino-like state 
as $\mu$ increases above $M_2$.  
The $\chi_1^+\chi_1^{-}$ and $\chi_2^{+}\chi_2^-$ 
rates follow the expectations for pure states in the appropriate limits,
while the rate for $\chi_1^{\pm}\chi_2^{\mp}$ 
is suppressed by the mixing factor it requires.

  Neutralino-neutralino production, shown in the middle panels of
Figs.~\ref{fig:domxsec0} and \ref{fig:domxsec1}, has a slightly
more complicated dependence on $\mu$.  Away from $\mu \sim M_1,\,M_2$, 
the physical mass eigenstates are closely aligned with pure gaugino 
($\widetilde{W}^0,\,\widetilde{B}^0$)
or Higgsino ($\widetilde{H}^0_{\pm}$) states.  
The hierarchy of production rates can be understood
by recalling that neutralino pair production occurs only through the 
$Z^0 \widetilde{H}_+^0 \widetilde{H}_-^0$ coupling, and that the 
$\widetilde{H}^0_-$-like state mixes much more
readily with gauginos than the $\widetilde{H}_+^0$-like state
(for $\mu$ and $M_{1,2}$ of the same sign). 
Thus, the largest production rate occurs for the pair 
of states coinciding with $\widetilde{H}_+^0\widetilde{H}_-^0$,
followed by gaugino-$\widetilde{H}^0_+$ pairs, and then
gaugino-$\widetilde{H}^0_-$.  The production of gaugino-gaugino
or $\widetilde{H}_{\pm}^0\widetilde{H}^0_{\pm}$ pairs requires
more small mixing factors and is further supressed.

  These considerations explain the $\mu$ dependence of neutralino pair
production seen in the middle panels of 
Figs.~\ref{fig:domxsec0} and \ref{fig:domxsec1}.
The $\widetilde{H}_+^0\widetilde{H}_-^0$-like combination
is $\n\nn$ for $\mu < M_< \equiv \min\{M_1,M_2\}$, 
$\nn\nnn$ for $M_< < \mu < M_> \equiv \max\{M_1,M_2\}$,
and $\nnn\nnnn$ for $\mu > M_{>}$, and these are seen to have
the largest rates (away from the gaugino masses).
A sharp crossover is seen in both panels for the rates 
of $\n\nn$ and $\n\nnn$ at a value of $\mu$ between $M_<$ and $M_>$.
For increasing $\mu$ in this range, 
mixing with the lighter gaugino tends
to push the $\widetilde{H}_-^0$ mass up relative to $\widetilde{H}_+^0$, 
while mixing with the heavier gaugino tends to push the $\widetilde{H}^0_-$
mass down.  This leads to a crossover where 
the $\widetilde{H}_-^0$-like state becomes lighter than 
the $\widetilde{H}_+^0$-like state, and the gauge contents 
of the mass-ordered $\nn$ and $\nnn$ states are suddenly exchanged
with each other.  At this point, $\n\nn$ goes from a moderately suppressed 
gaugino-$\widetilde{H}_+^0$ process to a highly suppressed 
gaugino-$\widetilde{H}_-^0$ process, with the opposite
occurring for $\n\nnn$.  A similar crossover is seen for the 
$\nn\nnnn$ and $\nnn\nnnn$ rates.  In both cases, the
physically relevant quantity is the inclusive neutralino pair
production rate, and this varies smoothly with $\mu$.

  The rightmost panels of Figs.~\ref{fig:domxsec0} and \ref{fig:domxsec1}
show the leading mixed neutralino-chargino production rates.
For $\mu < M_2$, the largest cross sections occur for pairs
of Higgsino-like states, such as $\chi_1^{\pm}\chi_1^0$.  
As $\mu$ grows larger than $M_2$, there is a smooth transition such 
that the largest rates occur for pairs of Wino-like states.
This corresponds to $\chi_1^{\pm}\nn$ for $M_1 < M_2$
and $\chi_1^{\pm}\n$ for $M_1> M_2$.

  We have also examined the effects of varying $\tan\beta$ over
the range $2\leq \tan\beta\leq 50$.  The dependence of the production
cross sections on $\tan\beta$ saturates at larger values,
with almost no variation between $\tan\beta=10$ and $\tan\beta=50$.
For smaller $\tan\beta \sim 2$, the variation can be stronger near 
a mass degeneracy although the net qualitative effect tends to be mild.

\subsection{Decay Branching Fractions}

  The collider signatures of the electroweakinos depend crucially
on how they decay.  When all the sfermions are very heavy, the dominant
decay channels are
\begin{eqnarray}
\begin{array}{rcrcr}
\nii \rightarrow \xj W^{\mp(*)}, &\;\;\;\;\;&
\nii \rightarrow \nj Z^{0(*)},\;\; &\;\;\;\;\;& 
\nii \rightarrow \nj h^{0(*)},\;\;    \\
\xii \rightarrow \nj W^{\pm(*)}, &\;\;\;\;\;&
\xii \rightarrow \xj Z^{0(*)}, &\;\;\;\;\;& 
\xii \rightarrow \xj h^{0(*)}, 
\end{array}
\end{eqnarray}
where $j<i$, and the $W^\pm$, $Z^0$, and $h^0$ can be potentially 
off-shell (as indicated by $(*)$).  A loop-mediated decay with a photon
is also possible, but we almost always find it to be highly suppressed
compared to the channels listed above.  

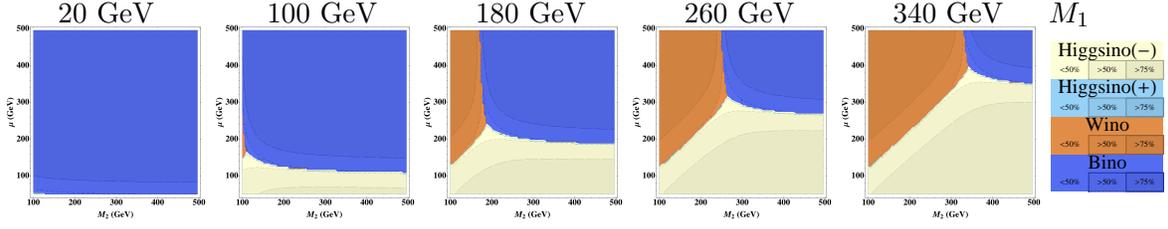
\begin{figure}[ttt]
	\begin{subfigure}{0.171\textwidth}
		\begin{picture}(120,100)
			\put(0,0){\includegraphics[width=\textwidth]{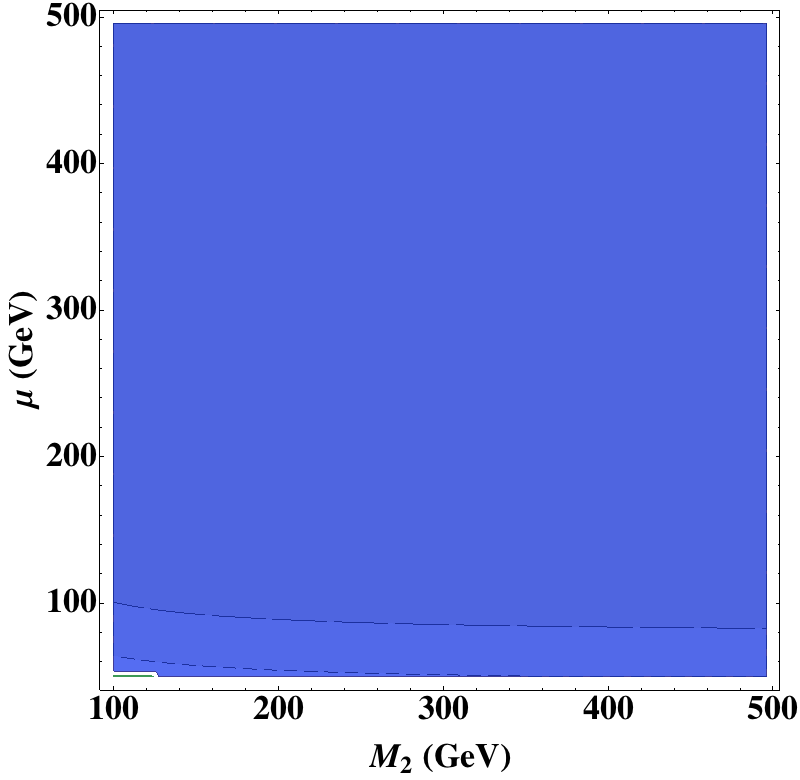}}
			\put(20,75){20 GeV}
		\end{picture}
		\label{}
	\end{subfigure}
	\begin{subfigure}{0.171\textwidth}
		\begin{picture}(120,100)
			\put(0,0){\includegraphics[width=\textwidth]{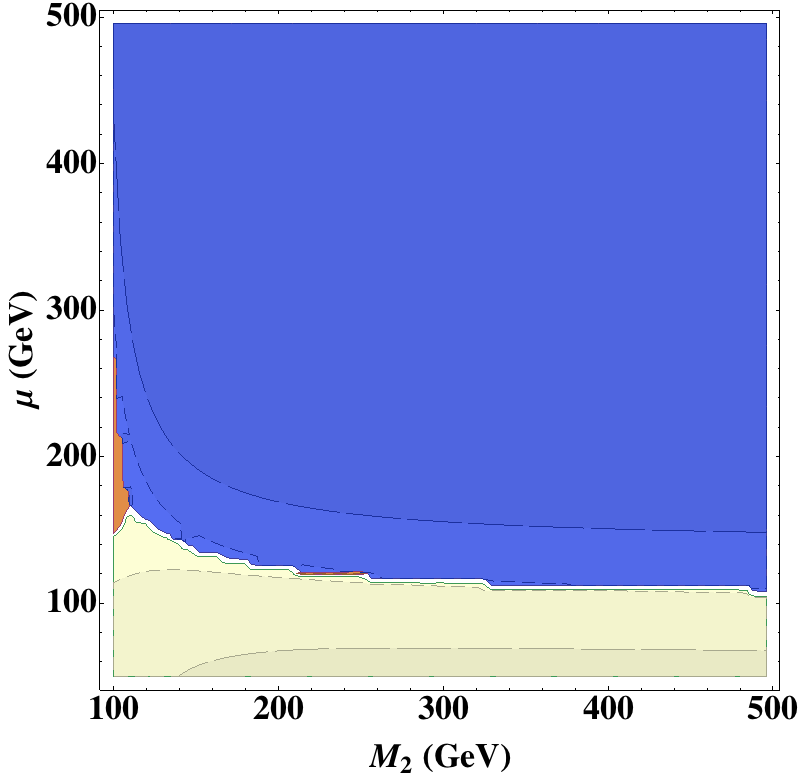}}
			\put(20,75){100 GeV}
		\end{picture}
		\label{}
	\end{subfigure}
	\begin{subfigure}{0.171\textwidth}
		\begin{picture}(120,100)
			\put(0,0){\includegraphics[width=\textwidth]{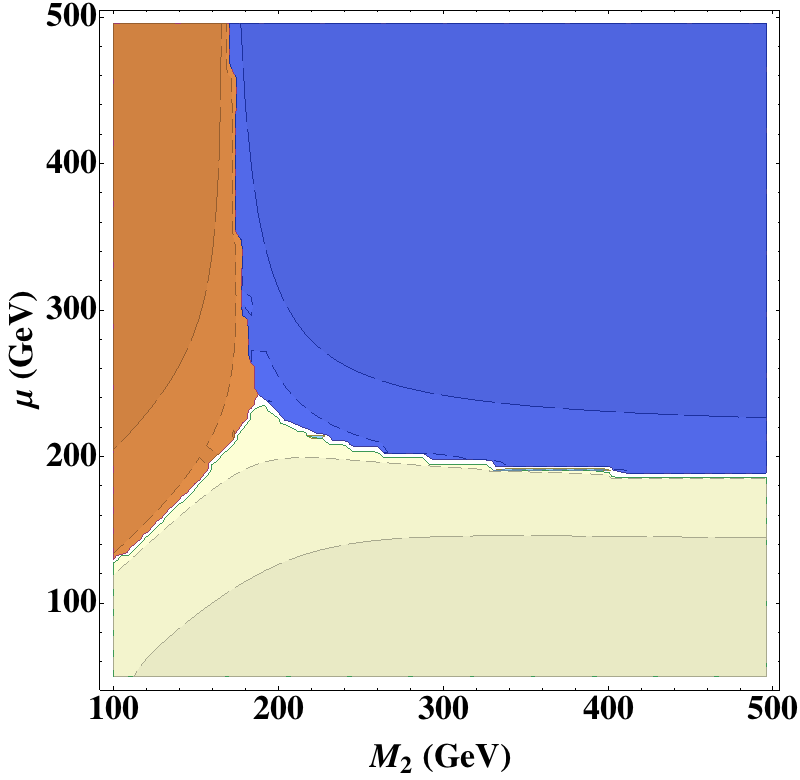}}
			\put(20,75){180 GeV}
		\end{picture}
		\label{}
	\end{subfigure}
	\begin{subfigure}{0.171\textwidth}
		\begin{picture}(120,100)
			\put(0,0){\includegraphics[width=\textwidth]{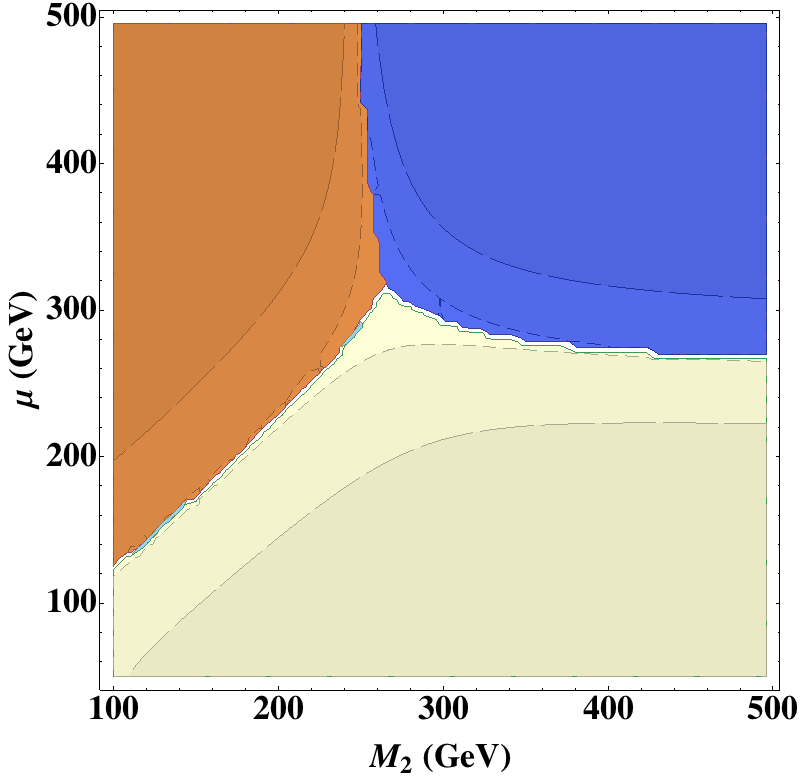}}
			\put(20,75){260 GeV}
		\end{picture}
		\label{}
	\end{subfigure}
	\begin{subfigure}{0.171\textwidth}
		\begin{picture}(120,100)
			\put(0,0){\includegraphics[width=\textwidth]{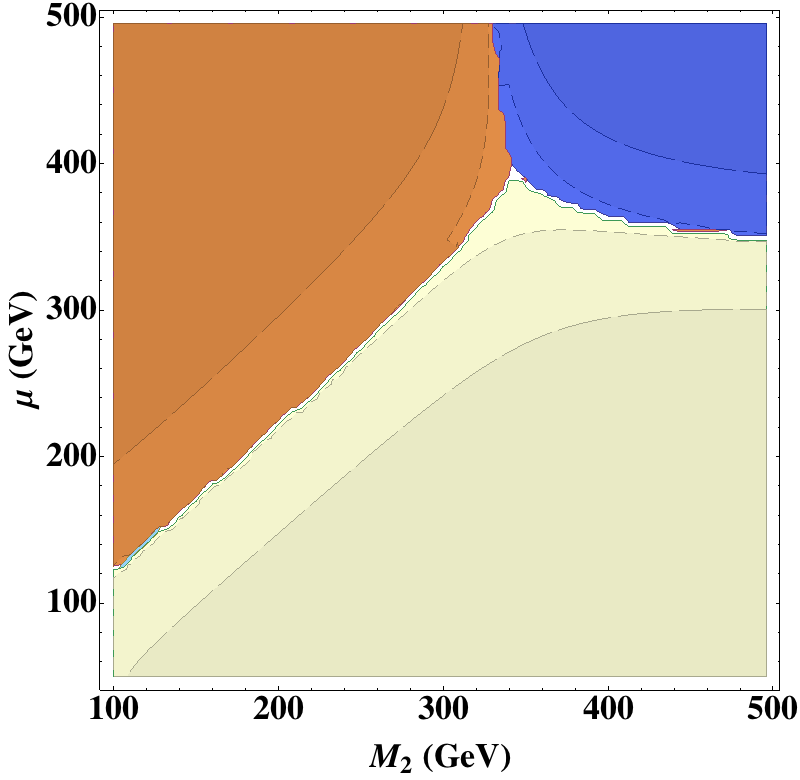}}
			\put(20,75){340 GeV}
		\end{picture}
		\label{}
	\end{subfigure}
	\begin{subfigure}{0.1\textwidth}
		\begin{picture}(120,100)
			\put(0,10){\includegraphics[width=\textwidth]{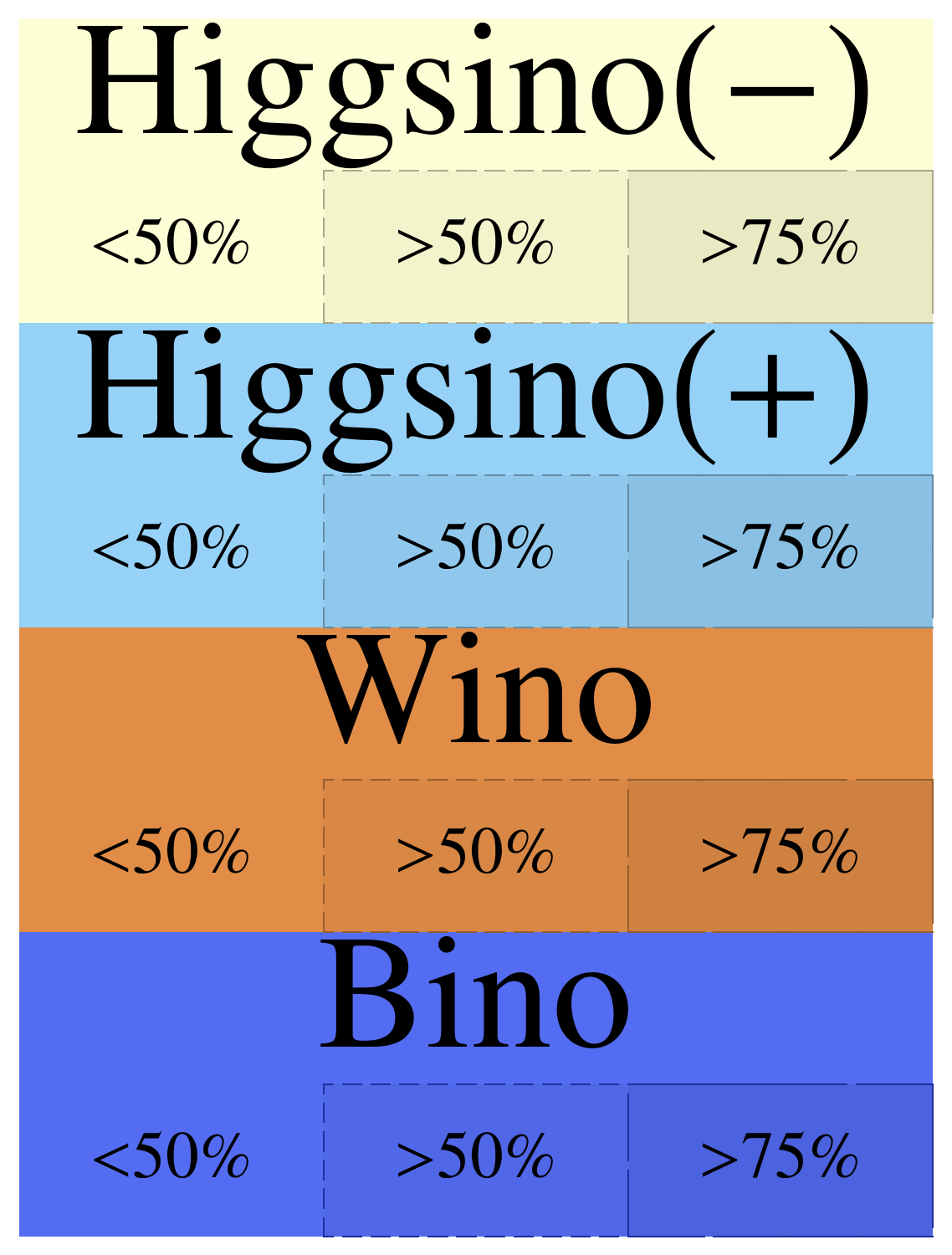}}
			\put(0,75){$M_1$}
		\end{picture}			
		\label{}
	\end{subfigure}
	\vspace{-4mm}	
	\caption{$\mathbf{\n:}$ 
Dominant gauge eigenstate content of the lightest
neutralino $\chi_1^0$ in the $M_2$--$\mu$ plane for various slices
of $M_1$ and $\tan\beta=10$. Shaded, dash-enclosed regions indicate
the boundary of 50\% and 75\% composition, as noted in the legend.}
	\label{fig:n1summary}
\vspace{-0.1cm}
\end{figure}

\begin{figure}[ttt]
	\begin{subfigure}{0.171\textwidth}
		\begin{picture}(100,100)
			\put(0,0){\includegraphics[width=\textwidth]{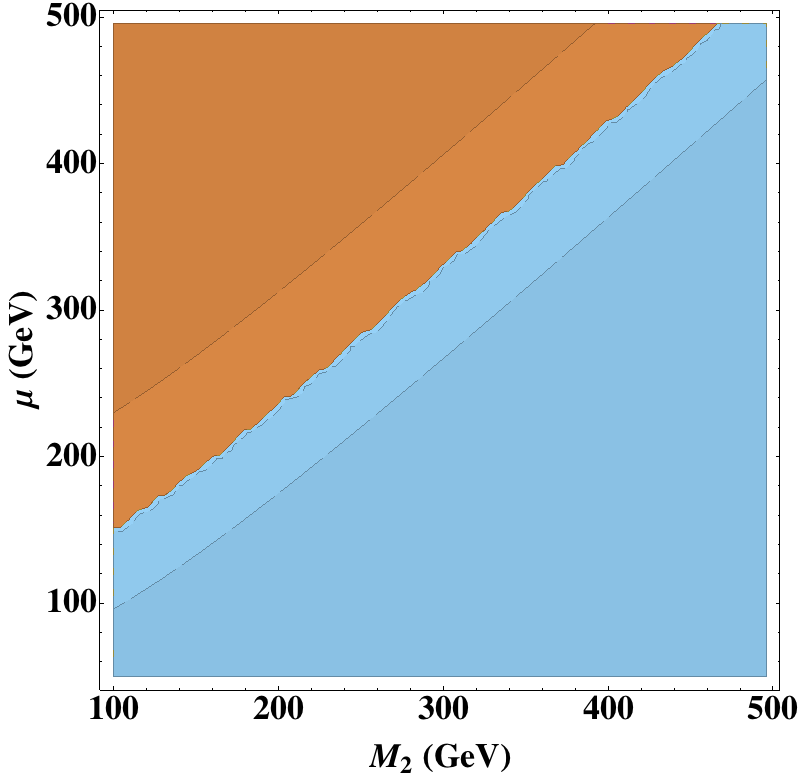}}
			\put(20,75){20 GeV}
		\end{picture}
		\label{}
	\end{subfigure}
	\begin{subfigure}{0.171\textwidth}
		\begin{picture}(100,100)
			\put(0,0){\includegraphics[width=\textwidth]{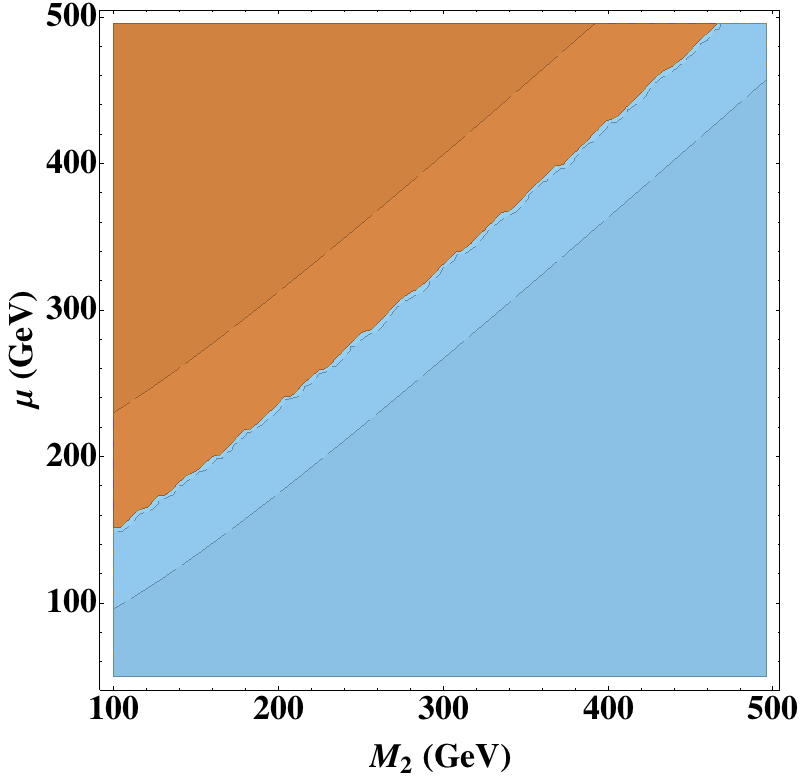}}
			\put(20,75){100 GeV}
		\end{picture}
		\label{}
	\end{subfigure}
	\begin{subfigure}{0.171\textwidth}
		\begin{picture}(100,100)
			\put(0,0){\includegraphics[width=\textwidth]{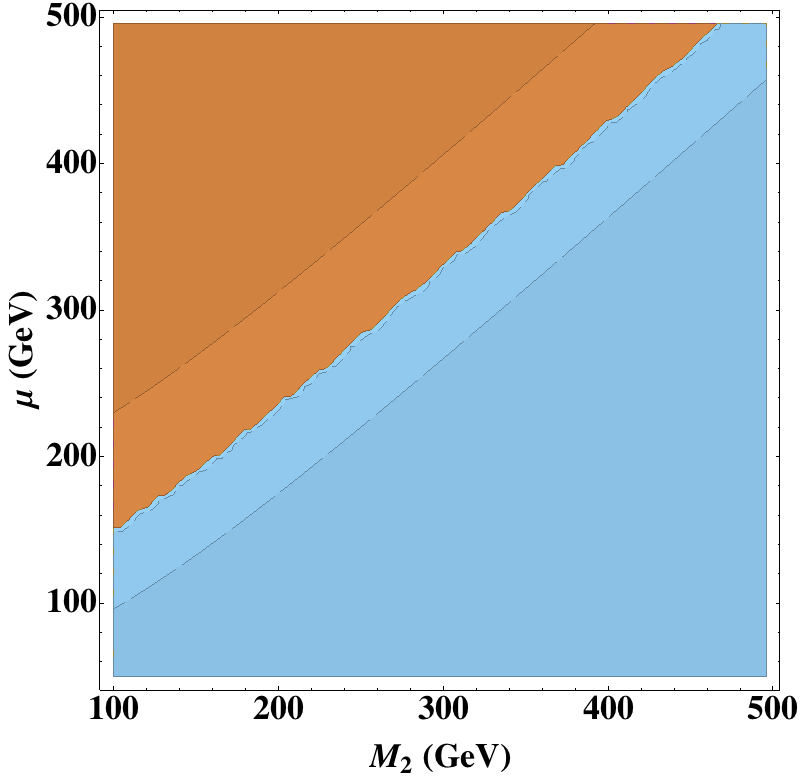}}
			\put(20,75){180 GeV}
		\end{picture}
		\label{}
	\end{subfigure}
	\begin{subfigure}{0.171\textwidth}
		\begin{picture}(100,100)
			\put(0,0){\includegraphics[width=\textwidth]{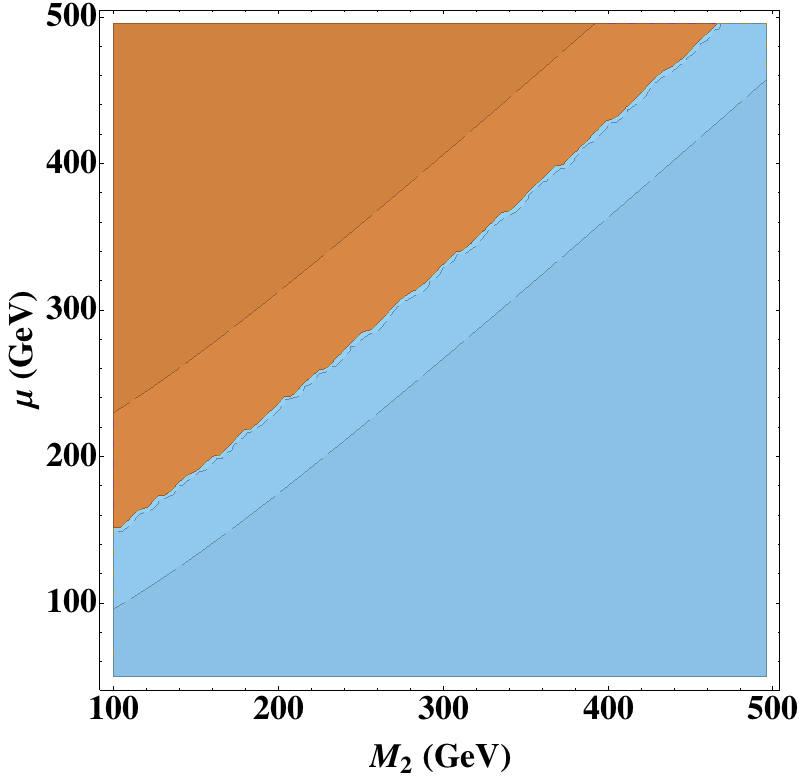}}
			\put(20,75){260 GeV}
		\end{picture}
		\label{}
	\end{subfigure}
	\begin{subfigure}{0.171\textwidth}
		\begin{picture}(100,100)
			\put(0,0){\includegraphics[width=\textwidth]{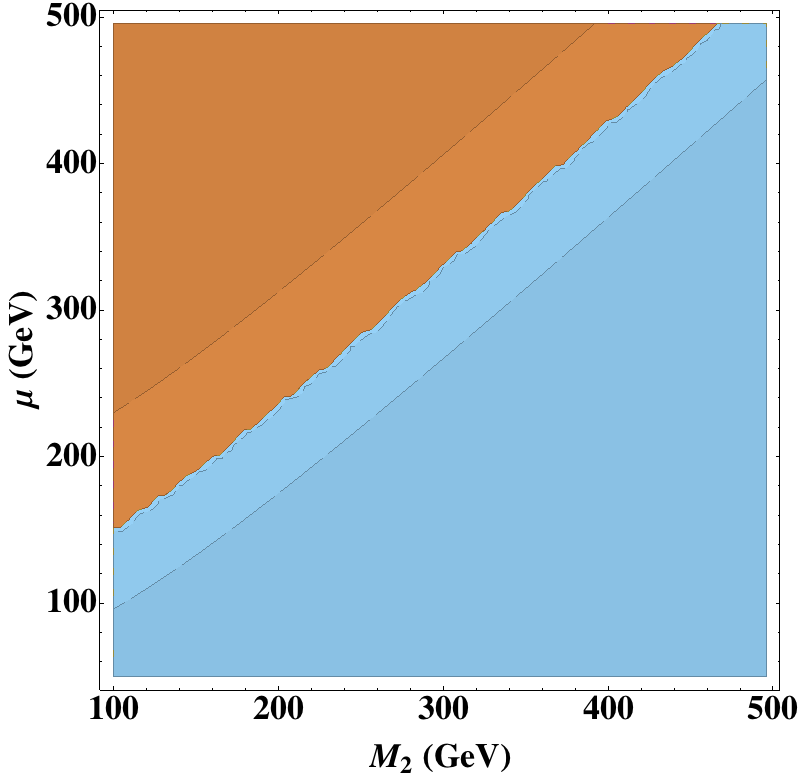}}
			\put(20,75){340 GeV}
		\end{picture}
		\label{}
	\end{subfigure}
	\begin{subfigure}{0.1\textwidth}
		\begin{picture}(100,100)
			\put(0,25){\includegraphics[width=\textwidth]{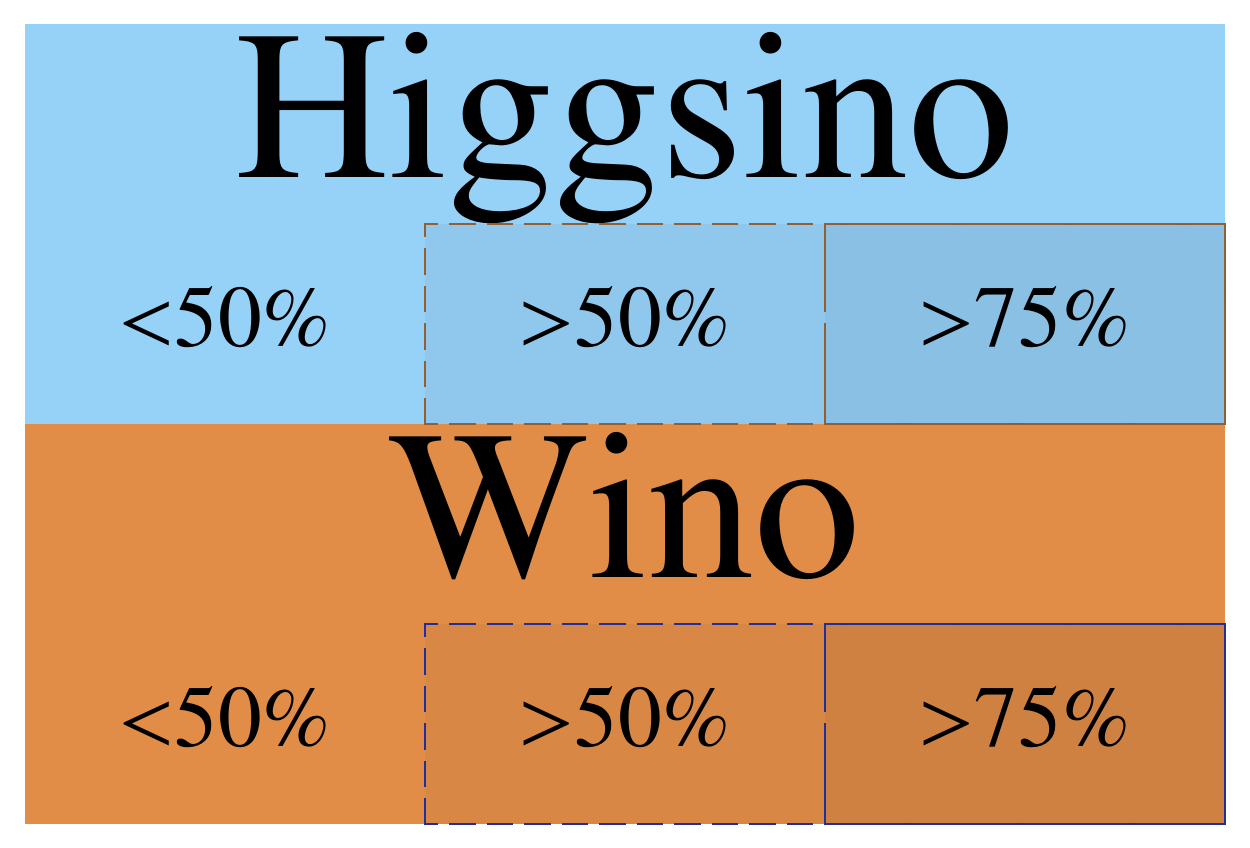}}
			\put(0,75){$M_1$}
		\end{picture}			
		\label{}
	\end{subfigure}
	\vspace{-2mm}  \\
	\begin{subfigure}{0.171\textwidth}
		\includegraphics[width=\textwidth]{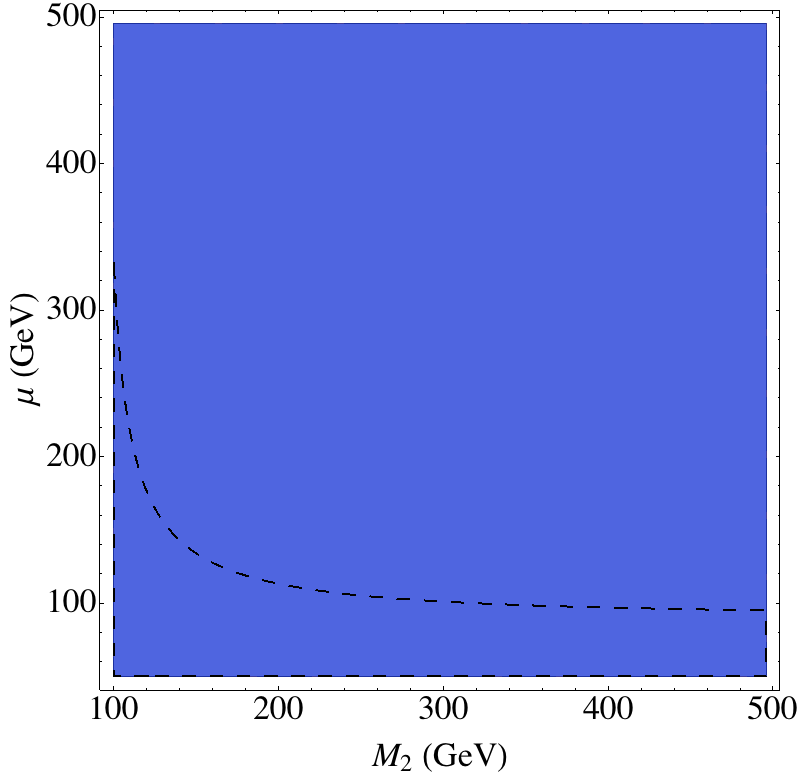}
		\label{}
	\end{subfigure}
	\begin{subfigure}{0.171\textwidth}
		\includegraphics[width=\textwidth]{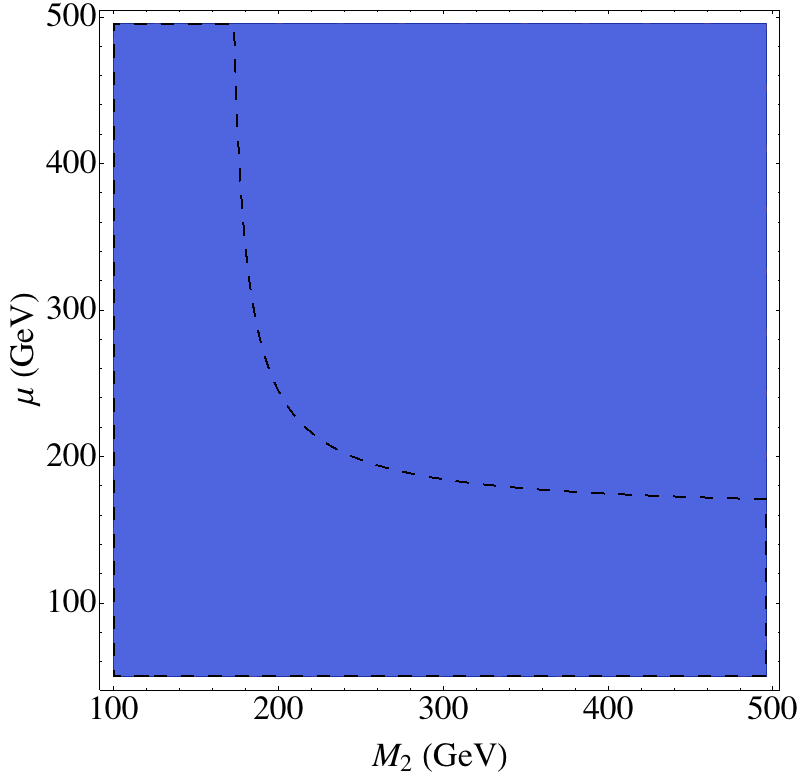}
		\label{}
	\end{subfigure}
	\begin{subfigure}{0.171\textwidth}
		\includegraphics[width=\textwidth]{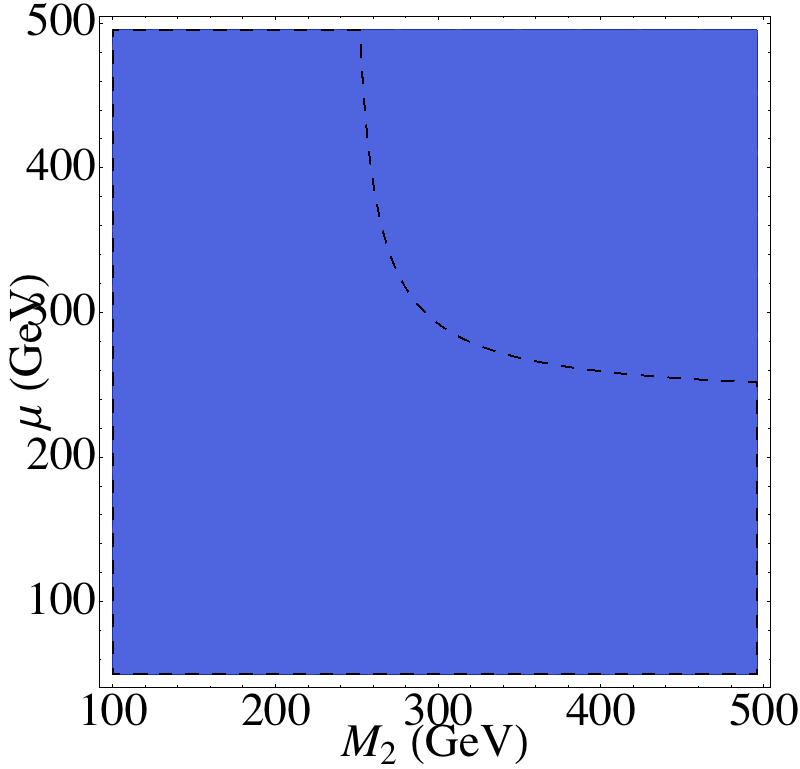}
		\label{}
	\end{subfigure}
	\begin{subfigure}{0.171\textwidth}
		\includegraphics[width=\textwidth]{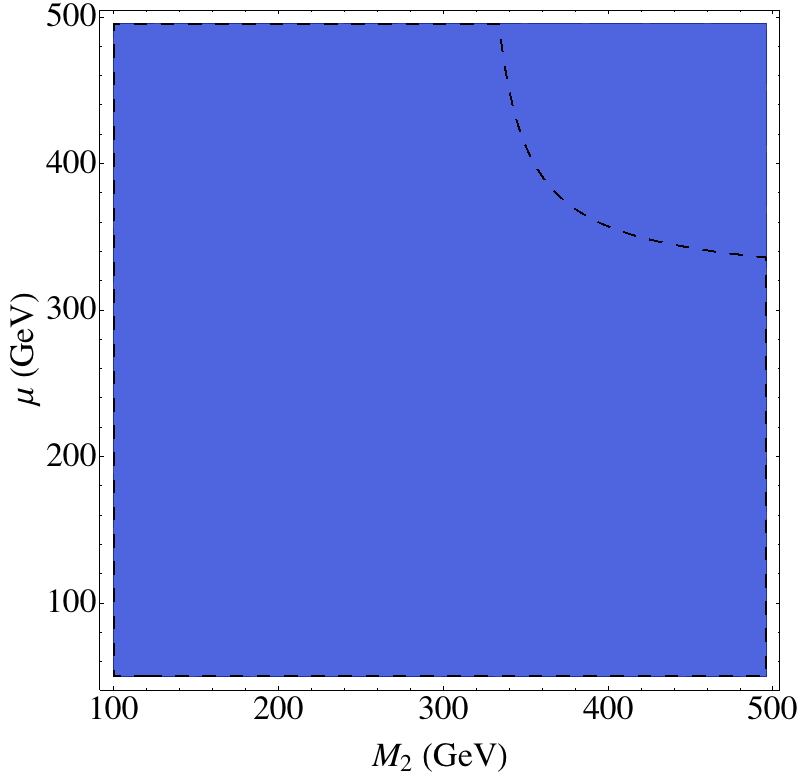}
		\label{}
	\end{subfigure}
	\begin{subfigure}{0.171\textwidth}
		\includegraphics[width=\textwidth]{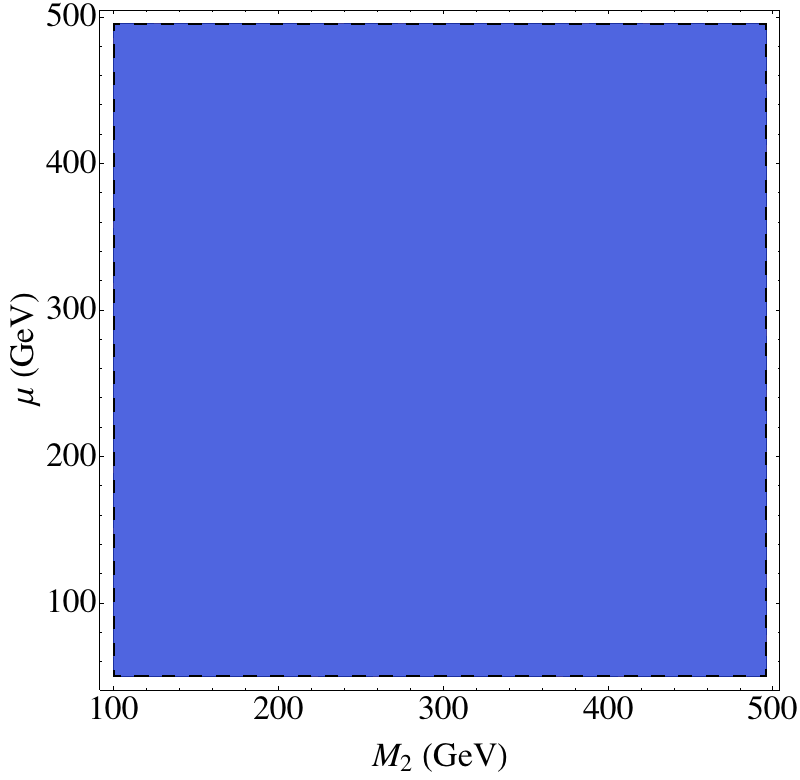}
		\label{}
	\end{subfigure}
	\begin{subfigure}{0.1\textwidth}
		\includegraphics[width=\textwidth]{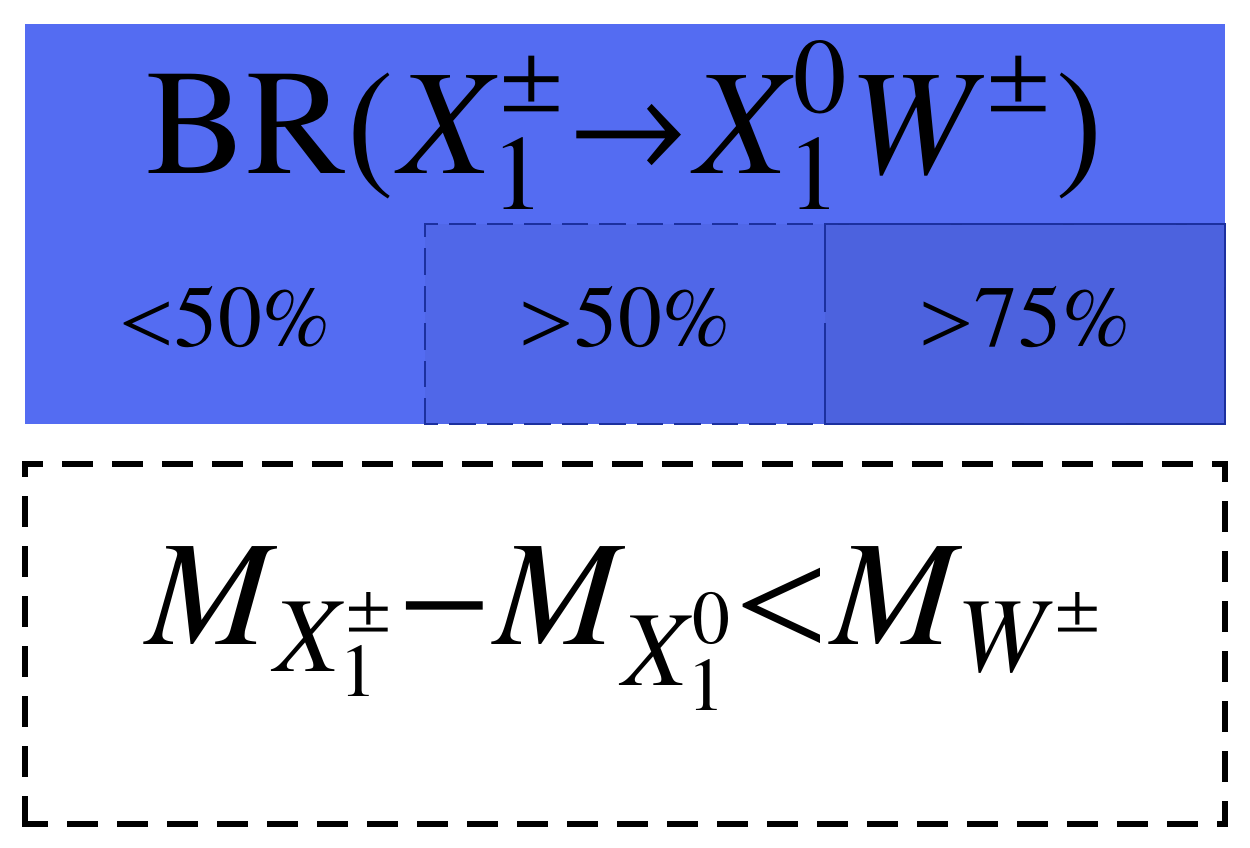}
		\label{}
	\end{subfigure}
	\vspace{-4mm}
	\caption{$\mathbf{\x:}$
Dominant gauge eigenstate content (top) and leading decay modes (bottom)
of the $\x$ chargino in the $M_2$--$\mu$ plane for various slices of 
$M_1$ and $\tan\beta=10$.  
The thick, dashed lines indicate where the corresponding
decay only occurs with an off-shell vector boson. 
Shaded, dash-enclosed regions indicate the boundary 
of 50\% and 75\% composition/branching ratio, as noted in the legend.
}
	\label{fig:x1summary}
\vspace{-0.1cm}
\end{figure}

\begin{figure}[ttt]
	\begin{subfigure}{0.171\textwidth}
		\begin{picture}(100,100)
			\put(0,0){\includegraphics[width=\textwidth]{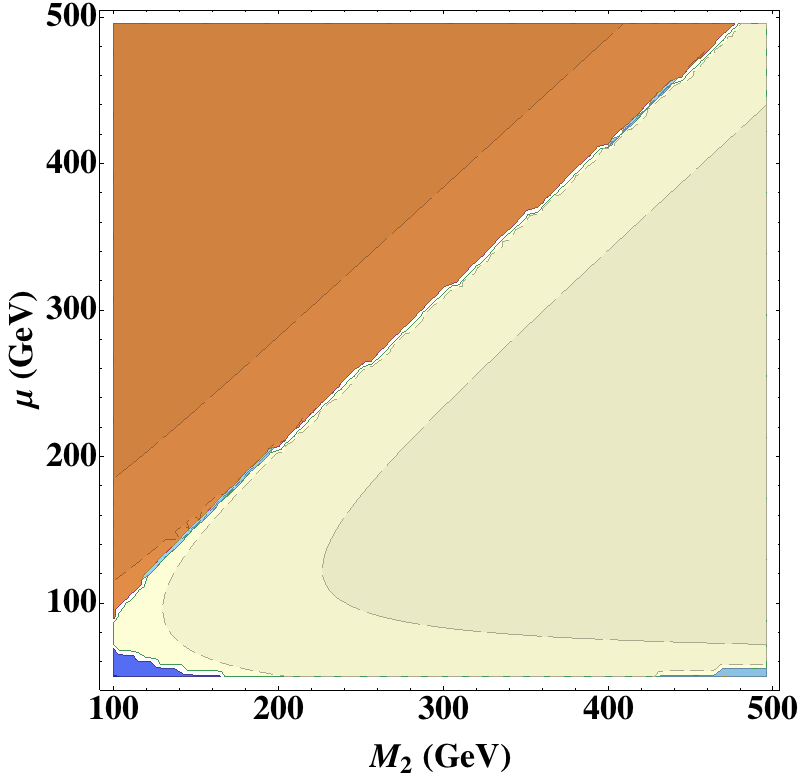}}
			\put(20,75){20 GeV}
		\end{picture}
		\label{}
	\end{subfigure}
	\begin{subfigure}{0.171\textwidth}
		\begin{picture}(100,100)
			\put(0,0){\includegraphics[width=\textwidth]{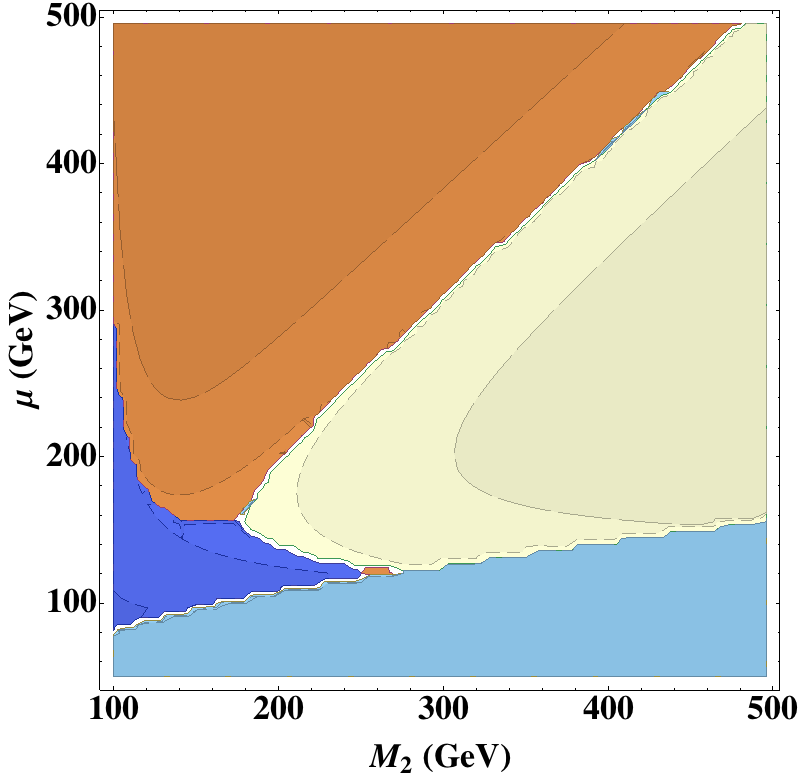}}
			\put(20,75){100 GeV}
		\end{picture}
		\label{}
	\end{subfigure}
	\begin{subfigure}{0.171\textwidth}
		\begin{picture}(100,100)
			\put(0,0){\includegraphics[width=\textwidth]{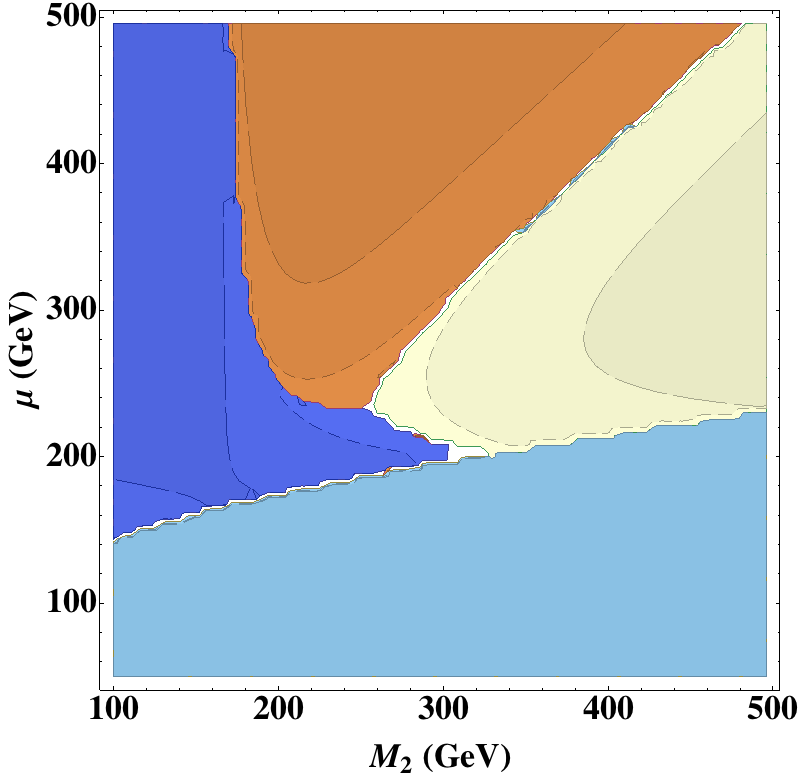}}
			\put(20,75){180 GeV}
		\end{picture}
		\label{}
	\end{subfigure}
	\begin{subfigure}{0.171\textwidth}
		\begin{picture}(100,100)
			\put(0,0){\includegraphics[width=\textwidth]{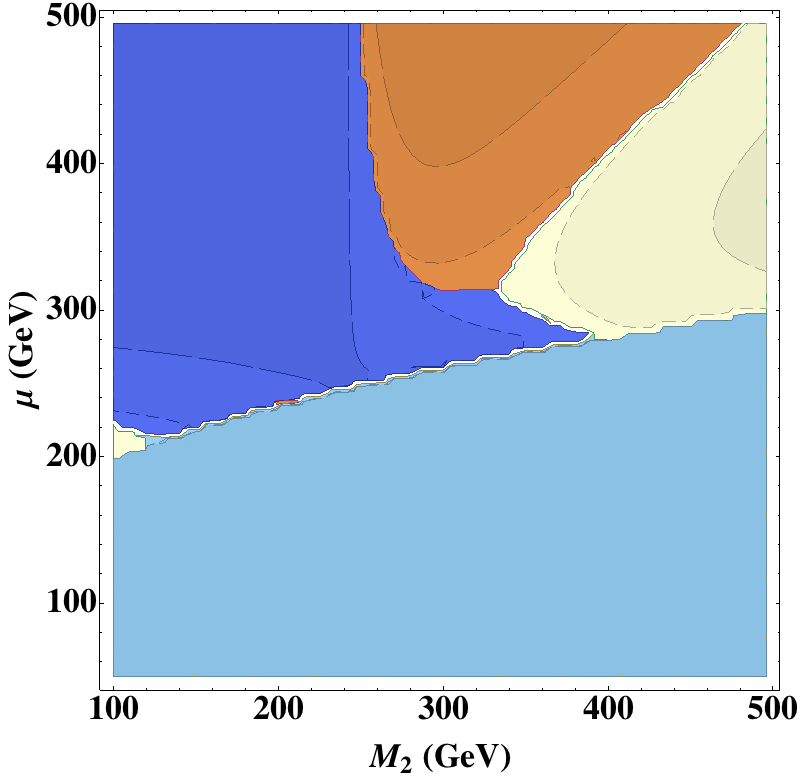}}
			\put(20,75){260 GeV}
		\end{picture}
		\label{}
	\end{subfigure}
	\begin{subfigure}{0.171\textwidth}
		\begin{picture}(100,100)
			\put(0,0){\includegraphics[width=\textwidth]{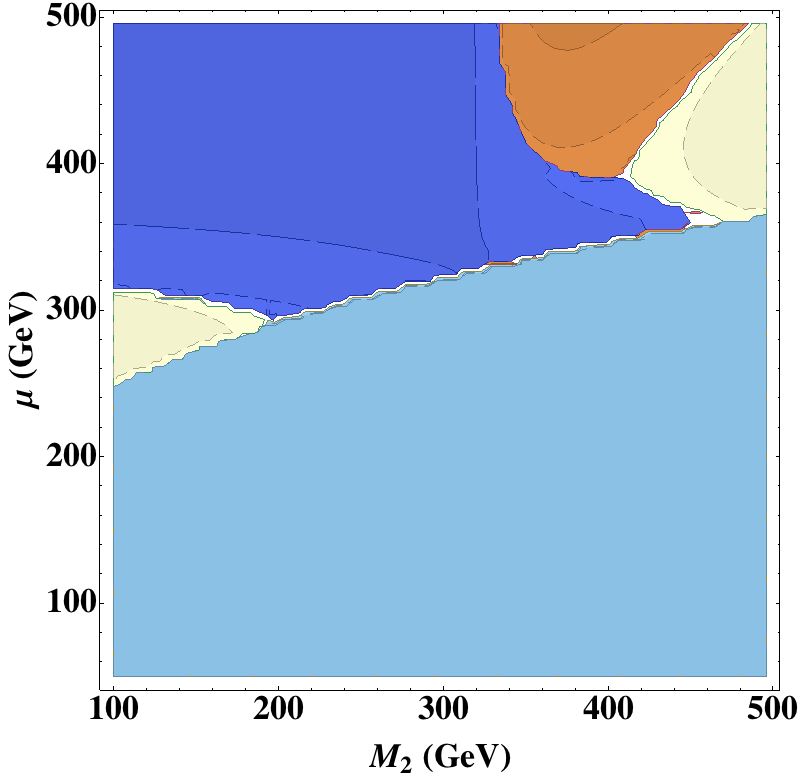}}
			\put(20,75){340 GeV}
		\end{picture}
		\label{}
	\end{subfigure}
	\begin{subfigure}{0.1\textwidth}
		\begin{picture}(100,100)
			\put(0,10){\includegraphics[width=\textwidth]{nstatelegend.pdf}}
			\put(0,75){$M_1$}
		\end{picture}			
		\label{}
	\end{subfigure}
	\vspace{-4mm} \\
	\begin{subfigure}{0.171\textwidth}
		\includegraphics[width=\textwidth]{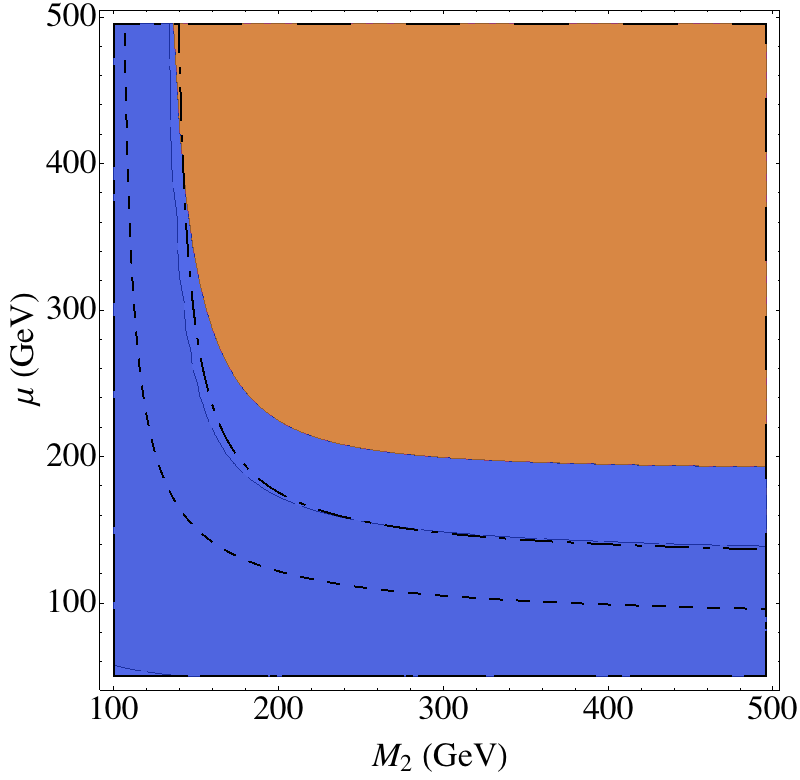}
		\label{}
	\end{subfigure}
	\begin{subfigure}{0.171\textwidth}
		\includegraphics[width=\textwidth]{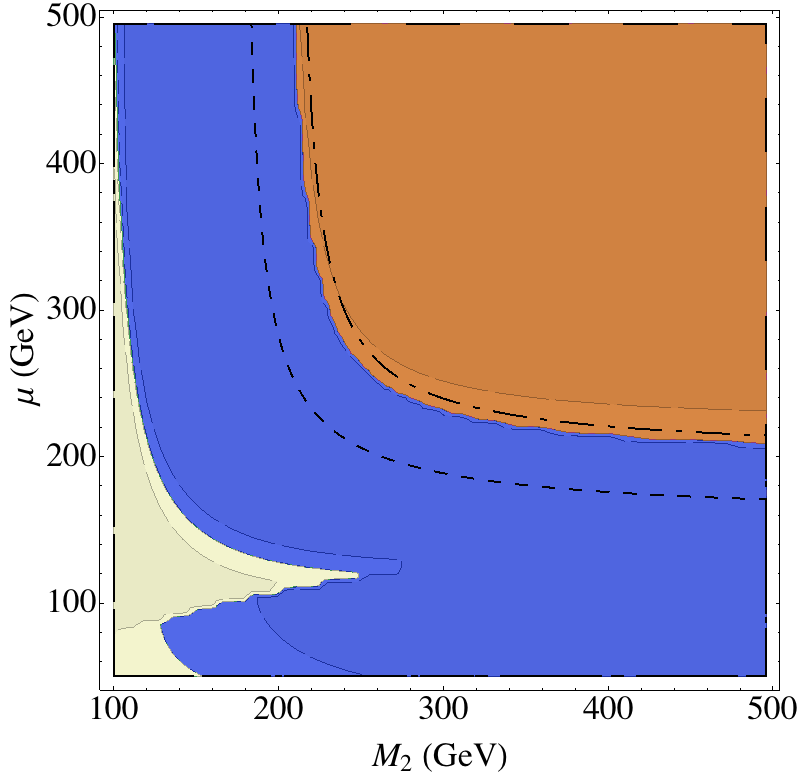}
		\label{}
	\end{subfigure}
	\begin{subfigure}{0.171\textwidth}
		\includegraphics[width=\textwidth]{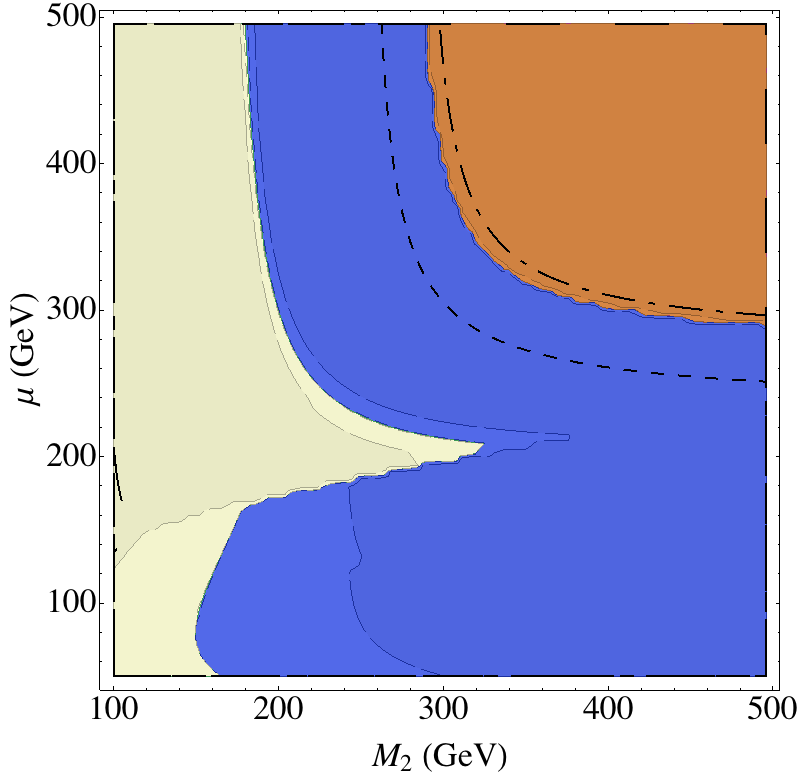}
		\label{}
	\end{subfigure}
	\begin{subfigure}{0.171\textwidth}
		\includegraphics[width=\textwidth]{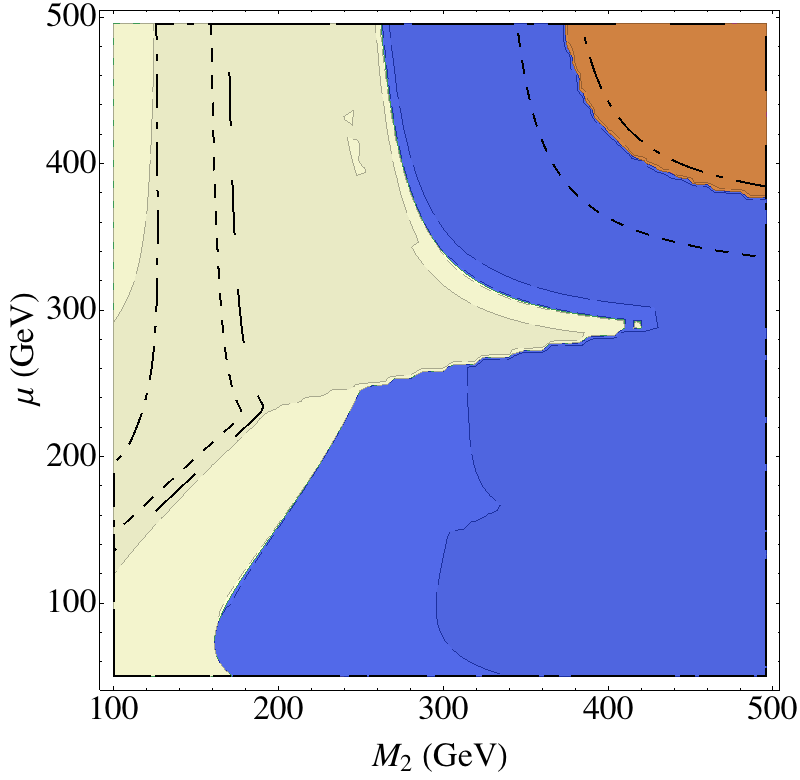}
		\label{}
	\end{subfigure}
	\begin{subfigure}{0.171\textwidth}
		\includegraphics[width=\textwidth]{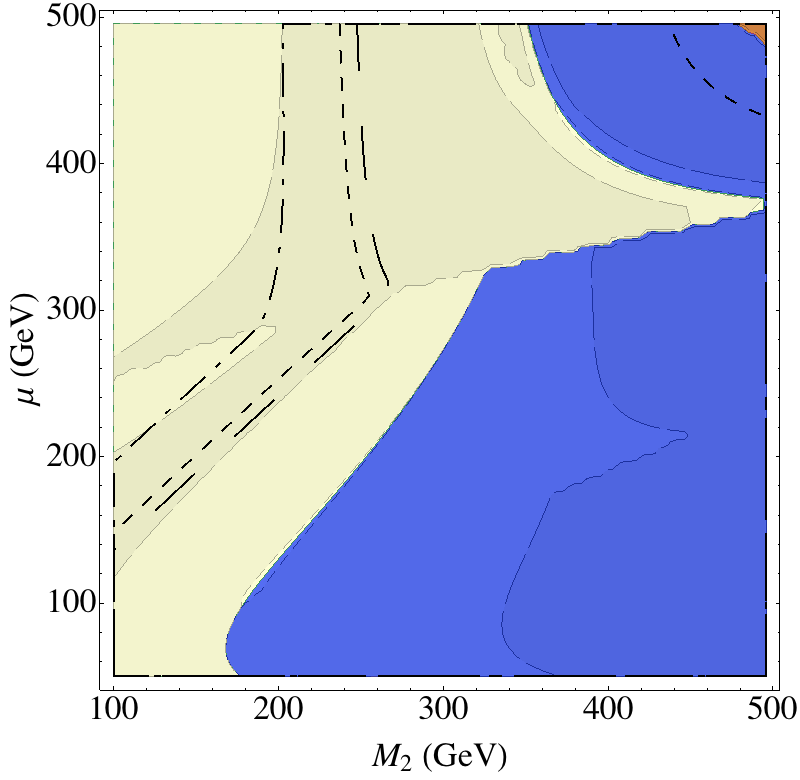}
		\label{}
	\end{subfigure}
	\begin{subfigure}{0.1\textwidth}
		\includegraphics[width=\textwidth]{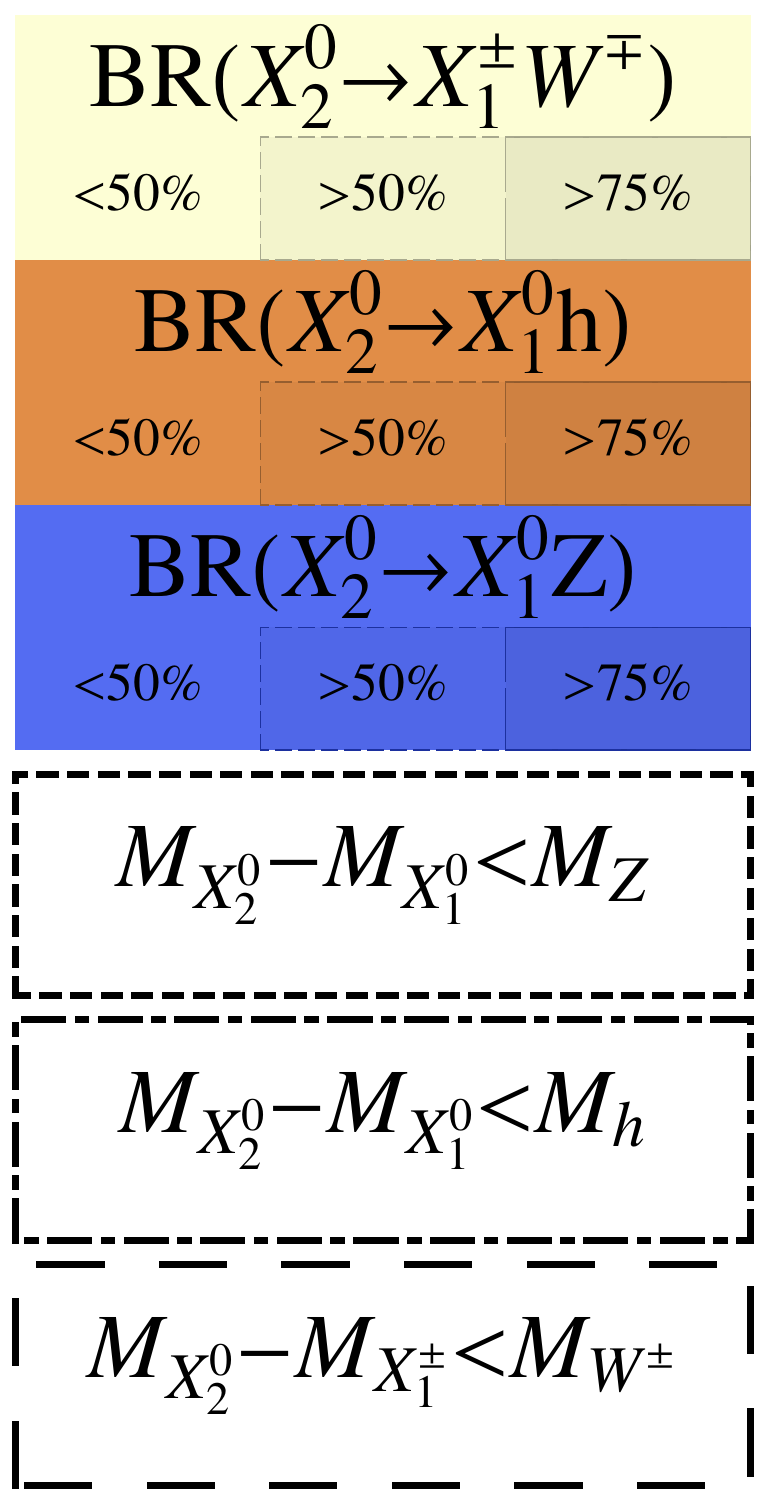}
		\label{}
	\end{subfigure}
	\vspace{-4mm}
	\caption{$\mathbf{\nn:}$ 
Dominant gauge eigenstate content (top) and leading decay modes (bottom)
of the $\nn$ neutralino in the $M_2$--$\mu$ plane for various slices of 
$M_1$ and $\tan\beta=10$.  The thick, dashed and dotted lines indicate where 
the corresponding decays only occur with an off-shell vector boson. Shaded, dash-enclosed regions indicate
the boundary of 50\% and 75\% composition/branching ratio, as noted in the legend.
}
	\label{fig:n2summary}
\vspace{-0.1cm}
\end{figure}

\begin{figure}[ttt]
	\begin{subfigure}{0.171\textwidth}
		\begin{picture}(100,100)
			\put(0,0){\includegraphics[width=\textwidth]{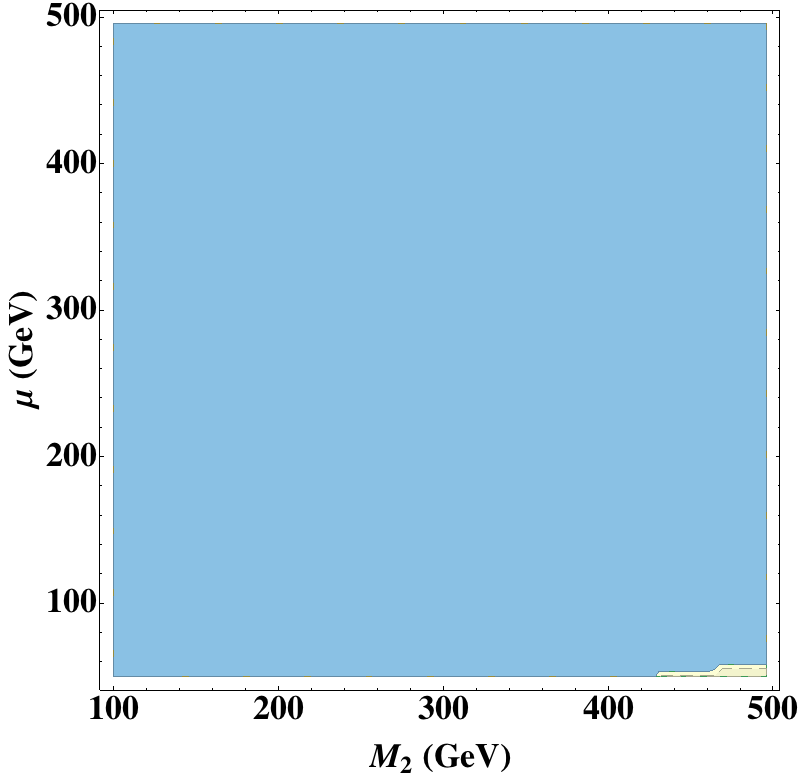}}
			\put(20,75){20 GeV}
		\end{picture}
		\label{}
	\end{subfigure}
	\begin{subfigure}{0.171\textwidth}
		\begin{picture}(100,100)
			\put(0,0){\includegraphics[width=\textwidth]{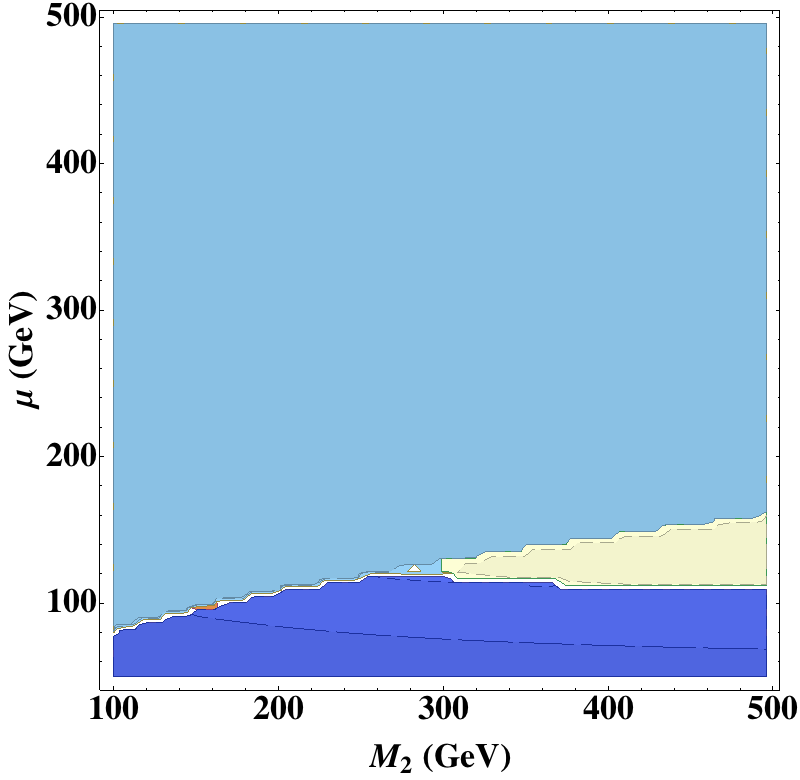}}
			\put(20,75){100 GeV}
		\end{picture}
		\label{}
	\end{subfigure}
	\begin{subfigure}{0.171\textwidth}
		\begin{picture}(100,100)
			\put(0,0){\includegraphics[width=\textwidth]{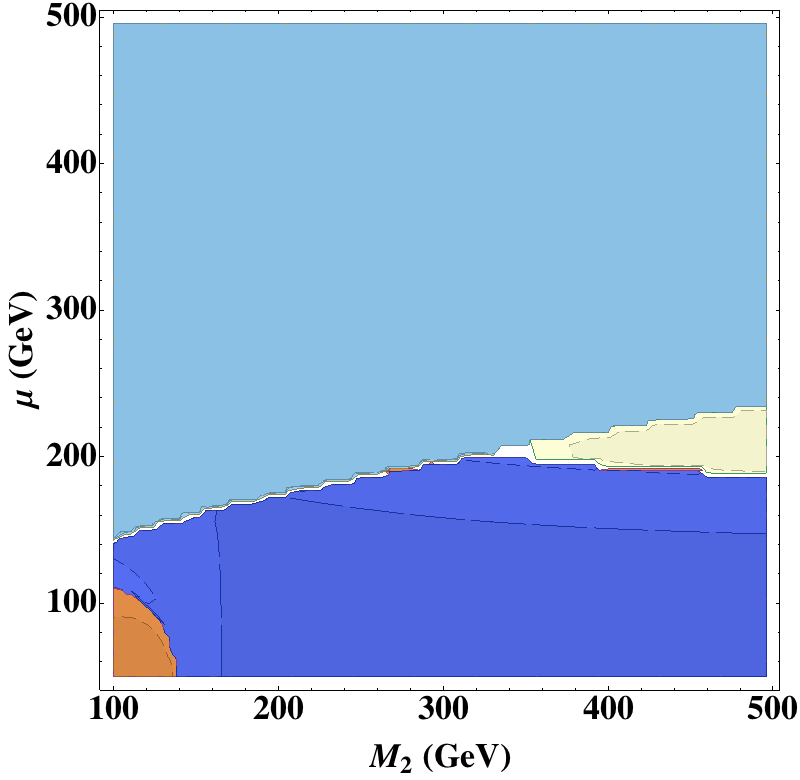}}
			\put(20,75){180 GeV}
		\end{picture}
		\label{}
	\end{subfigure}
	\begin{subfigure}{0.171\textwidth}
		\begin{picture}(100,100)
			\put(0,0){\includegraphics[width=\textwidth]{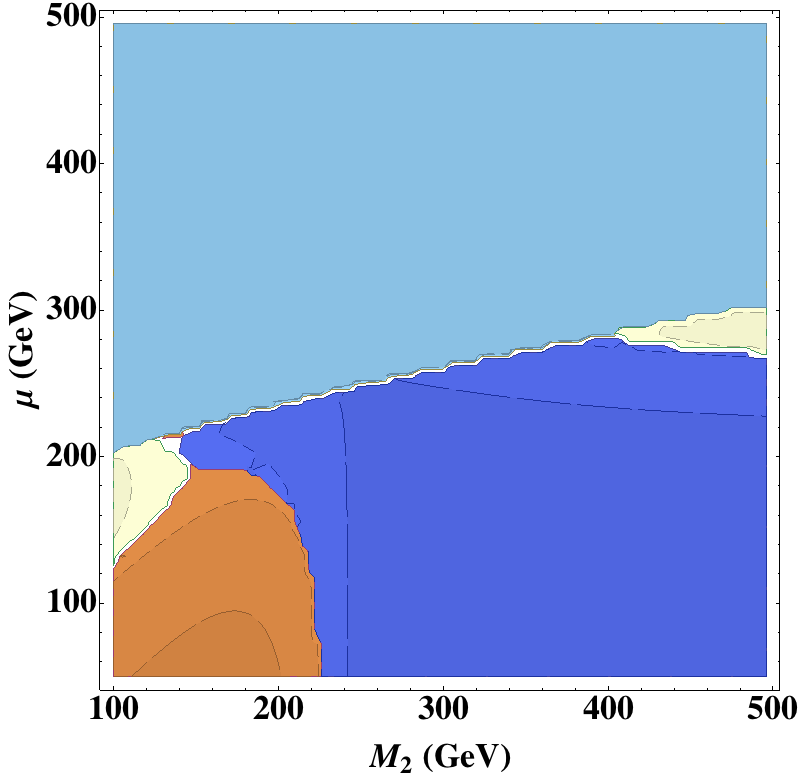}}
			\put(20,75){260 GeV}
		\end{picture}
		\label{}
	\end{subfigure}
	\begin{subfigure}{0.171\textwidth}
		\begin{picture}(100,100)
			\put(0,0){\includegraphics[width=\textwidth]{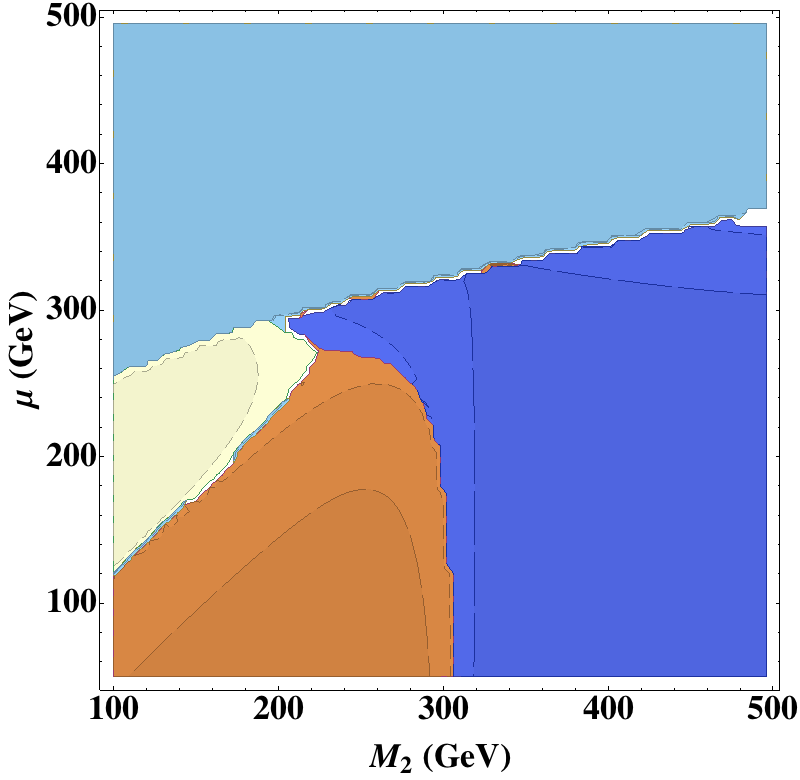}}
			\put(20,75){340 GeV}
		\end{picture}
		\label{}
	\end{subfigure}
	\begin{subfigure}{0.1\textwidth}
		\begin{picture}(100,100)
			\put(0,10){\includegraphics[width=\textwidth]{nstatelegend.pdf}}
			\put(0,75){$M_1$}
		\end{picture}			
		\label{}
	\end{subfigure}	
	\vspace{-4mm}\\
	\begin{subfigure}{0.171\textwidth}
		\includegraphics[width=\textwidth]{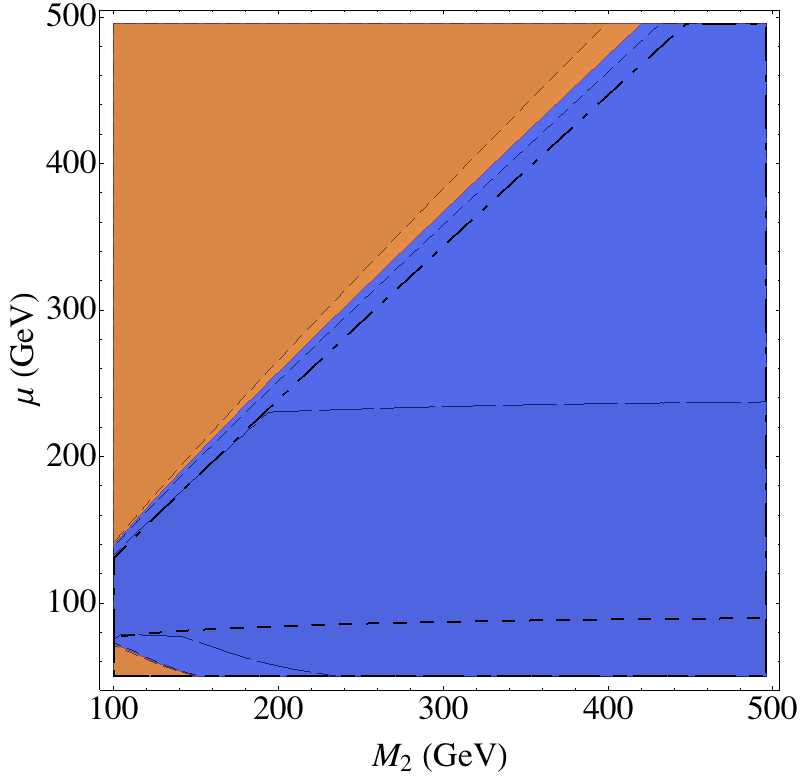}
		\label{}
	\end{subfigure}
	\begin{subfigure}{0.171\textwidth}
		\includegraphics[width=\textwidth]{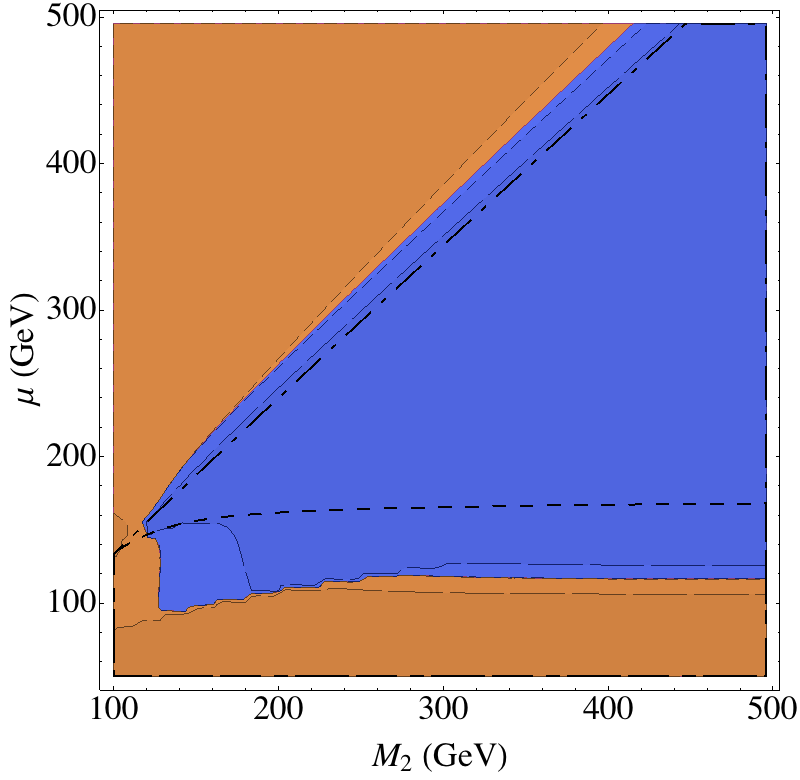}
		\label{}
	\end{subfigure}
	\begin{subfigure}{0.171\textwidth}
		\includegraphics[width=\textwidth]{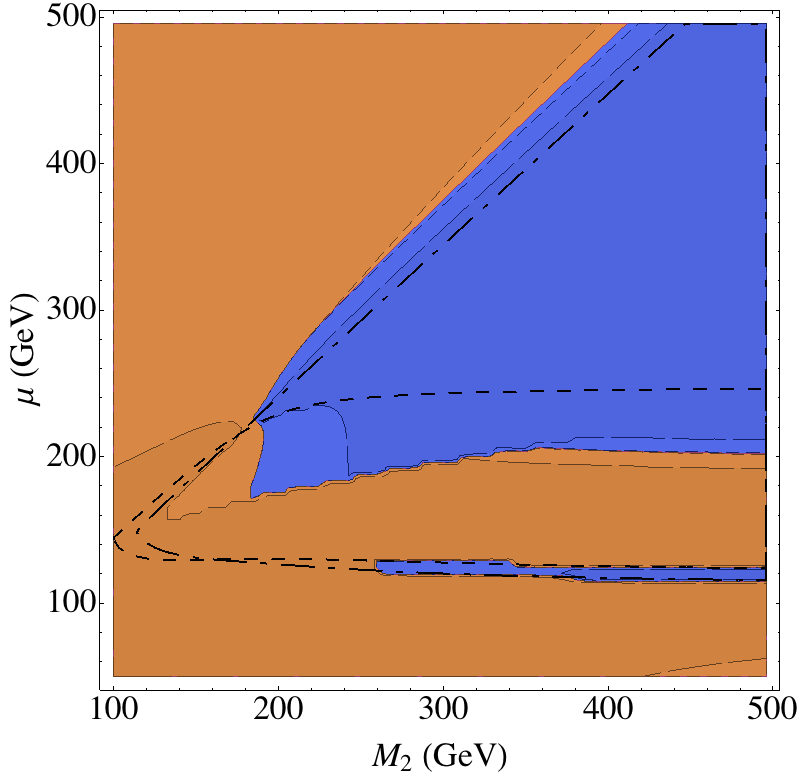}
		\label{}
	\end{subfigure}
	\begin{subfigure}{0.171\textwidth}
		\includegraphics[width=\textwidth]{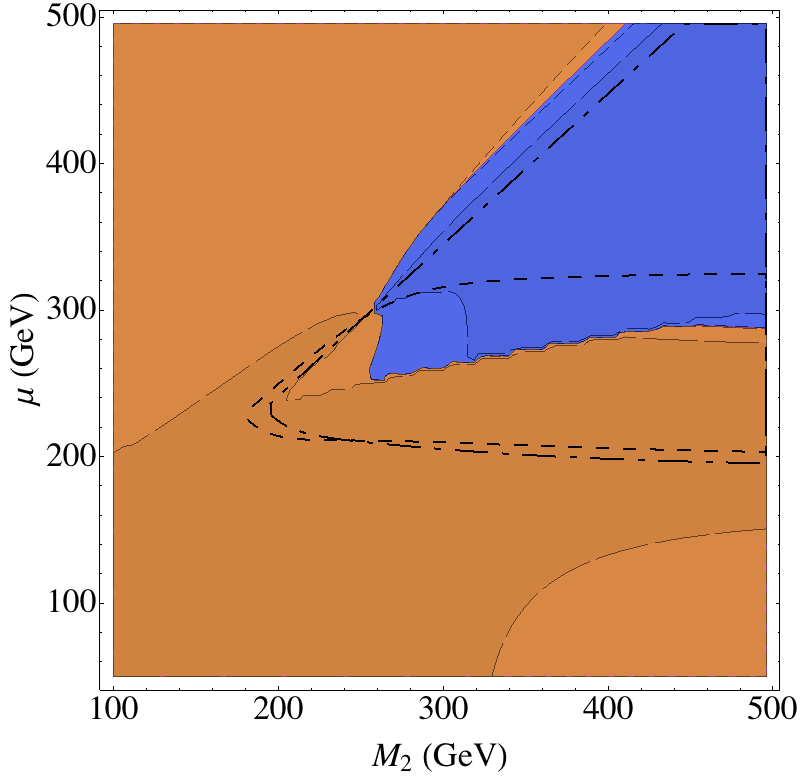}
		\label{}
	\end{subfigure}
	\begin{subfigure}{0.171\textwidth}
		\includegraphics[width=\textwidth]{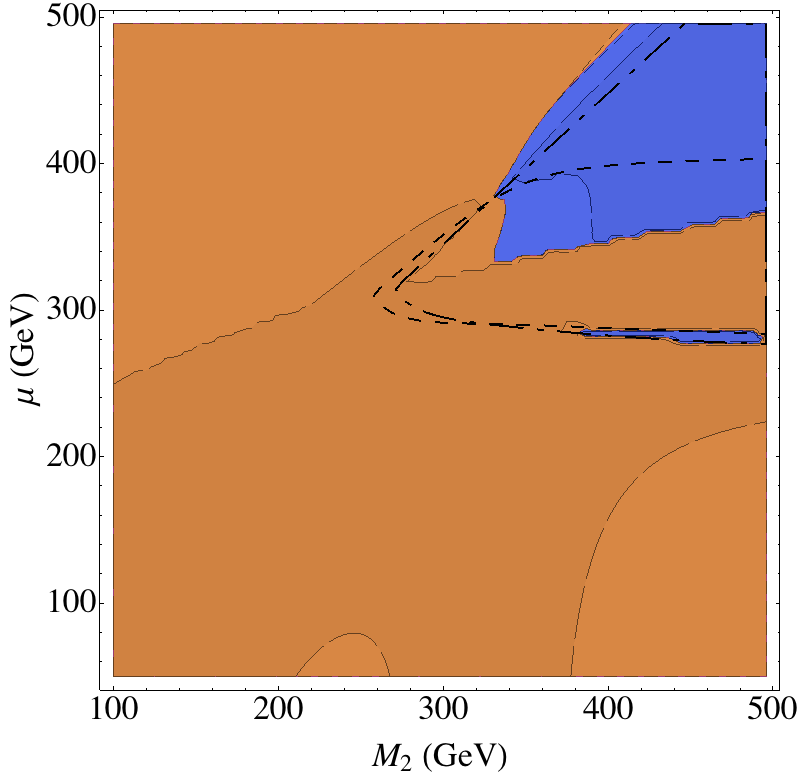}
		\label{}
	\end{subfigure}
	\begin{subfigure}{0.1\textwidth}
		\includegraphics[width=\textwidth]{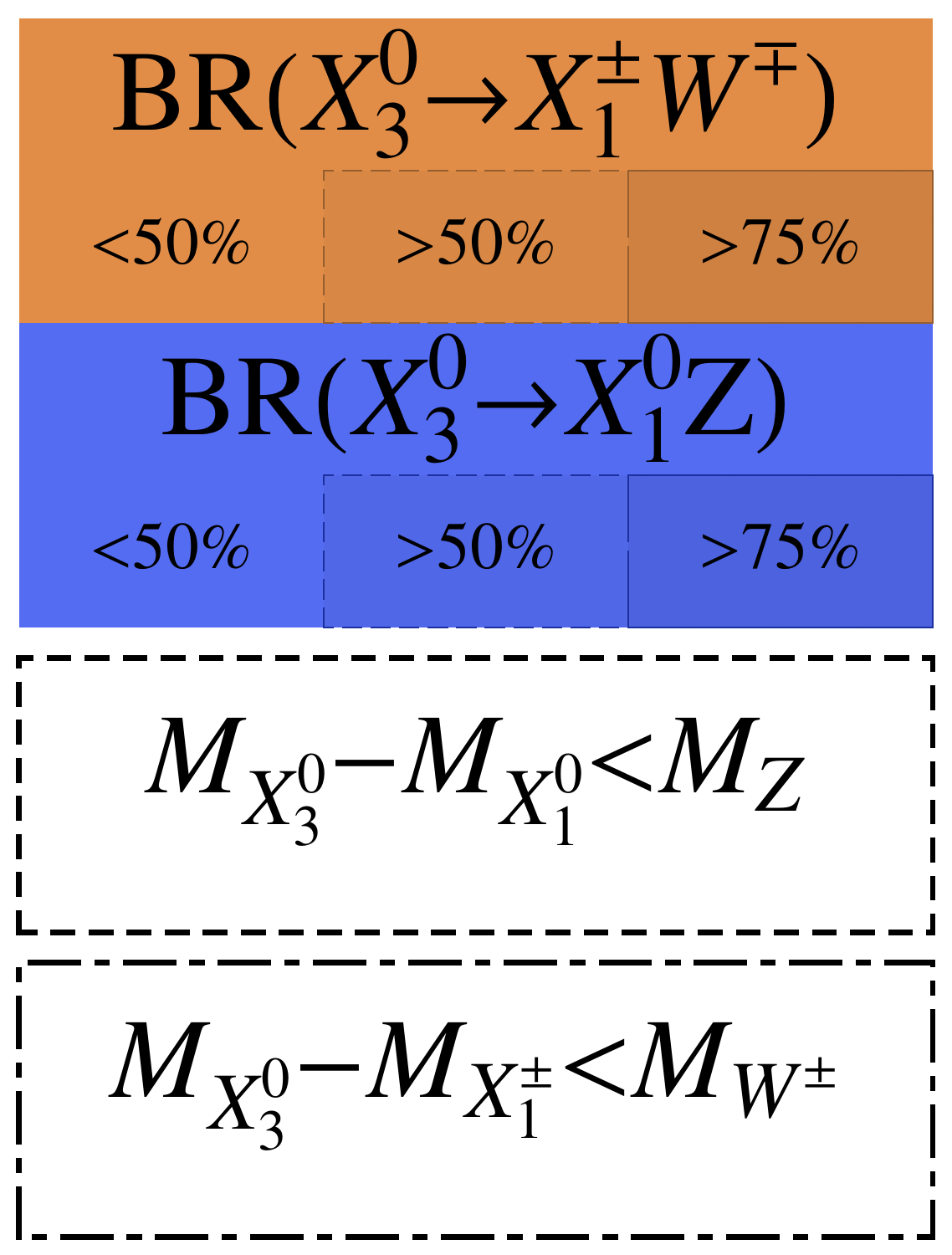}
		\label{}
	\end{subfigure}
	\vspace{-4mm}
	\caption{$\mathbf{\nnn:}$
Dominant gauge eigenstate content (top) and leading decay modes (bottom)
of the $\nnn$ neutralino in the $M_2$--$\mu$ plane for various slices of 
$M_1$ and $\tan\beta=10$.  The thick, dashed and dotted lines indicate where 
the corresponding decays only occur with an off-shell vector boson. Shaded, dash-enclosed regions indicate
the boundary of 50\% and 75\% composition/branching ratio, as noted in the legend.
}
\vspace{-0.3cm} 
	\label{fig:n3summary}
\end{figure}

\begin{figure}[ttt]
	\begin{subfigure}{0.171\textwidth}
		\begin{picture}(100,100)
			\put(0,0){\includegraphics[width=\textwidth]{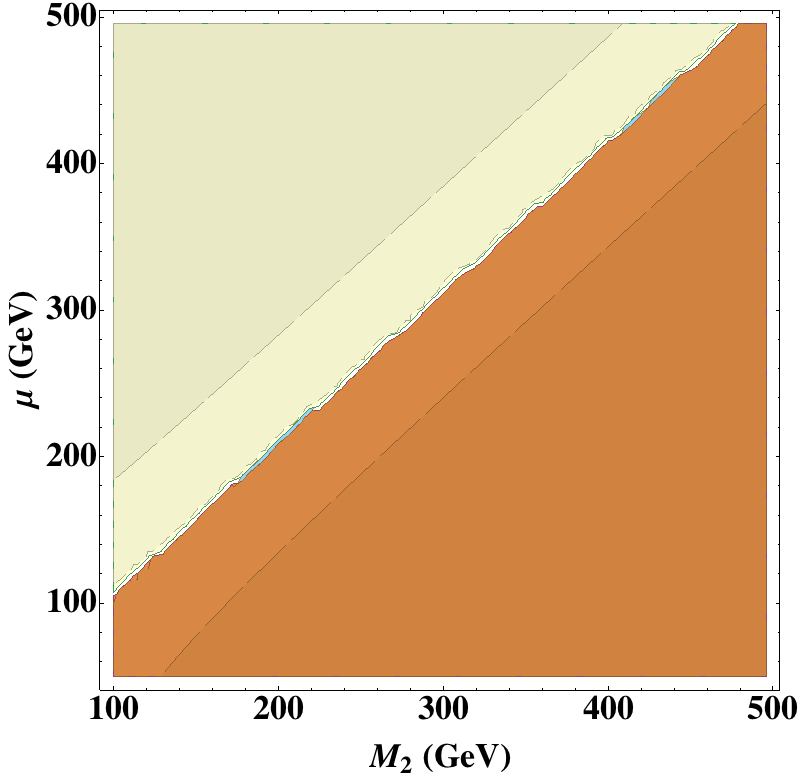}}
			\put(20,75){20 GeV}
		\end{picture}
		\label{}
	\end{subfigure}
	\begin{subfigure}{0.171\textwidth}
		\begin{picture}(100,100)
			\put(0,0){\includegraphics[width=\textwidth]{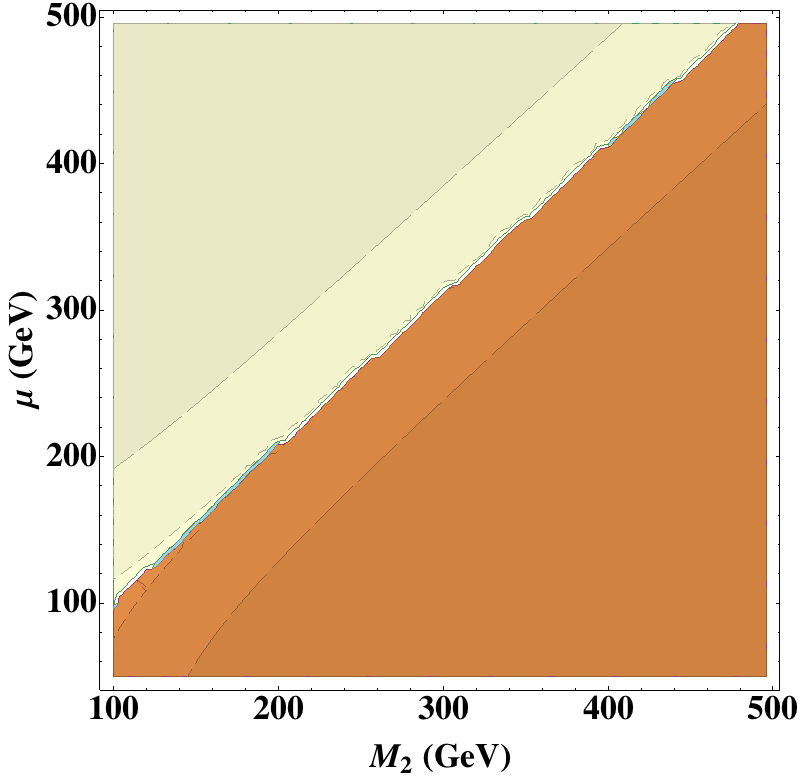}}
			\put(20,75){100 GeV}
		\end{picture}
		\label{}
	\end{subfigure}
	\begin{subfigure}{0.171\textwidth}
		\begin{picture}(100,100)
			\put(0,0){\includegraphics[width=\textwidth]{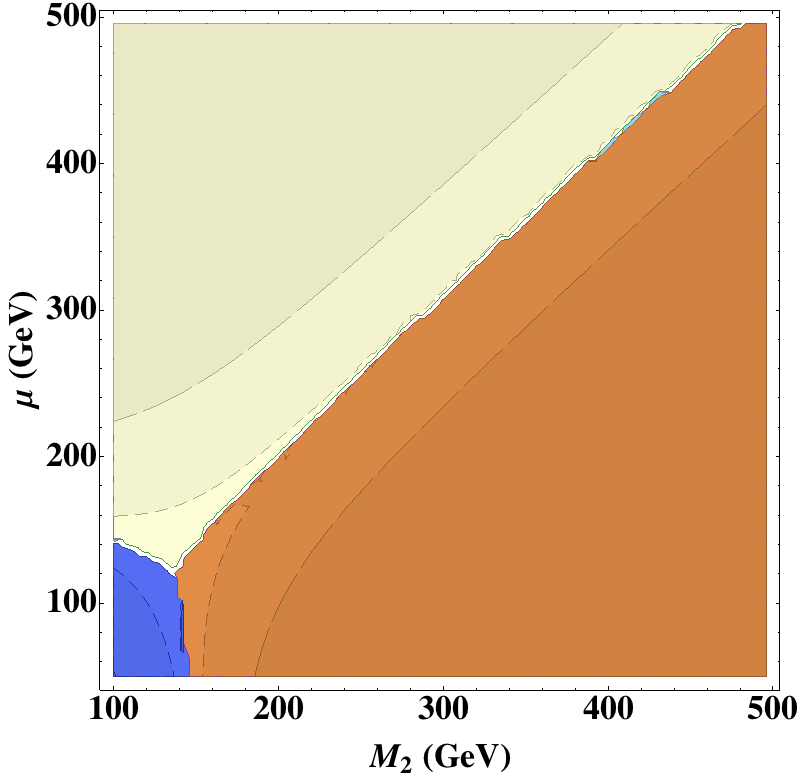}}
			\put(20,75){180 GeV}
		\end{picture}
		\label{}
	\end{subfigure}
	\begin{subfigure}{0.171\textwidth}
		\begin{picture}(100,100)
			\put(0,0){\includegraphics[width=\textwidth]{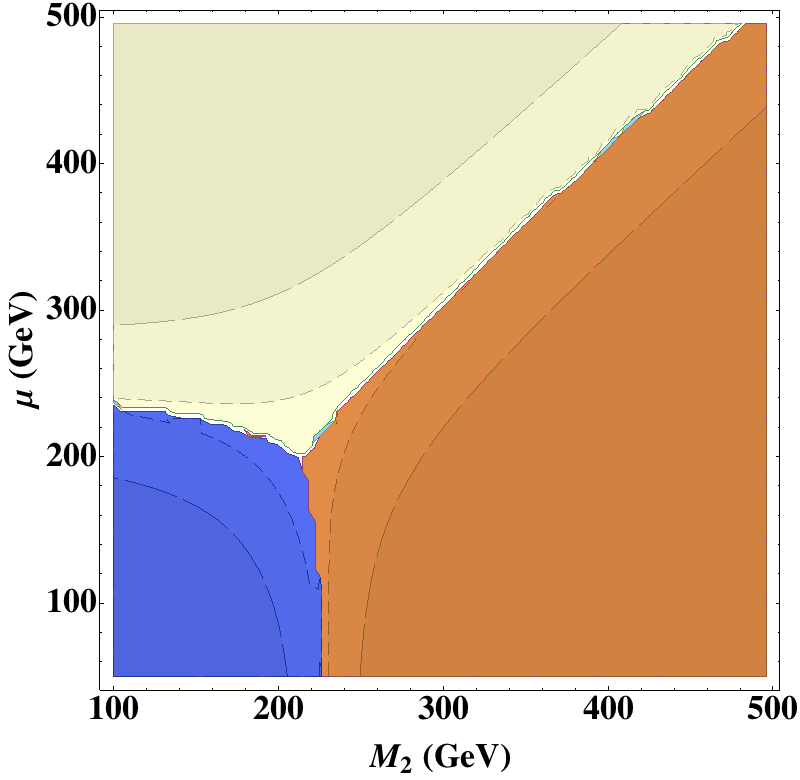}}
			\put(20,75){260 GeV}
		\end{picture}
		\label{}
	\end{subfigure}
	\begin{subfigure}{0.171\textwidth}
		\begin{picture}(100,100)
			\put(0,0){\includegraphics[width=\textwidth]{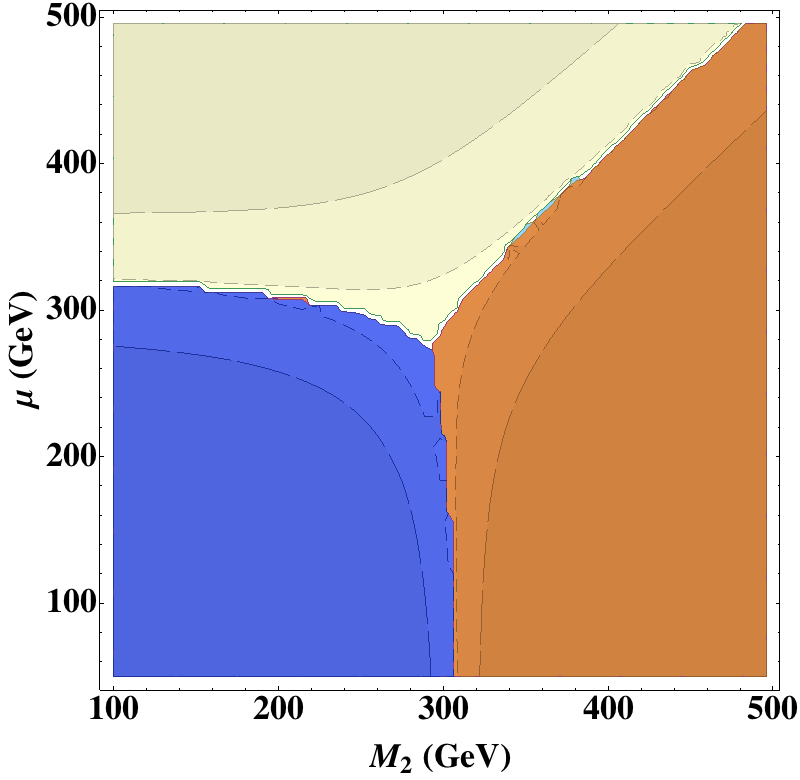}}
			\put(20,75){340 GeV}
		\end{picture}
		\label{}
	\end{subfigure}
	\begin{subfigure}{0.1\textwidth}
		\begin{picture}(100,100)
			\put(0,10){\includegraphics[width=\textwidth]{nstatelegend.pdf}}
			\put(0,75){$M_1$}
		\end{picture}			
		\label{}
	\end{subfigure}
	\vspace{-4mm} \\
	\begin{subfigure}{0.171\textwidth}
		\includegraphics[width=\textwidth]{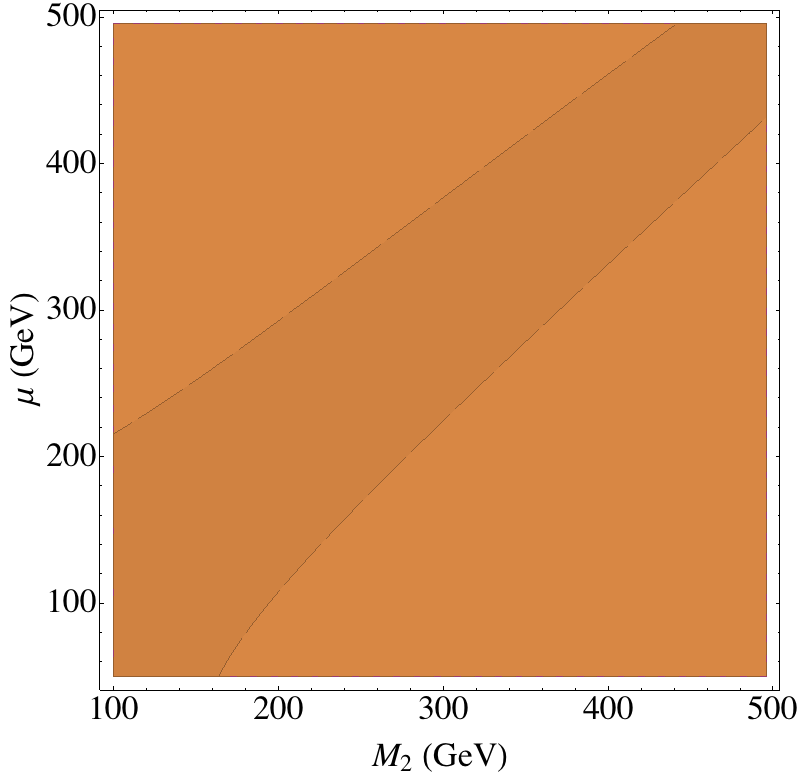}
		\label{}
	\end{subfigure}
	\begin{subfigure}{0.171\textwidth}
		\includegraphics[width=\textwidth]{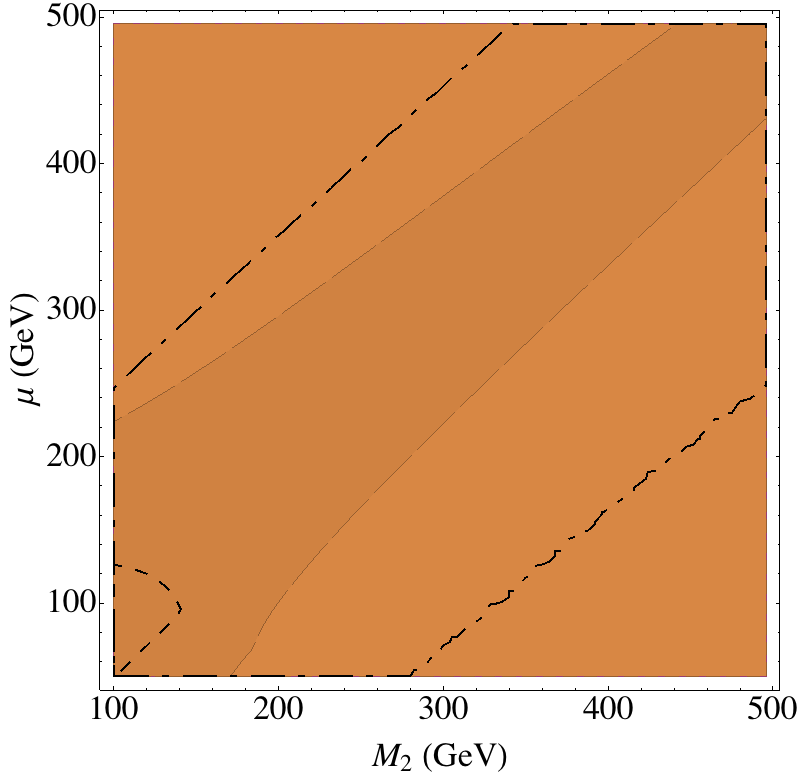}
		\label{}
	\end{subfigure}
	\begin{subfigure}{0.171\textwidth}
		\includegraphics[width=\textwidth]{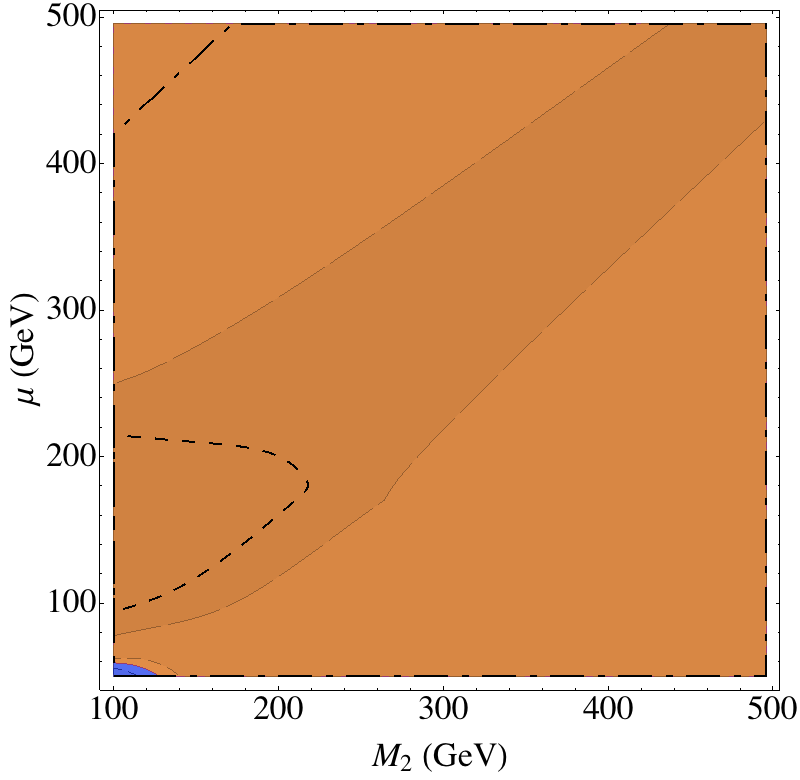}
		\label{}
	\end{subfigure}
	\begin{subfigure}{0.171\textwidth}
		\includegraphics[width=\textwidth]{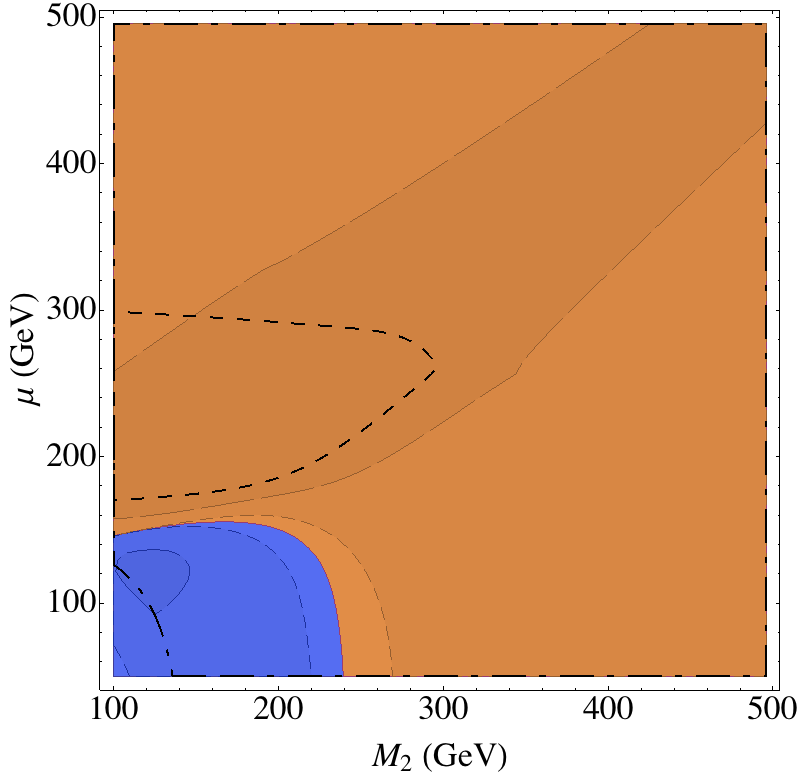}
		\label{}
	\end{subfigure}
	\begin{subfigure}{0.171\textwidth}
		\includegraphics[width=\textwidth]{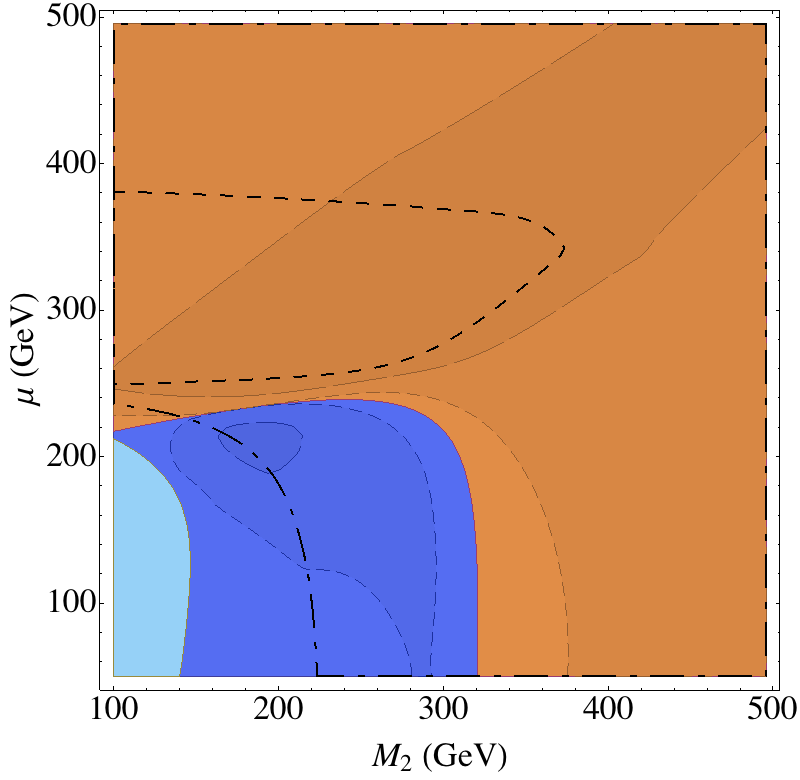}
		\label{}
	\end{subfigure}
	\begin{subfigure}{0.1\textwidth}
		\includegraphics[width=\textwidth]{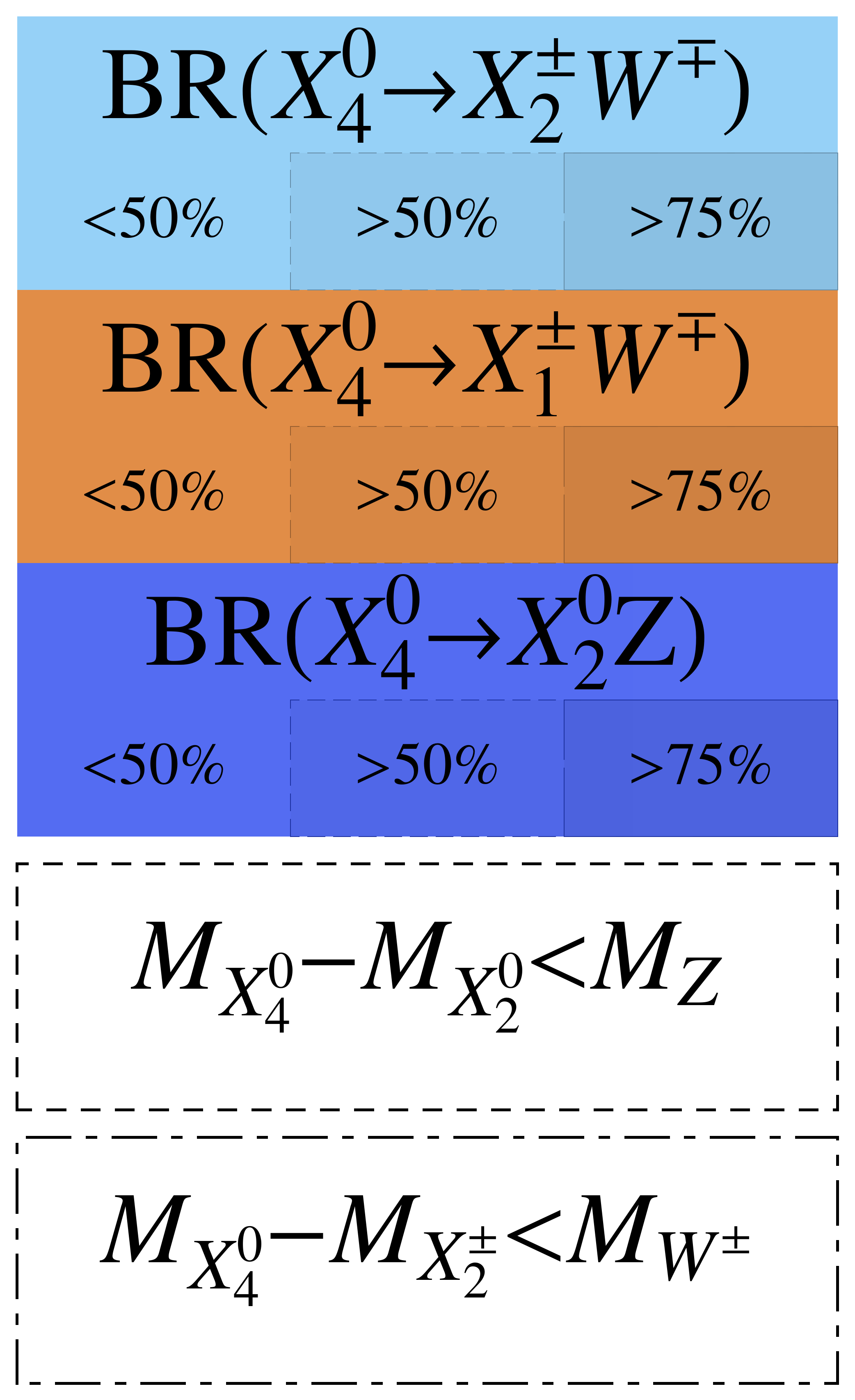}
		\label{}
	\end{subfigure}
	\vspace{-4mm}
	\caption{$\mathbf{\nnnn:}$
Dominant gauge eigenstate content (top) and leading decay modes (bottom)
of the $\nnnn$ neutralino in the $M_2$--$\mu$ plane for various slices of 
$M_1$ and $\tan\beta=10$.  The thick, dashed and dotted lines indicate where 
the corresponding decays only occur with an off-shell vector boson. Shaded, dash-enclosed regions indicate
the boundary of 50\% and 75\% composition/branching ratio, as noted in the legend.
}
	\label{fig:n4summary}
\vspace{-0.3cm}
\end{figure}

\begin{figure}
	\begin{subfigure}{0.171\textwidth}
		\begin{picture}(100,100)
			\put(0,0){\includegraphics[width=\textwidth]{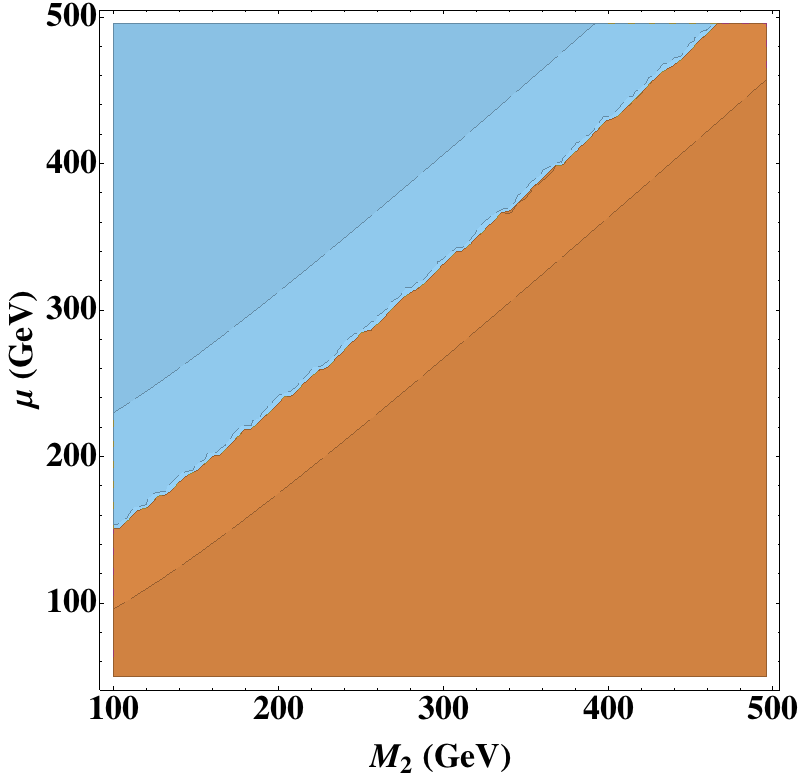}}
			\put(20,75){20 GeV}
		\end{picture}
		\label{}
	\end{subfigure}
	\begin{subfigure}{0.171\textwidth}
		\begin{picture}(100,100)
			\put(0,0){\includegraphics[width=\textwidth]{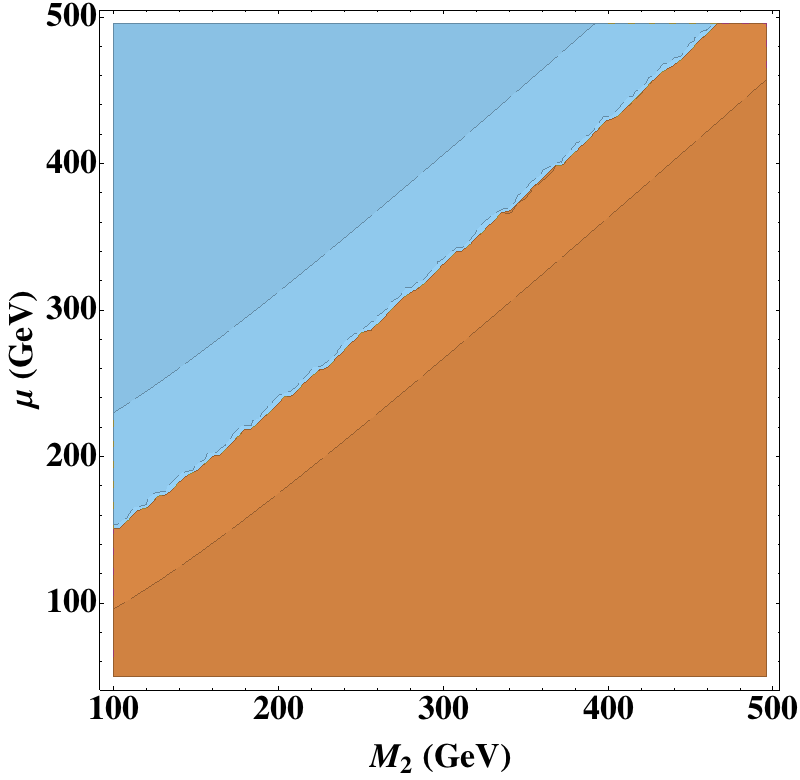}}
			\put(20,75){100 GeV}
		\end{picture}
		\label{}
	\end{subfigure}
	\begin{subfigure}{0.171\textwidth}
		\begin{picture}(100,100)
			\put(0,0){\includegraphics[width=\textwidth]{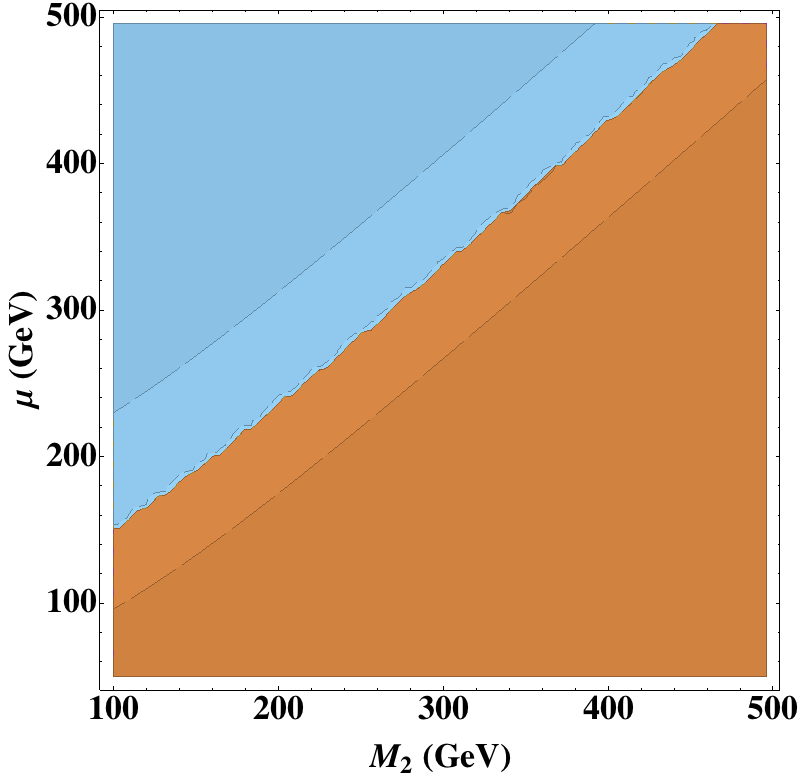}}
			\put(20,75){180 GeV}
		\end{picture}
		\label{}
	\end{subfigure}
	\begin{subfigure}{0.171\textwidth}
		\begin{picture}(100,100)
			\put(0,0){\includegraphics[width=\textwidth]{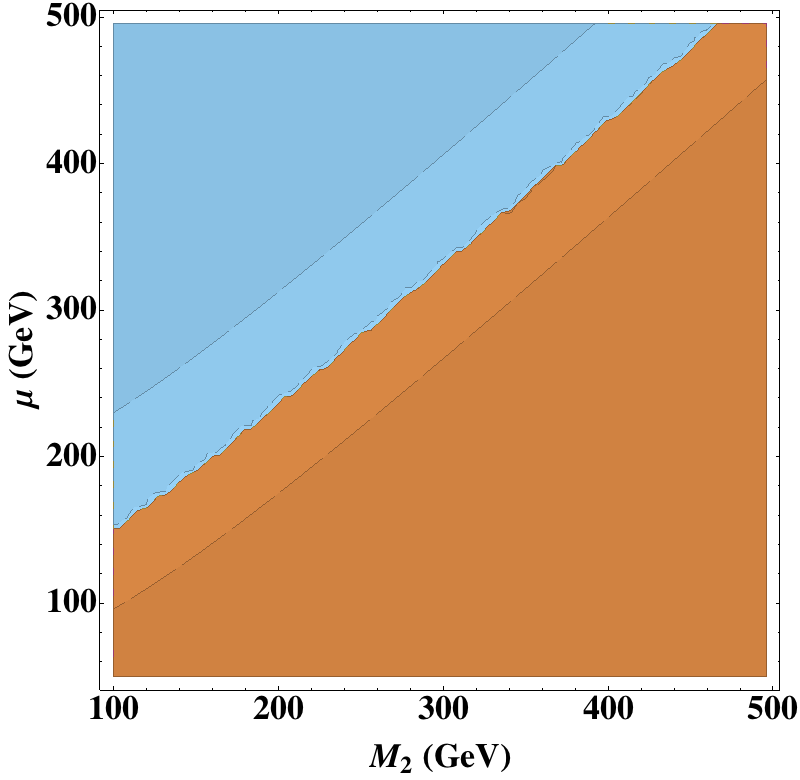}}
			\put(20,75){260 GeV}
		\end{picture}
		\label{}
	\end{subfigure}
	\begin{subfigure}{0.171\textwidth}
		\begin{picture}(100,100)
			\put(0,0){\includegraphics[width=\textwidth]{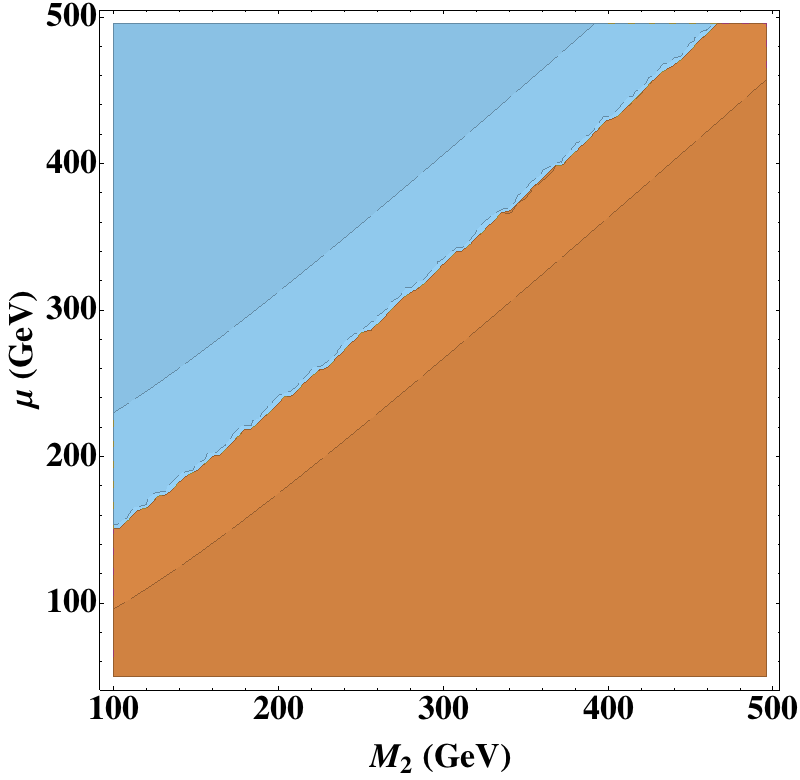}}
			\put(20,75){340 GeV}
		\end{picture}
		\label{}
	\end{subfigure}
	\begin{subfigure}{0.1\textwidth}
		\begin{picture}(100,100)
			\put(0,25){\includegraphics[width=\textwidth]{cstatelegend.pdf}}
			\put(0,75){$M_1$}
		\end{picture}			
		\label{}
	\end{subfigure}
	\vspace{-4mm}  \\
	\begin{subfigure}{0.171\textwidth}
		\includegraphics[width=\textwidth]{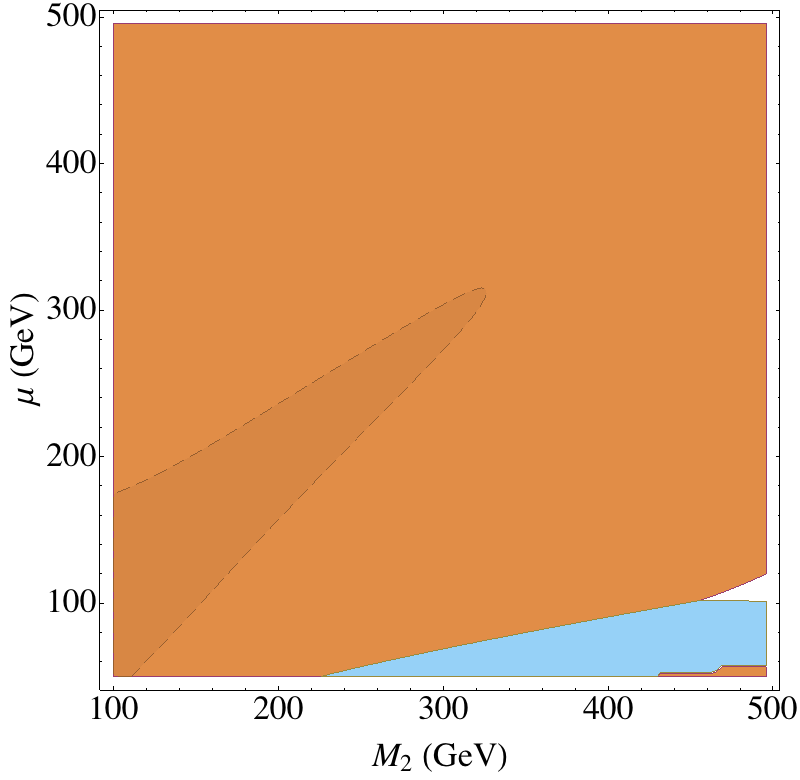}
		\label{}
	\end{subfigure}
	\begin{subfigure}{0.171\textwidth}
		\includegraphics[width=\textwidth]{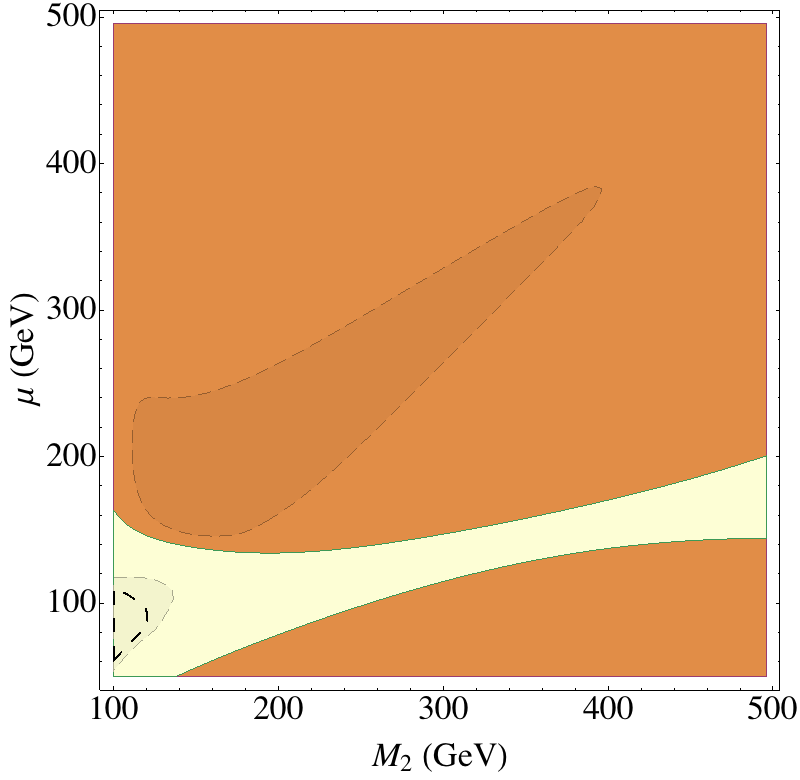}
		\label{}
	\end{subfigure}
	\begin{subfigure}{0.171\textwidth}
		\includegraphics[width=\textwidth]{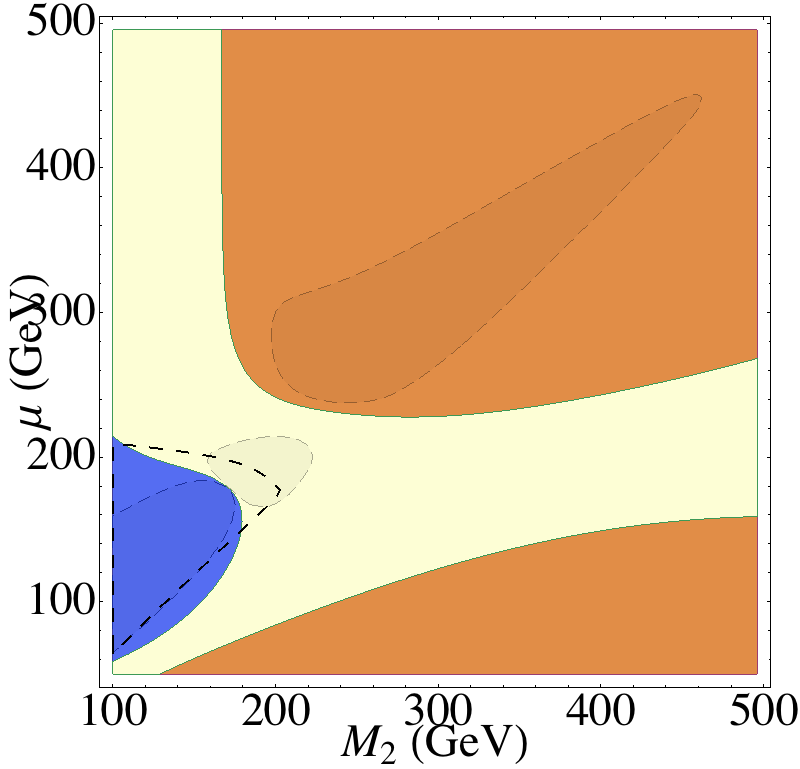}
		\label{}
	\end{subfigure}
	\begin{subfigure}{0.171\textwidth}
		\includegraphics[width=\textwidth]{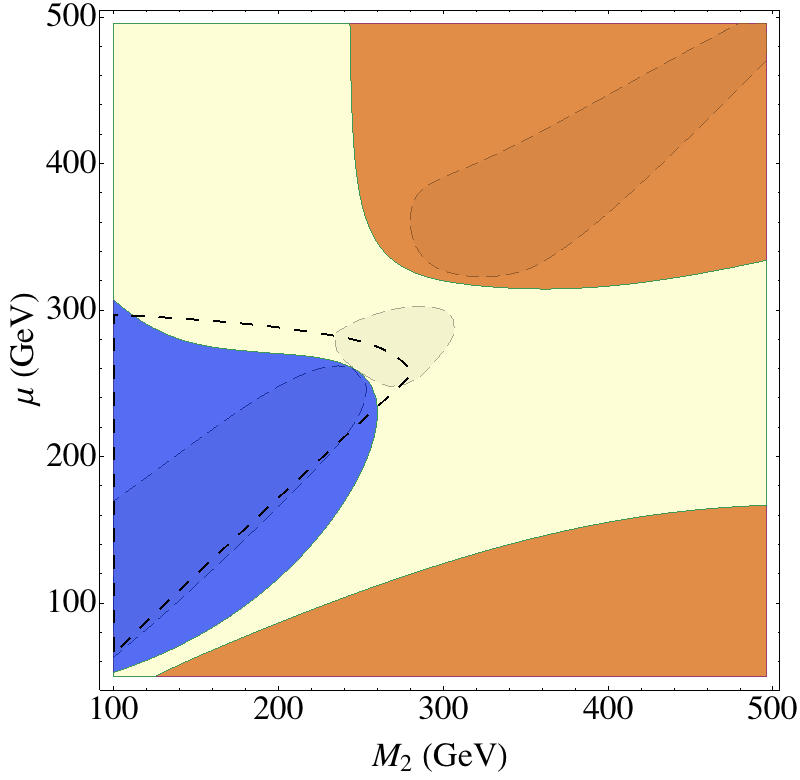}
		\label{}
	\end{subfigure}
	\begin{subfigure}{0.171\textwidth}
		\includegraphics[width=\textwidth]{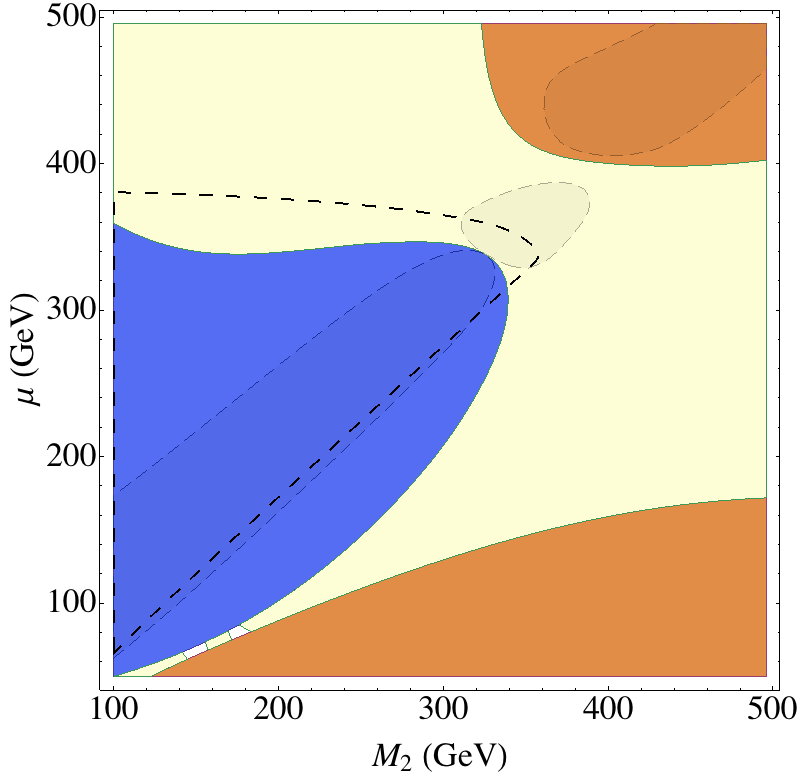}
		\label{}
	\end{subfigure}
	\begin{subfigure}{0.1\textwidth}
		\includegraphics[width=\textwidth]{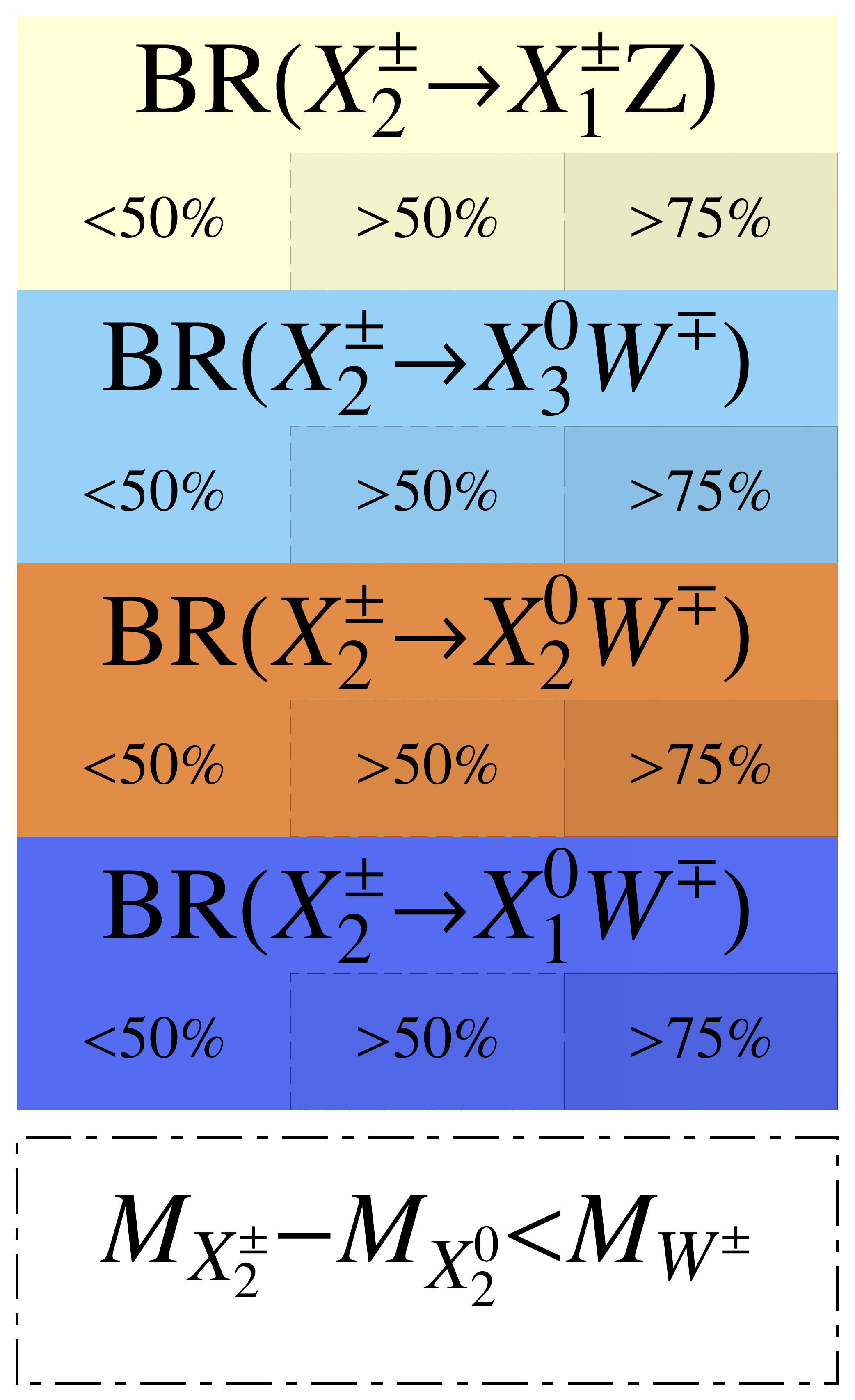}
		\label{}
	\end{subfigure}
	\vspace{-4mm}
	\caption{$\xx$ 
Dominant gauge eigenstate content (top) and leading decay modes (bottom)
of the $\xx$ chargino in the $M_2$--$\mu$ plane for various slices of 
$M_1$ and $\tan\beta=10$.  The thick, dashed and dotted lines indicate where 
the corresponding decays only occur with an off-shell vector boson. Shaded, dash-enclosed regions indicate
the boundary of 50\% and 75\% composition/branching ratio, as noted in the legend.
}
	\label{fig:x2summary}
\end{figure}

  The branching ratios of these decays depend on the
gauge-eigenstate content and the mass splittings among the states.
In Figs.~\ref{fig:n1summary}--
\ref{fig:x2summary} we show the dominant gauge eigenstate components and 
the leading decay modes for all the neutralino and chargino states
in the $\mu$--$M_2$ plane at fixed slices of 
$M_1 = 20,\,100,\,180,\,260,\,340\,\gev$ with $\tan\beta=10$.
In all cases, the mixing factors and branching ratios
were computed with SUSY-HIT~1.3~\cite{Djouadi:2006bz} interfaced with 
SuSpect~2.41~\cite{Djouadi:2002ze} or SoftSusy~3.3.10~\cite{Allanach:2001kg}.
The upper panels in Figs.~\ref{fig:n1summary}--\ref{fig:x2summary}
indicate where the dominant neutralino gauge component is 
$\widetilde{H}^0_+$~(light yellow), $\widetilde{H}^0_-$~(light-medium blue),
$\widetilde{W}^0$~(dark-medium orange), or $\widetilde{B}^0$~(dark blue).
The variations in shading in these panels show where the fraction
of the corresponding state exceeds $50\,\%$ or $75\,\%$.
The lower panels of Figs.~\ref{fig:n2summary}--\ref{fig:x2summary}
show the dominant decay fractions.  The dotted, dot-dashed, and dashed
lines in these figures indicate boundaries where the decay modes
can occur on shell.  
While we only show results for positive values of $\mu$, $M_1$, and $M_2$,
we find similar results for other relative signs.

  The relative importance of the decay channels 
shown in Figs.~\ref{fig:x1summary}--\ref{fig:x2summary}
can be understood by counting the number of mixings required for
each to occur while also taking into account the mass splitting
between the initial and final states.  Recall that the
mixing goes like ${m_Z/|M_a\pm\mu|}$.  As listed in Appendix~\ref{sec:appb},
couplings to $W^{\pm}$ involve Wino-Wino or Higgsino-Higgsino,
couplings to $Z^0$ involve only Higgsino-Higgsino, 
and couplings to $h^0$ involve Higgsino-Wino or Higgsino-Bino.  
The mass matrices of Appendix~\ref{sec:appa} also show that the 
mass splitting between two relatively pure Wino-like 
or Higgsino-like states is less than about $m_Z$.  

  To illustrate this counting, and an additional subtlety associated
with it, consider the decay of a Bino-like neutralino into a much lighter 
Wino-like neutralino or chargino.
The gauge modes $\widetilde{B}^0\to W^{\pm}\widetilde{W}^{\mp}$
and $\widetilde{B}^0\to Z^0\widetilde{W}^0$ both require two
mixings in the decay amplitude while $\widetilde{B}^0\to h^0\widetilde{W}^0$ 
requires only one.  While this would seem to favour the Higgs mode when
all three can occur on-shell, the gauge modes are found to be comparable 
or even more likely.
This follows from the Goldstone boson equivalence 
theorem~\cite{Cornwall:1974km,Vayonakis:1976vz,Lee:1977eg}.
For large mass splittings $\Delta M$, the decay rates to massive vectors
are enhanced by a factor on the order of $(\Delta M/m_Z)^2$
relative to the Higgs channel, and this effectively cancels the 
additional mixing factor appearing in the amplitudes 
for the gauge modes~\cite{Cheung:2009su}.
Note that this enhancement is present only when the intial and final
states have a mass splitting parametrically larger than $m_Z$.
In particular, no such enhancement occurs for vector boson decays
involving Higgsino to Higgsino or Wino to Wino states.

 The dominant gauge-eigenstate component of the $\chi_1^0$ neutralino 
is shown in Fig.~\ref{fig:n1summary}.  Unsurprisingly, it nearly 
always corresponds to the smallest of the underlying neutralino 
mass parameters.  When the $\n$ is mostly Higgsino, it coincides
with the $\widetilde{H}_-^0$ linear combination (for $\mu$, $M_1$, and
$M_2$ of the same sign).
This state is stable by assumption, and there are no decay modes to be shown.

  In Fig.~\ref{fig:x1summary}, we show the gauge content and the
leading decay modes of the lighter chargino, $\chi_1^{\pm}$.
For our choice of positive signs for $\mu$, $M_1$, and $M_2$,
we find that it is always the next-to-lightest state in the
spectrum.  For this reason, the only available decay mode
is $\x\to W^{\pm\,(*)}\n$, as can be seen in the lower panels of the
figure.  The dashed lines in these plots show the boundary between 
where this decay occurs with the $W^{\pm}$ off or on shell.  
This line lies slightly above the contour in Fig.~\ref{fig:n1summary}
where the lightest $\n$ state is Bino-like.  When this is not the case,
the $\x$ and $\n$ modes are typically both Wino-like or Higgsino-like
and the mass splitting between them is less than $m_W$.

  The content and decays of the $\nn$ neutralino are shown
in Fig.~\ref{fig:n2summary}.  Three distinct decay modes
are now possible, and the thick dashed and dotted lines in the lower
panels illustrate where they can occur on-shell.
The decay $\chi_2^0\to h^0\chi_1^0$ is seen to dominate in the
upper right corner of these plots when $\chi_2^0$ is Wino- or Higgsino-like
and $\chi_1^0$ is Bino-like.  The related decay with a $Z^0$ typically
has a similar (but smaller) branching in this region.  It requires
an additional mixing factor relative to the Higgs mode, but can also
receive a Goldstone boson enhancement.  On the other hand, 
no such enhancement occurs for $\nn\to W^{\mp}\x$ in this region, since
both states are close in mass, and the corresponding branching ratio
is negligible.  Note as well that the Higgs decay dominates only 
when it is two-body due to the very small width of the Higgs.  
Vector boson modes are dominant in the rest of the parameter space.
When the $\nn$ state is Wino-like, it tends to be very degenerate
with the $\x$, and so the $Z^0\n$ channel dominates due to the larger
available phase space.  For a Bino- or Higgsino-like $\nn$ state,
the neutral and charged vector modes tend to have similar branchings,
with the Higgs mode contributing at a significant (but sub-leading)
level when it can occur on-shell.

  The leading components and decay channels of $\nnn$ are shown
in Fig.~\ref{fig:n3summary}.  The $\nnn\to Z^0\n$ mode dominates
when it can occur on-shell but $\nnn\to W^{\mp}\x$ cannot.
This occurs when $|M_1| < |\mu| < |M_2|$.  Otherwise, the $W^{\mp}$
is dominant, although $\nnn \to h^0\n$ can be significant as well
when it can occur on-shell.

  In Fig.~\ref{fig:n4summary}, we show the content and leading decay modes
of the heaviest neutralino $\nnnn$.  The dominant decay channel
is to the kinematically unsuppressed $\nnnn \to \x W^{\mp}$ in the regions
where the $\nnnn$ has a significant Higgsino or Wino components. For intermediate
values of $M_1$, where the LSP is either Wino-like or Higgsino-like and either
$\nn$ or $\nnn$ is Bino-like, the production of $\nnnn \xx$ is important, as 
lighter modes will be either suppressed (Bino production) or result in
soft decays with low acceptance rates. 
In the region where $\nnnn$ is significantly Bino-like, 
the branching ratio is split between all unsuppressed modes with one mixing 
($\xx W^\mp$, $\x W^\mp$, $\nn Z$), with the largest component 
(though $<50\%$) to the Wino-like $\xx$. 
As the mass of the $\xx$ increases, the $\xx W^\mp$
mode becomes kinematically suppressed, and the Higgsino-like $\nn Z^0$ mode
dominates over the Higgsino-like $\x W^\mp$ for the remainder of the region
with a Bino-like $\nnnn$. 

Finally, we show the dominant components and leading decay modes of
the heavier chargino $\xx$ in Fig.~\ref{fig:x2summary}.  Production of the
$\xx$ is important where the LSP is either Higgsino-like or Wino-like, since the
$\x$ state will decay to soft leptons in this region, as indicated in 
Fig.~\ref{fig:x1summary}. The decays of the $\xx$ are relatively uniformly split
between $\x Z$, $\nn W^\pm$ and $\n W^\pm$, as very little of the
parameter space shows branching ratios larger than $50\%$. 

We have also examined the dependence of these decay fractions on $\tan\beta$
in the range $2\leq\tan\beta\leq 50$.  The results throughout this wide
range are qualitatively very similar to the $\tan\beta=10$ case that we
have studied in detail.

\subsection{Implications for LHC Signals}

  Before turning to a detailed analysis of the sensitivity of LHC
searches to electroweakinos, let us briefly emphasize three points
that will be important in the analysis to follow.
First, production rates tend to be greatest for the lightest pairs
of states with significant Higgsino or Wino components, and the subsequent
cascade decays are usually fairly short.  This motivates searches for
relatively simple decay topologies.  Second, in a very significant
fraction of the parameter space, the leading decay modes
occur between states with mass splittings less than $m_Z$ or $m_W$.
As a result, the decay products frequently have low $p_T$,
and invariant mass pairings that do not reconstruct a resonance
(or a kinematic edge).  This limits the sensitivity of searches
that attempt to reconstruct specific kinematic features characteristic
of on-shell vector boson decays or large missing energy.
And third, many states are found to have multiple relevant
decay modes.  This implies that the full inclusive signals of 
MSSM electroweakinos can be much richer and more complicated 
than the simplified-model realizations that are frequently 
applied (\emph{e.g.} Ref.~\cite{ATLAS:2013rla}).

\section{Methodology of LHC Sensitivity Estimates \label{sec:search}}

  We turn next to investigate the sensitivity of ATLAS and CMS searches
to neutralinos and charginos.  Both collaborations have explored
a wide variety of possible SUSY signals, including specific searches
geared towards the electroweakinos.  In this section we describe the
techniques we used to apply these and more general searches to the MSSM.
Our results will be presented in the section to follow.

  Signal events were generated independently for all 21 possible
production pairings using MadGraph5~\cite{Alwall:2011uj}
interfaced with Pythia~6.4~\cite{Sjostrand:2006za}.
Hard scattering processes with zero or one additional jets
($pp \rightarrow {\chi}_i {\chi}_j + \{0,1\}j$) 
were obtained from MadGraph5 and passed to Pythia~6.4
to be decayed, showered, and hadronized, with the inclusion
of MLM matching between additional hard jets 
and the parton shower~\cite{Alwall:2008qv}.
For each MSSM parameter point and each inclusive production channel,
50000 events were generated.  
These events were then passed to the Delphes~3 detector 
simulator~\cite{deFavereau:2013fsa}, with triggers, 
jet reconstruction (anti-kT), and hadronic/leptonic tagging 
efficiencies modified to match the specifications for each experimental
search channel considered.  The results from all 21 production channels 
were combined for each search to obtain the inclusive MSSM signal 
by weighting each channel by its net cross section 
after matching and cuts.

  The cuts implemented in each search channel in each analysis were 
reproduced from the information provided by Delphes.
All analyses were vetted against cut-flow tables where provided by the 
experimental groups.  To account for pile-up, the $\met$ values extracted
from Delphes were smeared in an additional post-processing step, 
which was found to be necessary in the vetting process.
Specifically, a Gaussian smearing was applied to the Delphes $\met$ values
with a standard deviation of $0.75$ times the value given in Ref.~\cite{metpub},
where the multiplicative factor compensates for the smearing already present
in Delphes.\footnote{Ref.~\cite{Whiteson} found that modifying the Delphes 
$\met$ smearing by a post-processed Gaussian smearing with a standard deviation 
of $\sim\!20$ GeV effectively reproduced the smearing effect at LHC8.} 
Values of the $m_{T2}$ variable used in some of the analyses were computed
using the MT2\_Bisect package \cite{Lester:1999tx,Barr:2003rg}, while
the Razor variables of Ref.~\cite{Rogan:2010kb} were calculated using 
the algorithm provided by the CMS collaboration.

Two superimposed grids of points were generated in the $M_2-\mu$ plane, with 
a $5 \times 5$ grid of $M_2$ (100--500 GeV) and $\mu$ (50--500 GeV), and a 
$4 \times 4$ grid of $M_2$ (140--440 GeV) and $\mu$ (95-433 GeV),
for seven slices of $M_1$ (20,\,60,\,100,\,180,\,240,\,320,\,420~GeV).
The $4 \times 4$ grid was critical in adding insight into the 
large regions between the rough $5 \times 5$ grid without significantly 
increasing the computation time, as would a more populated, uniform grid.
The signals calculated at each grid point were then extended to form a uniform 
$9 \times 9$ grid using linear interpolation of the logarithm of the event rates,
following which a three-dimensional order-three polynomial interpolation 
was performed over the entire $7 \times 9 \times 9$ dataset, 
again on the logarithm of the event rates.
Exclusion regions were then determined from comparison of the calculated
number events to the 95\% confidence level (C.L.) limit on the number of
signal events ($N_{i}^{95}$). For the ATLAS studies, the $N_{i}^{95}$ were
provided, while for the CMS studies, the $N_{i}^{95}$ were calculated using
the $CL_s$ method~\cite{Junk:1999kv} with Gaussian-distributed 
uncertainties as implemented in RooStats~\cite{Moneta:2010pm}. 
In our analysis, we combine the exclusion regions from each separate
signal region in a boolean fashion.

\begin{flushleft}
From ATLAS, we investigated the following searches:
\vspace{-0.15cm}
\begin{itemize}
  \item opposite-sign dileptons with $\met$ and no jets~\cite{TheATLAScollaboration:2013hha}
\vspace{-0.15cm}
  \item trilepton plus $\met$~\cite{ATLAS:2013rla,Aad:2014nua};
\vspace{-0.15cm}
  \item four or more leptons~\cite{ATLAS:2013qla};
\vspace{-0.15cm}
  \item dileptons with razor variables~\cite{TheATLAScollaboration:2013via}
\vspace{-0.15cm}
  \item hadronic di-$\tau$ plus $\met$~\cite{ATLAS:2013yla}
\vspace{-0.15cm}
  \item same-sign dileptons plus jets~\cite{ATLAS:2013tma}
\vspace{-0.15cm}
\item monojet~\cite{ATLAS:2012zim,TheATLAScollaboration:2013aia}.
\vspace{-0.15cm}
  \item jets plus $\met$~\cite{TheATLAScollaboration:2013fha};
\vspace{-0.15cm}
\item disappearing charged tracks~\cite{Aad:2013yna}
\end{itemize}
\end{flushleft}

\begin{flushleft}
From CMS, we considered the following studies:
\begin{itemize}
\item leptons (dilepton, trilepton, multi-lepton) with $\met$~\cite{CMS:2013dea}
\vspace{-0.15cm}
  \item chargino and neutralino search using $h\rightarrow \bar{b}b$ decays \cite{1596278}.
\vspace{-0.15cm}
  \item monojet~\cite{CMS:rwa}.
\end{itemize}
\end{flushleft}

Many of the LHC searches use similar final states
but different specific search strategies. 
In particular, ATLAS searches tend to focus on a small 
number of cut regions that are designed to enhance the signal 
from a specific simplified model, 
whereas CMS searches tend to be more broadly focused, 
arranging a grid of cut regions over a larger phase space. 
As a result, these searches are frequently complementary.

\section{Limits from the LHC \label{sec:constraint}}

  Following the methods described above, we have derived exclusions
on the parameter space of MSSM charginos and neutralinos using existing
LHC searches.  Many of the searches we apply were designed to find other
superpartners (or forms of new physics), while others have only been
used by the collaborations to constrain specific simplified models
of the electroweakinos.  We find that some of these searches also give new
constraints on the more general MSSM electroweakino sector.

\begin{figure}[ttt]
	\begin{subfigure}{0.23\textwidth}
		\begin{picture}(100,100)
			\put(0,0){\includegraphics[width=\textwidth]{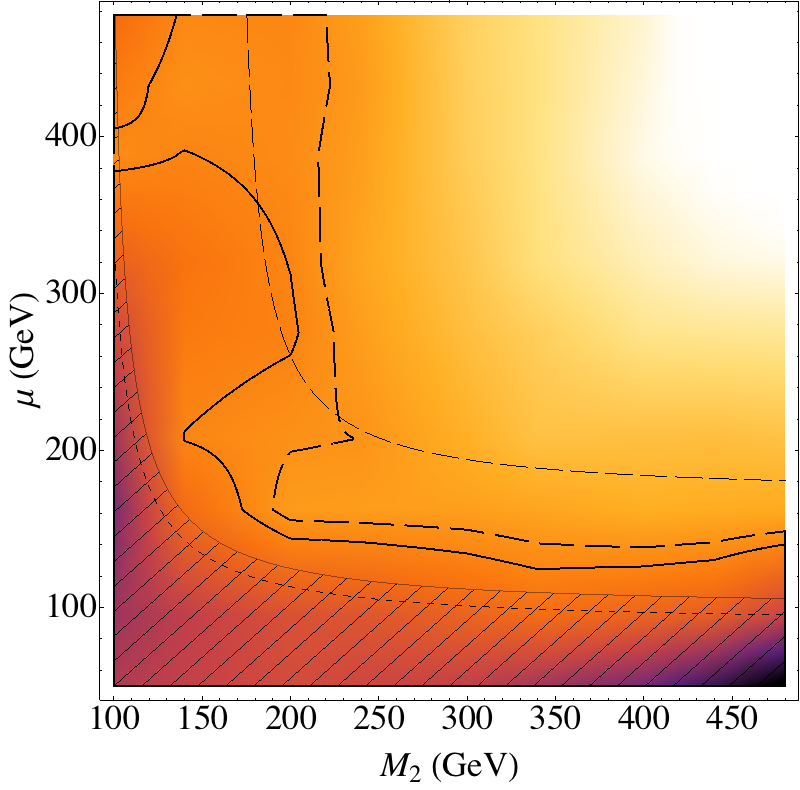}}
			\put(40,105){20 GeV}
			\put(0,105){$M_1=$}
		\end{picture}
		\label{}
	\end{subfigure}
	\begin{subfigure}{0.23\textwidth}
		\begin{picture}(100,100)
			\put(0,0){\includegraphics[width=\textwidth]{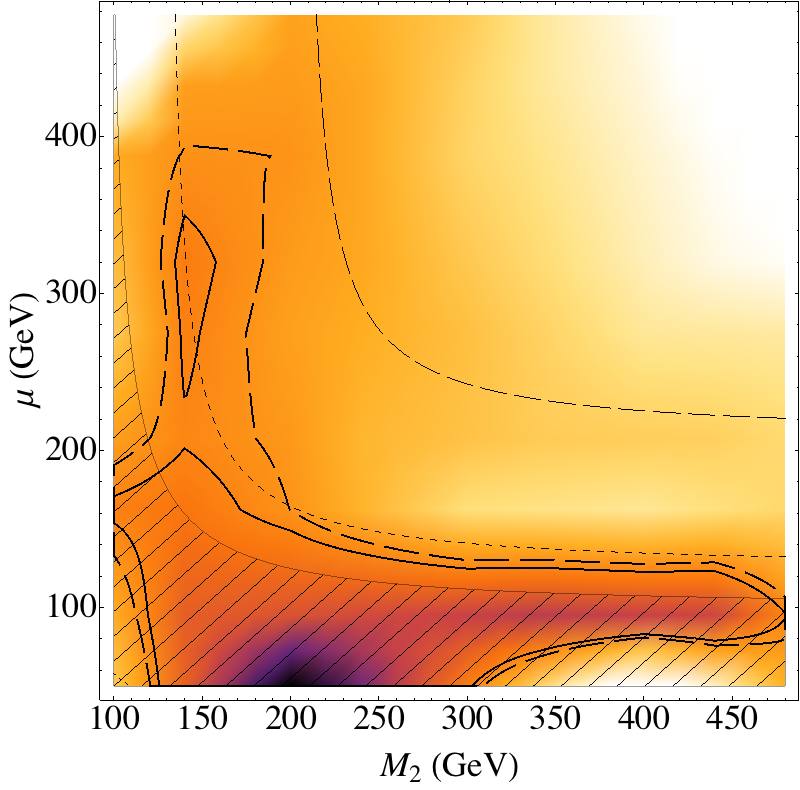}}
			\put(40,105){60 GeV}
		\end{picture}
		\label{}
	\end{subfigure}
	\begin{subfigure}{0.23\textwidth}
		\begin{picture}(100,100)
			\put(0,0){\includegraphics[width=\textwidth]{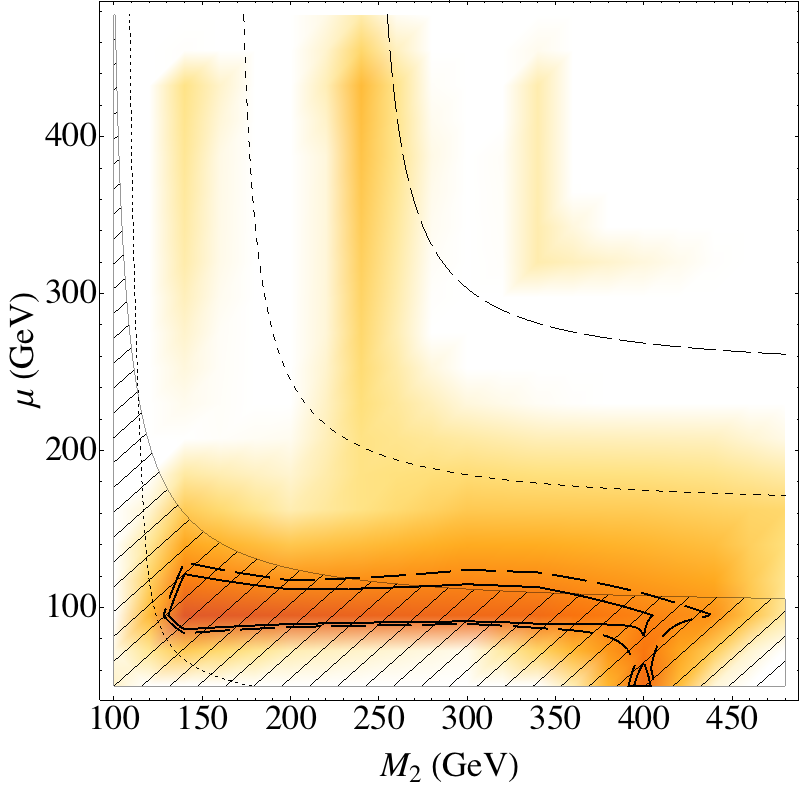}}
			\put(40,105){100 GeV}
		\end{picture}
		\label{}
	\end{subfigure}
	\begin{subfigure}{0.23\textwidth}
		\begin{picture}(100,100)
			\put(0,0){\includegraphics[width=\textwidth]{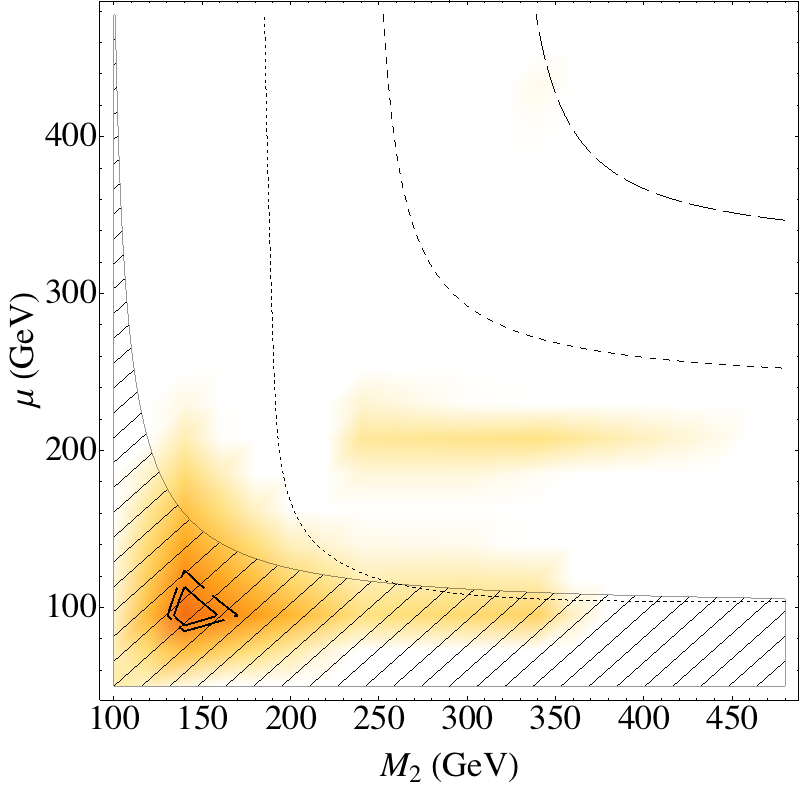}}
			\put(40,105){180 GeV}
		\end{picture}
		\label{}
	\end{subfigure}
	\begin{subfigure}{0.055\textwidth}
		\begin{picture}(100,100)
			\put(0,0){\includegraphics[width=\textwidth]{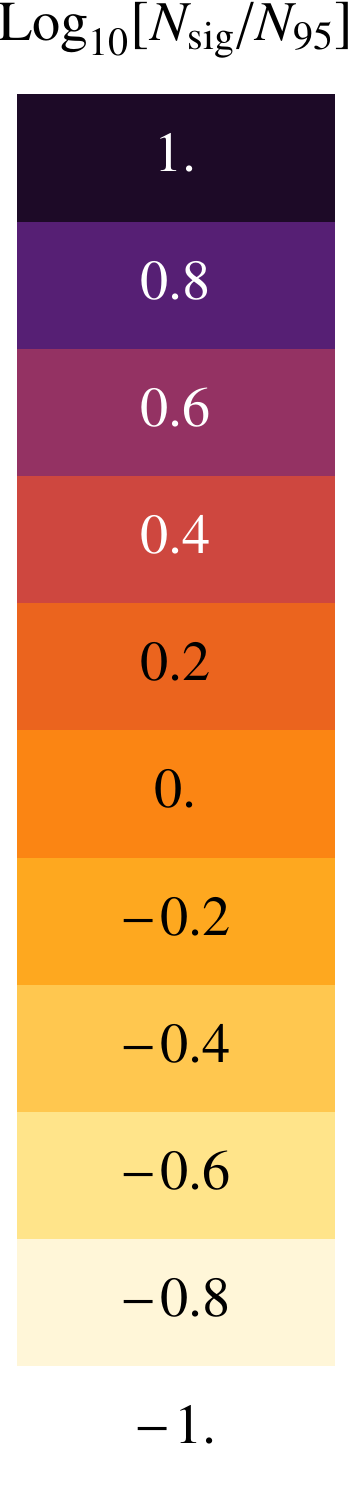}}
		\end{picture}
		\label{}
	\end{subfigure}
	\vspace{-4mm}	
	\caption{Parameter exclusions from the ATLAS opposite-sign dilepton search of Ref.~\cite{TheATLAScollaboration:2013hha} in the $M_2$--$\mu$ plane for several fixed values of $M_1$.  The boundaries of the 95\,\%~c.l. excluded regions are denoted by the thick black solid lines (thick black dashed lines) assuming a $K$ factor of 1.0 (1.2).  Colour shading indicates the number of predicted signal events relative to the number excluded by the experimental analysis.  The hatched area shows the 95\,\%~c.l. exclusion from LEP.   Contours of constant $\Delta M = m_{\x}-m_{\n}$ are indicated by the thin dashed lines -- long dash: $\Delta M=2 m_W$; mid dash: $\Delta M=m_W$; short dash: $\Delta M=15$ GeV.}
	\label{fig:a2lep}
\vspace{0.5cm}
\end{figure}

\begin{figure}[ttt]
	\begin{subfigure}{0.23\textwidth}
		\begin{picture}(100,100)
			\put(0,0){\includegraphics[width=\textwidth]{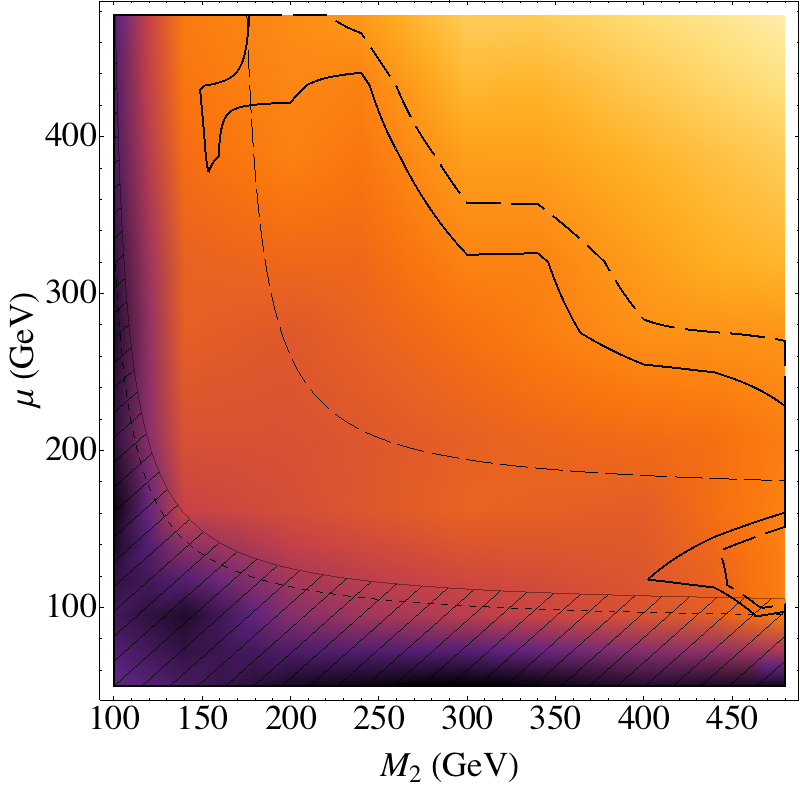}}
			\put(40,105){20 GeV}
			\put(0,105){$M_1=$}
		\end{picture}
		\label{}
	\end{subfigure}
	\begin{subfigure}{0.23\textwidth}
		\begin{picture}(100,100)
			\put(0,0){\includegraphics[width=\textwidth]{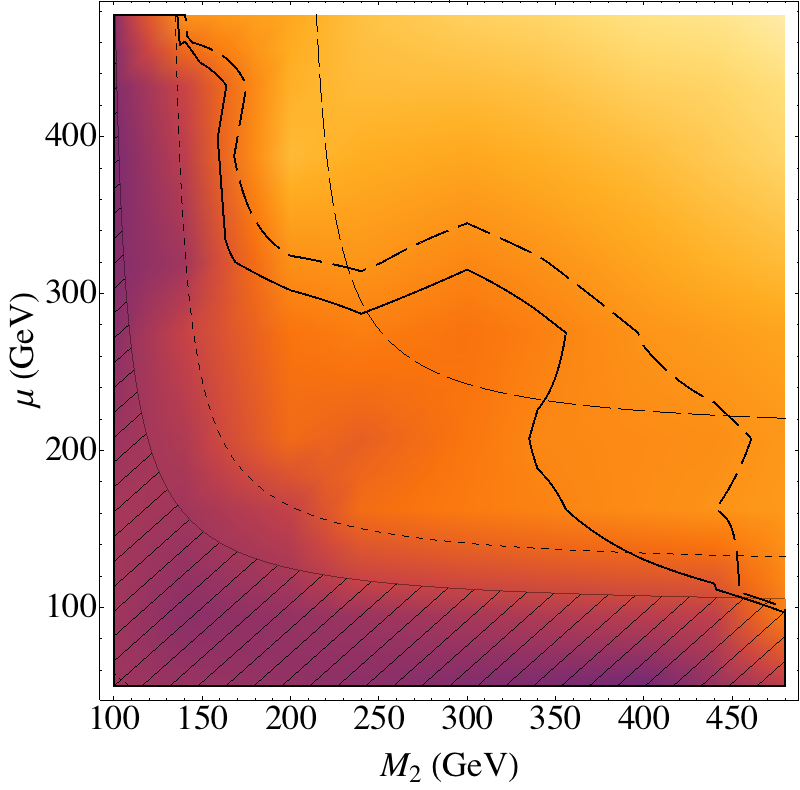}}
			\put(40,105){60 GeV}
		\end{picture}
		\label{}
	\end{subfigure}
	\begin{subfigure}{0.23\textwidth}
		\begin{picture}(100,100)
			\put(0,0){\includegraphics[width=\textwidth]{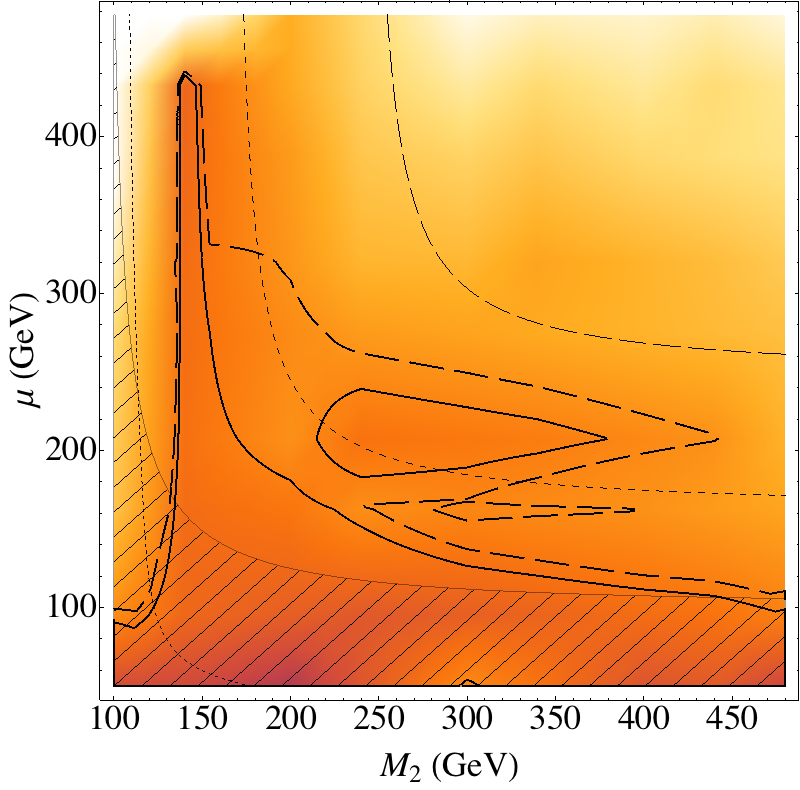}}
			\put(40,105){100 GeV}
		\end{picture}
		\label{}
	\end{subfigure}
	\begin{subfigure}{0.23\textwidth}
		\begin{picture}(100,100)
			\put(0,0){\includegraphics[width=\textwidth]{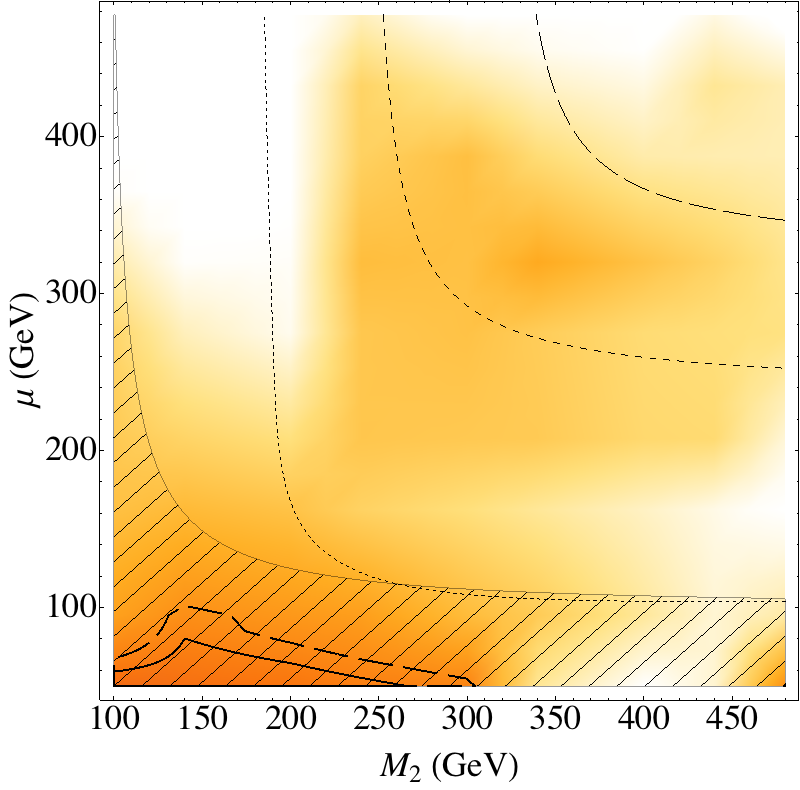}}
			\put(40,105){180 GeV}
		\end{picture}
		\label{}
	\end{subfigure}
	\begin{subfigure}{0.055\textwidth}
		\begin{picture}(100,100)
			\put(0,0){\includegraphics[width=\textwidth]{./Legend.pdf}}
		\end{picture}
		\label{}
	\end{subfigure}
	\vspace{-4mm}	
	\caption{Parameter exclusions from the ATLAS trilepton search of Ref.~\cite{ATLAS:2013rla} in the $M_2$--$\mu$ plane for several fixed values of $M_1$.  The boundaries of the 95\,\%~c.l. excluded regions are denoted by the thick black solid lines (thick black dashed lines) assuming a $K$ factor of 1.0 (1.2).  Colour shading indicates the number of predicted signal events relative to the number excluded by the experimental analysis.  The hatched area shows the 95\,\%~c.l. exclusion from LEP.   Contours of constant $\Delta M = m_{\x}-m_{\n}$ are indicated by the thin dashed lines -- long dash: $\Delta M=2 m_W$; mid dash: $\Delta M=m_W$; short dash: $\Delta M=15$ GeV.}
	\label{fig:a3lep}
\vspace{0.5cm}
\end{figure}

  The most important LHC exclusions are shown in 
Figs.~\ref{fig:a2lep}--\ref{fig:clbb}.  
These correspond to the ATLAS opposite-sign dilepton, trilepton,
and four-plus lepton studies, together with the CMS lepton plus
$b\bar{b}$ search, and will be discussed in more detail below.
The thick black lines in the figures show the boundaries of the
combined 95\,\%~confidence level~(c.l.) exclusions in the $M_2$-$\mu$ 
plane for $\tan\beta=10$ and several values of $M_1$.  
The colour shading indicates the number of predicted signal events 
(after cuts) relative to the number that are excluded by the corresponding 
experimental analyses.  The hatched regions 
in Figs.~\ref{fig:a2lep}--\ref{fig:clbb} indicate
the 95\%~c.l. exclusions from the LEP experiments~\cite{lep:chg1},
which are close to $m_{\chi_1^{\pm}} > 103.4\,\gev$ for $\Delta M = 
m_{\x}\!-\!m_{\n} > 3\,\gev$
or $\Delta M < 0.15\,\gev$~\cite{lep:chg1} but can fall to as low as
$92.4\,\gev$ for mass differences between these 
boundaries~\cite{lep:chg2,Batell:2013bka}.
In these figures we also show contours of constant mass differences
$\Delta M$ with thin dashed lines: long dash for $\Delta M=2 m_x$; mid dash
for $\Delta M=m_x$; short dash for $\Delta M=15$ GeV -- where $m_x = m_W$
and $\Delta M=m_{\x}-m_{\n}$ for the lepton analyses 
in Fig.~\ref{fig:a2lep}--\ref{fig:a3lep}, $m_x=m_Z$ with 
$\Delta M = m_{\nn}-m_{\n}$ in Fig.~\ref{fig:a4lep},
and $m_x=m_h$ with $\Delta M = m_{\nn}-m_{\n}$ 
for the Higgs-motivated $b\bar{b}$ analysis in Fig.~\ref{fig:clbb}.  
These lines are useful for understanding constraints, as acceptance
rates typically depend on whether decays occur on- or off-shell.
 
  In setting these exclusions, we used the leading-order
production cross sections obtained from MadGraph5.  These were generally
found to lie between the LO and NLO cross sections derived
from Prospino2.1~\cite{Beenakker:1996ed,Beenakker:1999xh}, and thus
our exclusions are somewhat conservative.  However, to illustrate the
effects of slightly larger cross sections, we also show with thick solid
dashed lines the boundaries of the regions excluded when a $K$-factor of $1.2$
is applied to the MadGraph LO signal cross sections.  This is typical of the
ratio of NLO to LO cross sections computed with Prospino2.1.

\subsection{ATLAS Opposite Sign Dileptons}

  The ATLAS opposite-sign~(OS) dilepton search of 
Ref.~\cite{TheATLAScollaboration:2013hha} was designed to
probe direct slepton and chargino production.  
Five distinct search regions were considered.  
All regions had a minimal requirement of two
isolated OS leptons with $p_T > 10\,\gev$ and $|\eta| \lesssim 2.4$, and no jets.
Additional requirements were imposed on lepton $p_T$, $\met$, relative lepton
flavour, and dileptonic kinematic variables.  
To suppress backgrounds, an effective $Z^0$ veto was imposed on all
five regions, either by rejecting events with the leading dilepton
invariant mass in the range $|m_{\ell\ell}-m_Z|< 10\,\gev$, or by demanding
that the leading OS leptons differ in flavour. In four of the five regions, 
a minimal requirement is imposed on the variable $m_{T2} > 90\,\gev$, 
based on the dilepton system~\cite{Lester:1999tx,Barr:2003rg}.\footnote{
In contrast to Refs.~\cite{Lester:1999tx,Barr:2003rg}, 
it is computed here under the assumption of massless decay products.}  
This is expected to have an endpoint at $m_W$ for SM backgrounds, 
while larger values can be obtained for chargino decays with 
$(m_{\x}-m_{\n}) \gg m_W$. 
The fifth signal region does not impose a cut on $m_{T2}$ but suffers 
from a much larger background rate.    

  The exclusions derived from this search for general electroweakino parameters
are shown in Fig.~\ref{fig:a2lep}.  The strongest bounds are obtained
for small values of $M_1$, and correspond mostly to the production of 
Wino- or Higgsino-like $\x$ followed by decays 
to a Bino-like LSP.  Lower $M_1$ gives larger mass differences
$\Delta M = m_{\x}-m_{\n}$ for a given value of $\mu$ or $M_2$, 
which leads to more $\met$, larger $m_{T2}$, and a higher fraction
of electroweakino events passing the acceptance cuts.
The larger production rate of Winos relative to Higgsinos 
(see Fig.~\ref{fig:pure}) leads to a stronger exclusion when $M_2 < \mu$.
For $\mu \sim M_2$ and $M_1 = 60$~GeV, the exclusions are increased
slightly over the $\mu \gg M_2$ or $\mu \ll M_2$ regions due to
contributions from $\n \nn$ production where the decay chain
$\nn \rightarrow \x W^\mp \rightarrow \n W^\pm W^\mp$ and off-shell
$\nn \rightarrow \n Z^0$ decays can also contribute to the signal regions.

  Very little new exclusion beyond the LEP limit is found for 
$M_1\gtrsim 100\,\gev$.  In this case, the LSP need not be Bino-like,
and there is the possibility of dominantly Wino-to-Higgsino or 
Higgsino-to-Wino transitions.  However, the LEP chargino bounds force 
$\mu$ and $M_2$ to each be larger than about $100\,\gev$.  
Together with the reduced production rate at higher mass and
the need for larger $\Delta M$ to pass the acceptance cuts,
there is not enough data to probe this possibility using 
the dilepton analysis.

\subsection{ATLAS Trilepton}

The ATLAS trilepton search~\cite{ATLAS:2013rla} was designed in part
to probe electroweakino production with decays through intermediate
sleptons or weak vector bosons.  Events with exactly three isolated
leptons were selected.  One pair must be same-flavour and opposite-sign~(SFOS)
with $m_{\ell\ell} > 12\,\gev$, and events with $b$ jets were vetoed
to suppress top backgrounds.   Six exclusive search regions were defined
with varying (but disjoint) requirements on the invariant mass
of the SFOS pairing that is closest to $m_Z$, the $\met$, the $p_{T}$ of the 
lepton not included in the SFOS pairing, and the
transverse mass $m_T$ of the unpaired lepton (for some signal regions).

  The combined exclusions derived from this analysis are shown 
in Fig.~\ref{fig:a3lep}.  As in the OS dilepton search discussed above,
the strongest limits are found for low $M_1$, where the dominant signal
processes involve Wino- or Higgsino-like states decaying to a much
lighter Bino-like LSP.  Sensitivity is lost at larger $\mu$ or $M_2$ 
due to reduced production rates and the opening of decays involving
Higgs bosons, which produce fewer leptons.  Also as above, 
the sensitivity of this search is greatest for larger $\Delta M$.
The interplay between production rates (smaller $M_2$ or $\mu$)
and signal acceptance (larger $\Delta M$) can be seen in the 
$M_1=100\,\gev$ slice. In this slice, disjoint regions are
excluded by separate signal regions that are sensitive to either on-shell
$Z^0$ decays (isolated exclusion region) or off-shell $Z^0$ decays (bulk 
exclusion region). The gap between these regions is indicative of the
reduced sensitivity of the study to the region where $m_{\nn}-m_{\n} \sim m_Z$,
which is also present in the results of~\cite{ATLAS:2013rla}.

For $M_1$ approaching $M_2$ and larger $\mu$, there is a rapid
drop in the sensitivity of this search as seen in the upper
portions of the $M_1=100,\,180\,\gev$ panels of Fig.~\ref{fig:a3lep}.
In this region, the $\n$ approaches the $\x$ state in mass,
leading to very soft leptons from $\x \to W^+\n$ decays
leading to a low acceptance in the trilepton search channels.

\begin{figure}[ttt]
	\begin{subfigure}{0.23\textwidth}
		\begin{picture}(100,100)
			\put(0,0){\includegraphics[width=\textwidth]{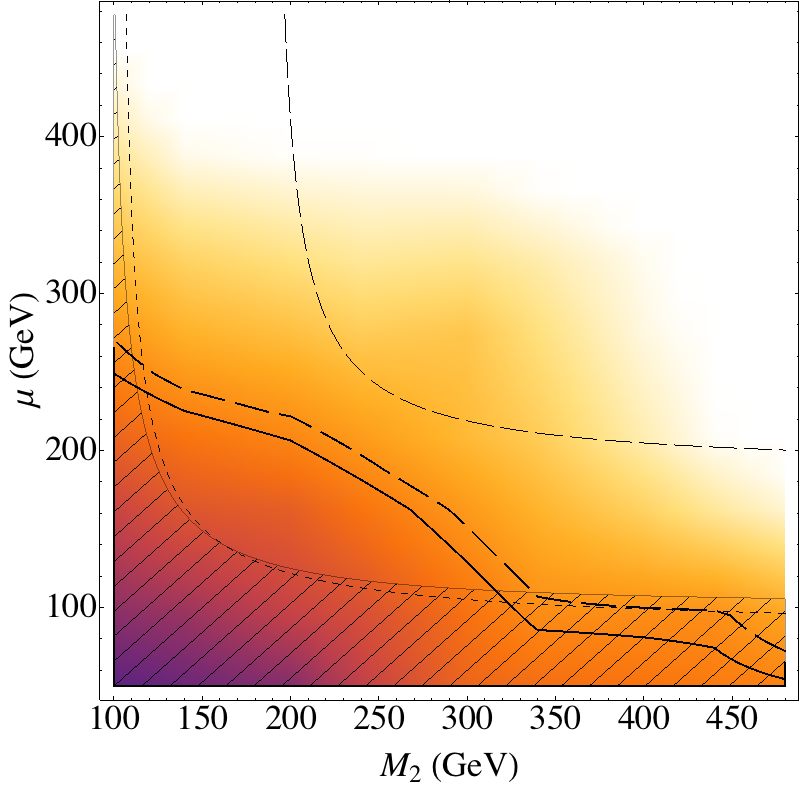}}
			\put(40,105){20 GeV}
			\put(0,105){$M_1=$}
		\end{picture}
		\label{}
	\end{subfigure}
	\begin{subfigure}{0.23\textwidth}
		\begin{picture}(100,100)
			\put(0,0){\includegraphics[width=\textwidth]{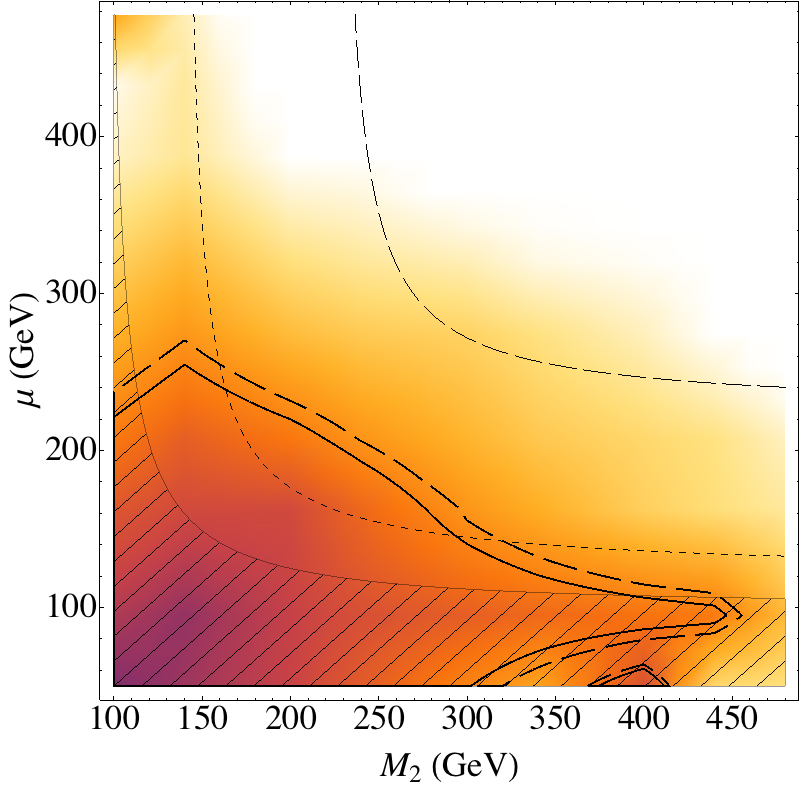}}
			\put(40,105){60 GeV}
		\end{picture}
		\label{}
	\end{subfigure}
	\begin{subfigure}{0.23\textwidth}
		\begin{picture}(100,100)
			\put(0,0){\includegraphics[width=\textwidth]{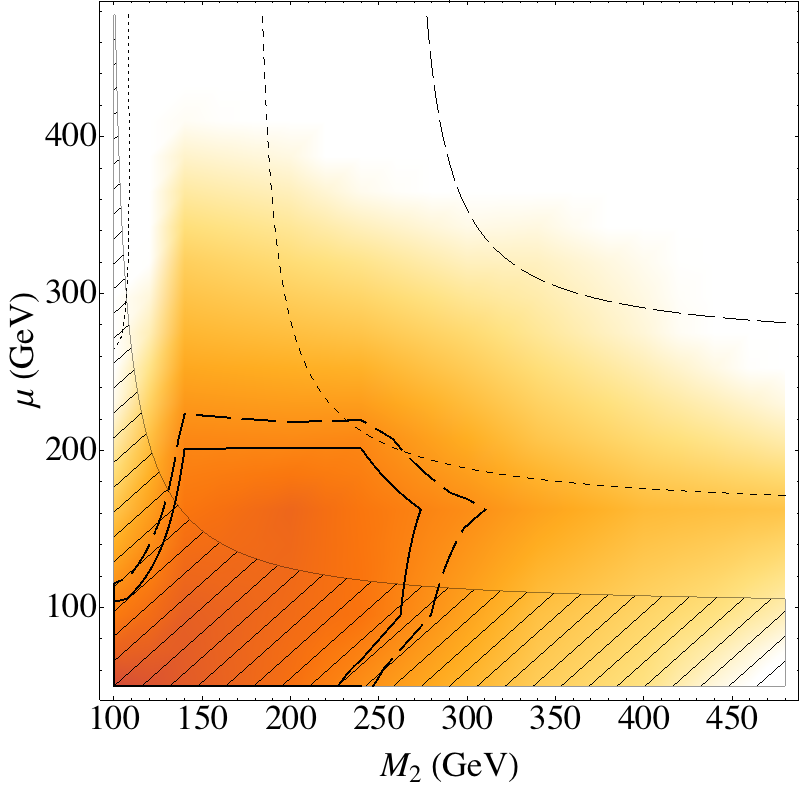}}
			\put(40,105){100 GeV}
		\end{picture}
		\label{}
	\end{subfigure}
	\begin{subfigure}{0.23\textwidth}
		\begin{picture}(100,100)
			\put(0,0){\includegraphics[width=\textwidth]{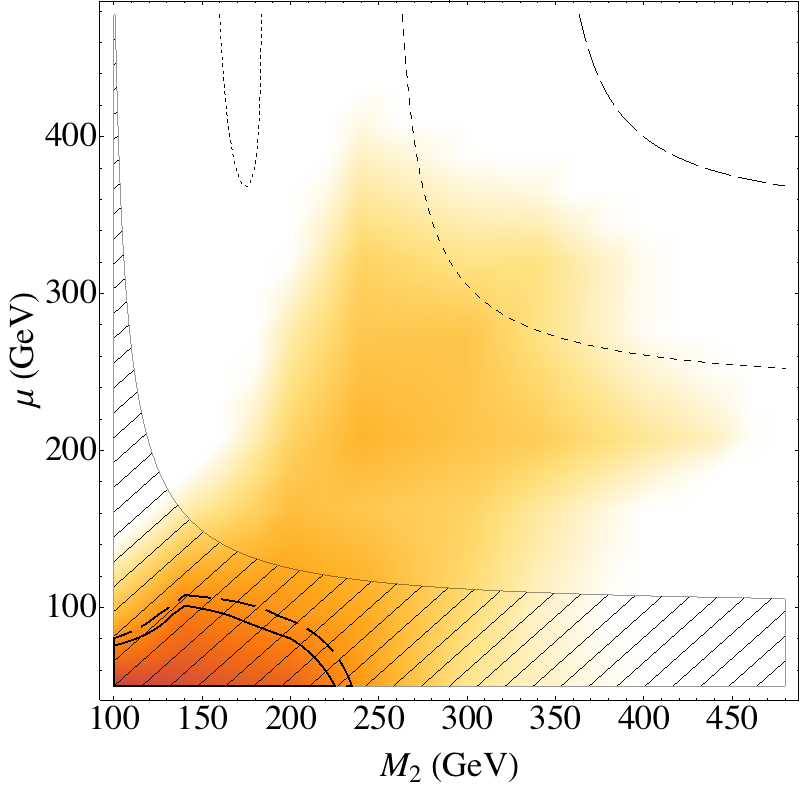}}
			\put(40,105){180 GeV}
		\end{picture}
		\label{}
	\end{subfigure}
	\begin{subfigure}{0.055\textwidth}
		\begin{picture}(100,100)
			\put(0,0){\includegraphics[width=\textwidth]{./Legend.pdf}}
		\end{picture}
		\label{}
	\end{subfigure}
	\vspace{-4mm}	
	\caption{Parameter exclusions from the ATLAS four-plus lepton search of Ref.~\cite{ATLAS:2013qla} in the $M_2$--$\mu$ plane for several fixed values of $M_1$.  The boundaries of the 95\,\%~c.l. excluded regions are denoted by the thick black solid lines (thick black dashed lines) assuming a $K$ factor of 1.0 (1.2).  Colour shading indicates the number of predicted signal events relative to the number excluded by the experimental analysis.  The hatched area shows the 95\,\%~c.l. exclusion from LEP.   Contours of constant $\Delta M = m_{\nn}-m_{\n}$ are indicated by the thin dashed lines -- long dash: $\Delta M=2 m_Z$; mid dash: $\Delta M=m_Z$; short dash: $\Delta M=15$ GeV.}
	\label{fig:a4lep}
\vspace{0.5cm}
\end{figure}

\subsection{ATLAS Four Lepton}

  The ATLAS four-lepton search in Ref.~\cite{ATLAS:2013qla} was motivated
by electroweakino production with decays through intermediate sleptons,
R-parity violation, or to a gravitino and $Z^0$ boson.  Four or more
well-identified leptons were required, with up to one tau included in the count.
Five search regions were defined, of which three have a $Z^0$ veto based
on the invariant masses of SFOS pairs, with the other two regions demanding
that a SFOS pair reconstruct a $Z^0$ to within $10\,\gev$.  Additional
requirements were imposed on $\met$ and $m_{eff}$ (defined to be the
scalar sum of jet, lepton, and missing $p_T$).

  The exclusions derived from this search are illustrated 
in Fig.~\ref{fig:a4lep}.  
The signal in this case can be generated by $\chi_i^0\chi_j^0$ production
with both $\chi^0_{i,j}\to Z^{0(*)}\chi_1^0$ and $Z^{0(*)}\to \ell\bar{\ell}$,
or through multistep cascades with $\chi^0_{i,j}\to W^{\pm(*)}\chi_1^{\mp}$.
We found that the most sensitive signal regions were SRnoZa and SRnoZb
defined in Ref.~\cite{ATLAS:2013qla}.  Recall that neutralino pair
production relies on the Higgsino components of the states,
and thus this study should be mainly sensitive to smaller values of $\mu$.
In addition, $\mu \sim M_2$ results in a number of states with similar 
masses and significant Higgsino components, which increases the multiplicity of
production modes that can contribute to the signal.
As for the previous analyses, the sensitivity of this search falls
off quickly with increasing $M_1$.

\begin{figure}[ttt]
	\begin{subfigure}{0.23\textwidth}
		\begin{picture}(100,100)
			\put(0,0){\includegraphics[width=\textwidth]{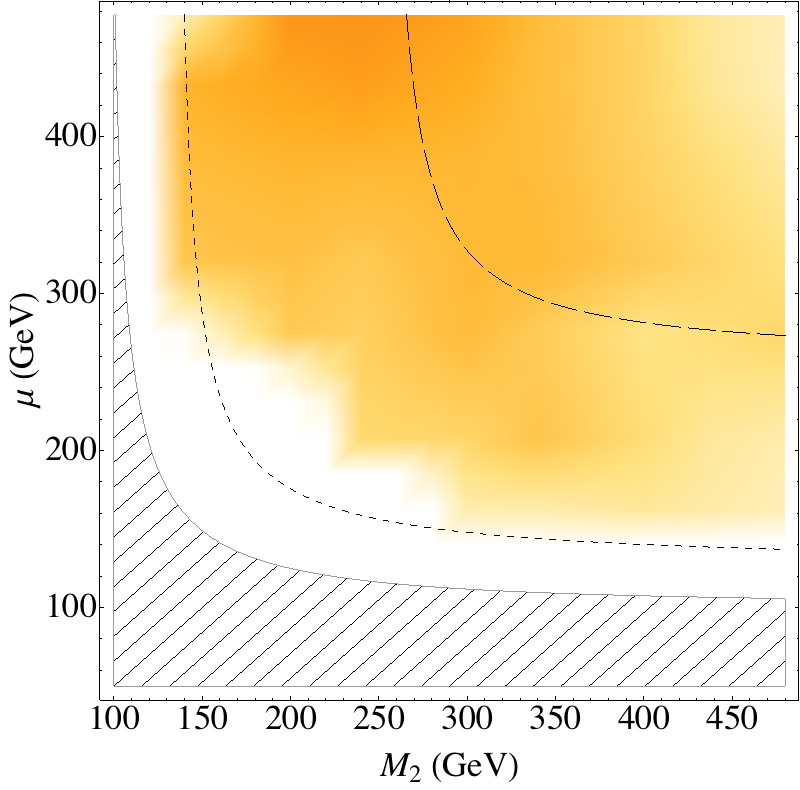}}
			\put(40,105){20 GeV}
			\put(0,105){$M_1=$}
		\end{picture}
		\label{}
	\end{subfigure}
	\begin{subfigure}{0.23\textwidth}
		\begin{picture}(100,100)
			\put(0,0){\includegraphics[width=\textwidth]{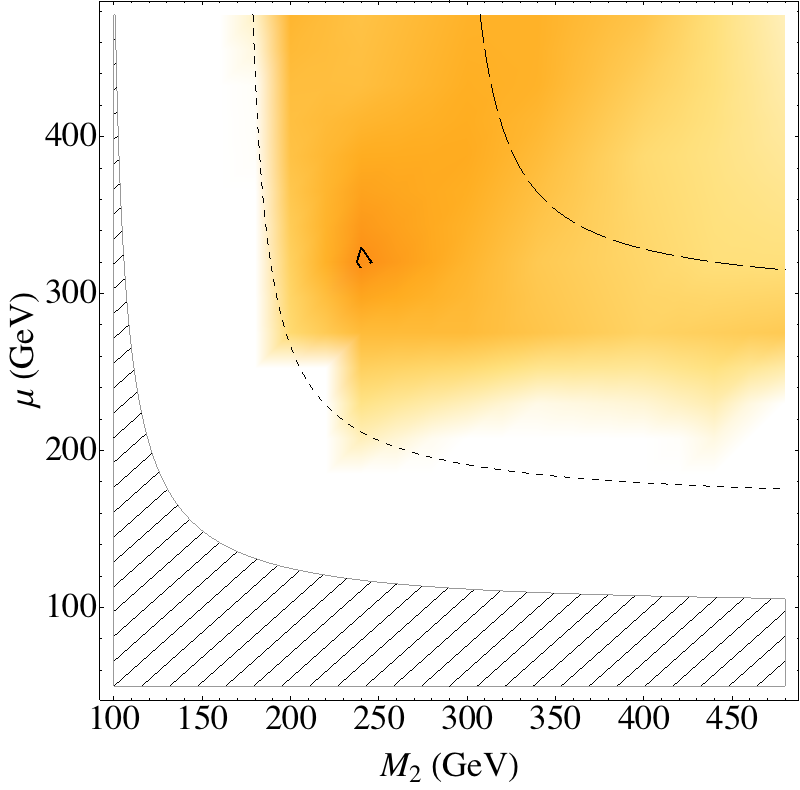}}
			\put(40,105){60 GeV}
		\end{picture}
		\label{}
	\end{subfigure}
	\begin{subfigure}{0.23\textwidth}
		\begin{picture}(100,100)
			\put(0,0){\includegraphics[width=\textwidth]{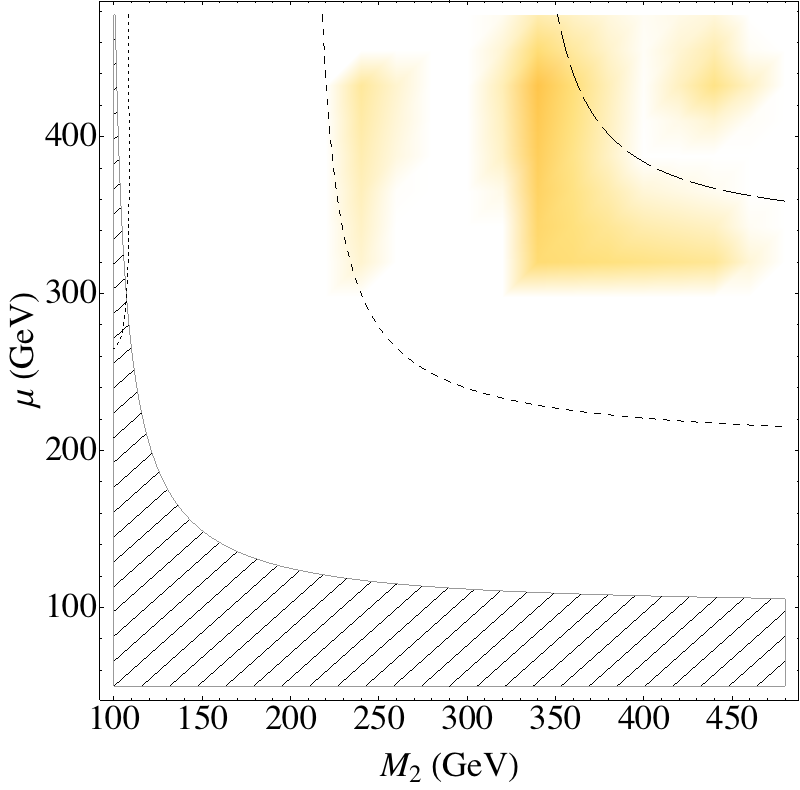}}
			\put(40,105){100 GeV}
		\end{picture}
		\label{}
	\end{subfigure}
	\begin{subfigure}{0.23\textwidth}
		\begin{picture}(100,100)
			\put(0,0){\includegraphics[width=\textwidth]{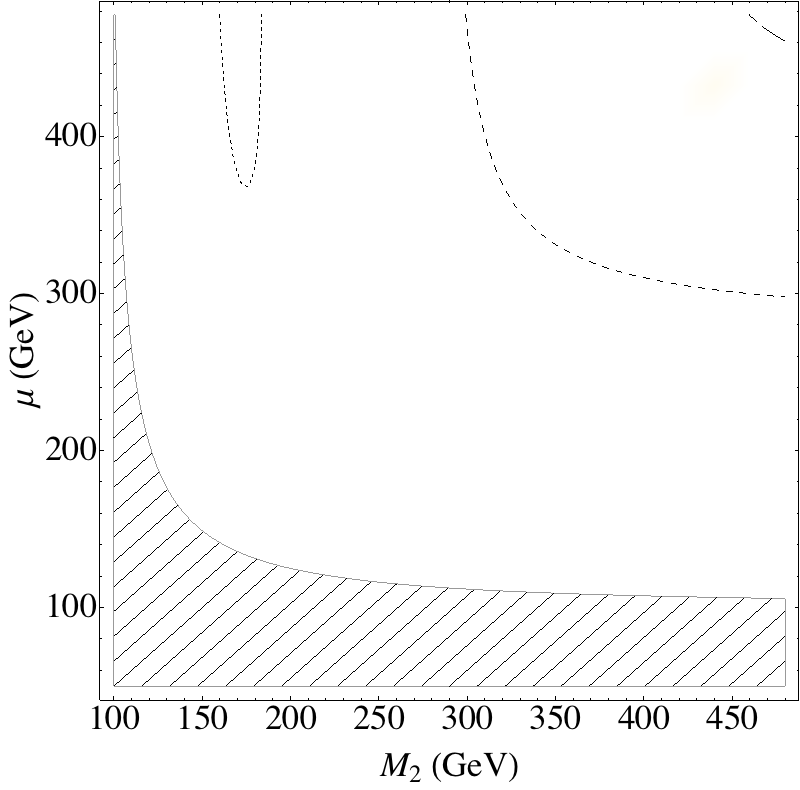}}
			\put(40,105){180 GeV}
		\end{picture}
		\label{}
	\end{subfigure}
	\begin{subfigure}{0.055\textwidth}
		\begin{picture}(100,100)
			\put(0,0){\includegraphics[width=\textwidth]{./Legend.pdf}}
		\end{picture}
		\label{}
	\end{subfigure}
	\vspace{-4mm}	
	\caption{Parameter exclusions from the CMS lepton plus bottom quarks search of Ref.~\cite{1596278} in the $M_2$--$\mu$ plane for several fixed values of $M_1$.  The boundaries of the 95\,\%~c.l. excluded regions are denoted by the thick black solid lines (thick black dashed lines) assuming a $K$ factor of 1.0 (1.2).  Colour shading indicates the number of predicted signal events relative to the number excluded by the experimental analysis.  The hatched area shows the 95\,\%~c.l. exclusion from LEP.   Contours of constant $\Delta M = m_{\nn}-m_{\n}$ are indicated by the thin dashed lines -- long dash: $\Delta M=2 m_h$; mid dash: $\Delta M=m_h$.}
	\label{fig:clbb}
\vspace{0.5cm}
\end{figure}
\subsection{CMS Lepton plus Bottom Quarks}

  The CMS lepton plus bottom quarks search of Ref.~\cite{1596278} 
was designed to probe $\nn \x$ production followed by $\nn \to h^0\n$
with $h^0\to b\bar{b}$ and $\x \to W^{+(*)}\n$ with $W^{+(*)}\to \ell\nu_{\ell}$.
Events with one lepton, two b-tagged jets, and missing energy were selected.
To suppress backgrounds from top quark production, a veto was imposed
on additional leptons or jets along with a kinematic cut.  
Other backgrounds involving a leptonic $W$ were suppressed by
demanding $m_T > 100\,\gev$ for the lepton.  The analysis also
required a $b\bar{b}$ invariant mass in the range 
$100\,\gev < m_{b\bar{b}}< 150\,\gev$ and applied a variable missing
energy cut of $\met > 100,\,125,\,150,\,175\,\gev$.

  The sensitivity of this search to the general electroweakino parameter
is shown in Fig.~\ref{fig:clbb}.  In contrast to Ref.~\cite{1596278},
we do not find any excluded regions.  The difference comes from our
use of the computed $\chi_2^0\to h^0\chi_1^0$ branching ratio, whereas
the CMS analysis assumes a branching fraction of one.  As expected,
a significant signal in this channel requires $\Delta M = m_{\nn}-m_{\n} > m_h$,
since off-shell decays involving the Higgs are very suppressed by its
narrow width.  Contours of $\Delta M = m_h~(2m_h)$ are indicated
by mid (long) dashed lines in Fig.~\ref{fig:clbb}.  For larger $M_1$ values,
$\Delta M > m_h$ requires significantly heavy $\mu$ and $M_2$ 
and thus the sensitivity of currect LHC searches drops off quickly.

\subsection{Other Searches}

  In addition to the four channels described above,
we have investigated the sensitivity of a number of other LHC searches
listed at the end of Sec.~\ref{sec:search}.  These give weaker exclusions,
and we will only comment on them briefly.  
  
  The CMS collaboration has performed searches for two, three, and four
leptons with missing energy in Ref.~\cite{CMS:2013dea} that are 
similar to the ATLAS studies considered above.  In the CMS studies
the signal region is subdivided into a large number of disjoint bins,
whereas ATLAS uses a small number of signal regions geared towards
specific decay cascades.  Since we do not attempt to combine signal bins
and only use boolean exclusions, the ATLAS limits are stronger.  
ATLAS has also performed a second trilepton analysis in Ref.~\cite{Aad:2014nua}
with slightly different signal requirements than Ref.~\cite{ATLAS:2013rla} 
discussed above.  We find similar bounds from Ref.~\cite{Aad:2014nua},
and our trilepton-excluded region matches fairly well with their limit 
in the $M_2$--$\mu$ plane with low $M_1$.

  We have also examined a broad range of searches that include
one or more hard jets and missing energy among the selection requirements.
These include the monojet~\cite{ATLAS:2012zim,
TheATLAScollaboration:2013aia,CMS:rwa} 
and Razor analyses~\cite{TheATLAScollaboration:2013via} 
that have been used to test dark matter production 
at the LHC~\cite{Rajaraman:2011wf,Fox:2011pm,Fox:2012ee,Buckley:2013kua},
as well as channels with both hard jets 
and leptons~\cite{ATLAS:2013tma}.  The limits obtained from these
are weaker than the lepton-centric studies above, with the typically
high requirements on jet $p_T$ greatly reducing the electroweakino signal.  
In particular, we do not find any exclusion beyond the LEP limit from monojet
searches, consistent with Refs.~\cite{Baer:2014cua,Han:2014kaa}.

  A qualitatively different analysis is the ATLAS search 
for disappearing charged tracks of Ref.~\cite{Aad:2013yna}.
This search is sensitive to charginos that decay slowly to the 
lightest neutralino.
Such long-lived charginos are expected to occur in the MSSM 
when $|M_2|\ll |M_1|,\,\mu$, as can occur in anomaly-mediated
supersymmetry breaking~\cite{Randall:1998uk,Giudice:1998xp}.  
In this limit, the tree-level splitting
between $\x$ and $\n$ is negligible, and the net mass splitting is 
dominated by loop effects that give 
$\Delta m \simeq 160\,\mev$~\cite{Feng:1999fu,Gherghetta:1999sw,Ibe:2012sx}
This leads to a dominant $\x\to \pi^{-}\n$ decay with a lifetime on
the order of 0.1\,ns~\cite{Ibe:2012sx}. 
For the moderate values of $\mu$ considered here,
we find that the mass splitting between $\x$ and $\n$ 
is larger than 200~MeV, leading to lifetimes below the
sensitivity of the ATLAS search.
Larger values of $\mu$ than are explored in this study
are needed to generate masses with a sufficiently compressed
spectrum to be sensitive to bounds from ATLAS, and
sufficiently large $M_1$ and $\mu$ can result in
sensitivity up to $M_2 \lesssim 260$~GeV.
We also find that the mass splitting can be smaller 
(or even negative~\cite{Kribs:2008hq}) when some of the mass parameters are
negative.

\subsection{Combined Exclusions}

\begin{figure}[ttt]
	\begin{subfigure}{0.23\textwidth}
		\begin{picture}(100,100)
			\put(0,0){\includegraphics[width=\textwidth]{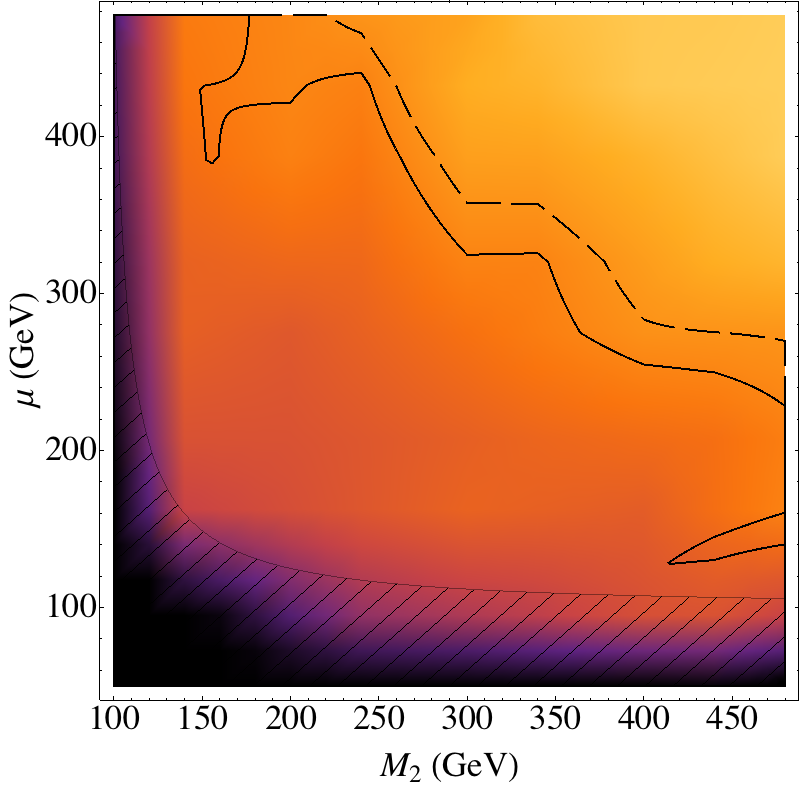}}
			\put(40,105){20 GeV}
			\put(0,105){$M_1=$}
		\end{picture}
		\label{}
	\end{subfigure}
	\begin{subfigure}{0.23\textwidth}
		\begin{picture}(100,100)
			\put(0,0){\includegraphics[width=\textwidth]{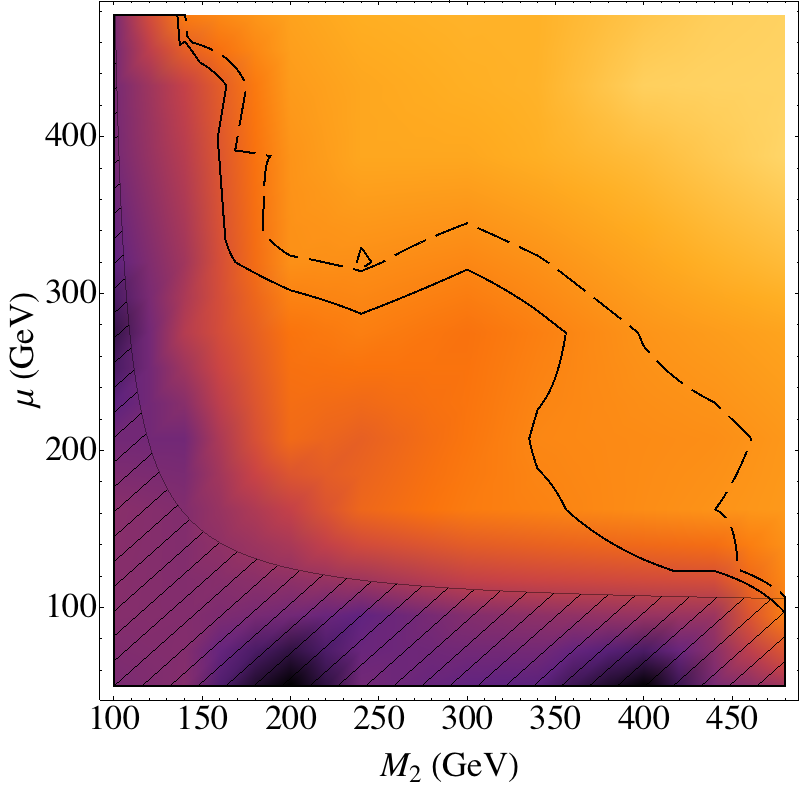}}
			\put(40,105){60 GeV}
		\end{picture}
		\label{}
	\end{subfigure}
	\begin{subfigure}{0.23\textwidth}
		\begin{picture}(100,100)
			\put(0,0){\includegraphics[width=\textwidth]{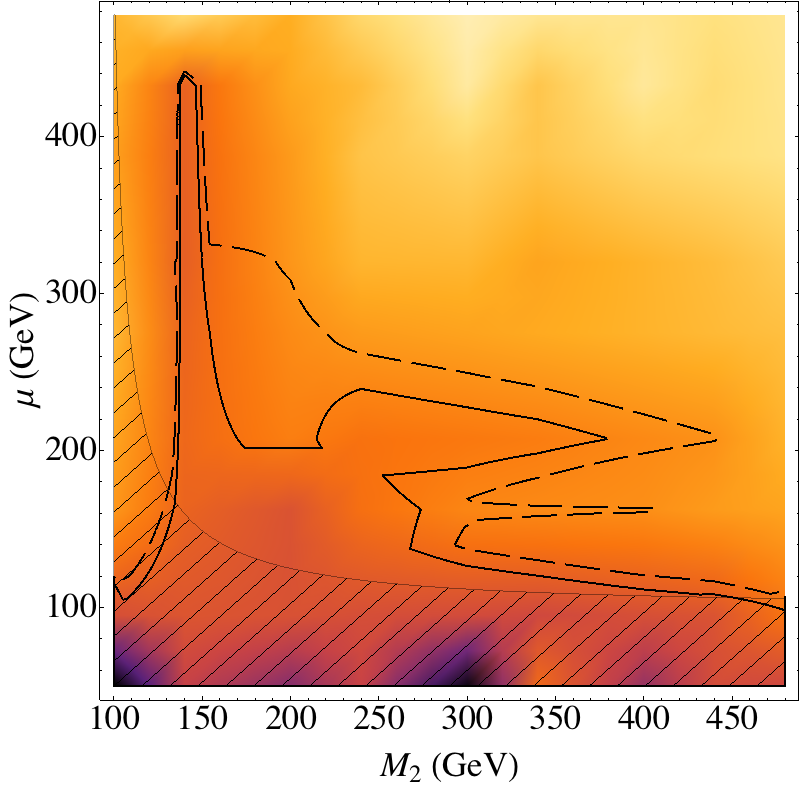}}
			\put(40,105){100 GeV}
		\end{picture}
		\label{}
	\end{subfigure}
	\begin{subfigure}{0.23\textwidth}
		\begin{picture}(100,100)
			\put(0,0){\includegraphics[width=\textwidth]{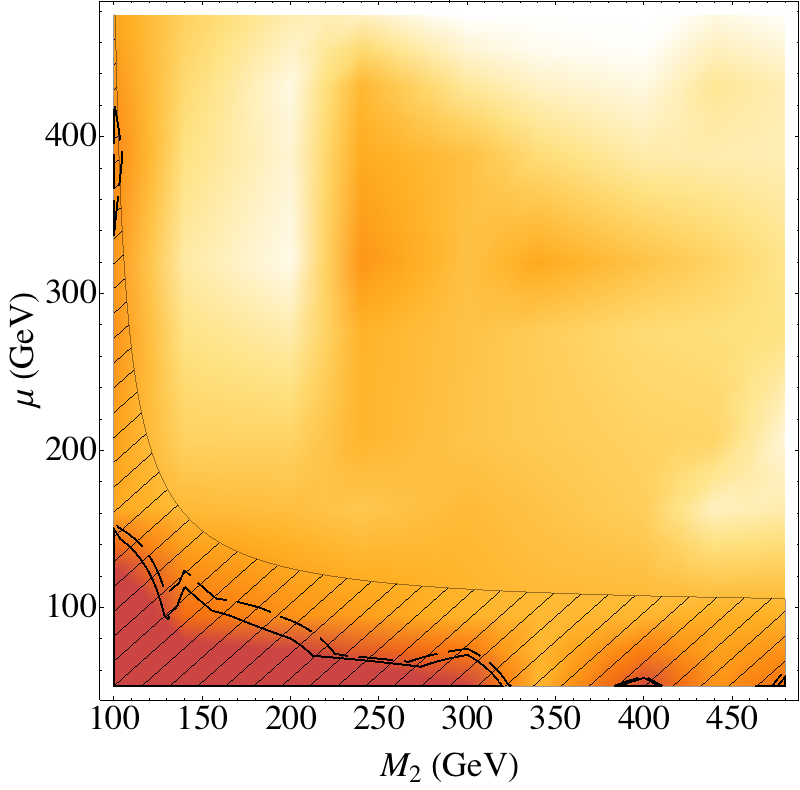}}
			\put(40,105){180 GeV}
		\end{picture}
		\label{}
	\end{subfigure}
	\begin{subfigure}{0.055\textwidth}
		\begin{picture}(100,100)
			\put(0,0){\includegraphics[width=\textwidth]{./Legend.pdf}}
		\end{picture}
		\label{}
	\end{subfigure}
	\vspace{-4mm}	
	\caption{Combined exclusions from the LHC analyses discussed in the text in the $M_2$--$\mu$ plane for several fixed values of $M_1$.  The boundaries of the 95\,\%~c.l. excluded regions are denoted by the thick black solid lines (thick black dashed lines) assuming a $K$ factor of 1.0 (1.2).  Colour shading indicates the number of predicted signal events relative to the number excluded by the experimental analysis.  The hatched area shows the 95\,\%~c.l. exclusion from LEP.}
	\label{fig:totalm1}
\vspace{-0.0cm}
\end{figure}

  Putting our results together, we show in Fig.~\ref{fig:totalm1} the 
combined sensitivity of all LHC searches considered in the $M_2$--$\mu$
plane for $M_1=20$, 60, 100, 180\,GeV.  The thick solid black line
shows the 95\%\,c.l. exclusion obtained using LO MadGraph production
cross sections, while the dashed black line gives the exclusion when
a signal $K$ factor of 1.2 is applied.  The hatched region is excluded
by LEP analyses.  As expected, the excluded region is significant
for small $M_1$, but shrinks quickly as $M_1$ is increased.

\begin{figure}[ttt]
	\begin{subfigure}{0.23\textwidth}
		\begin{picture}(100,100)
			\put(0,0){\includegraphics[width=\textwidth]{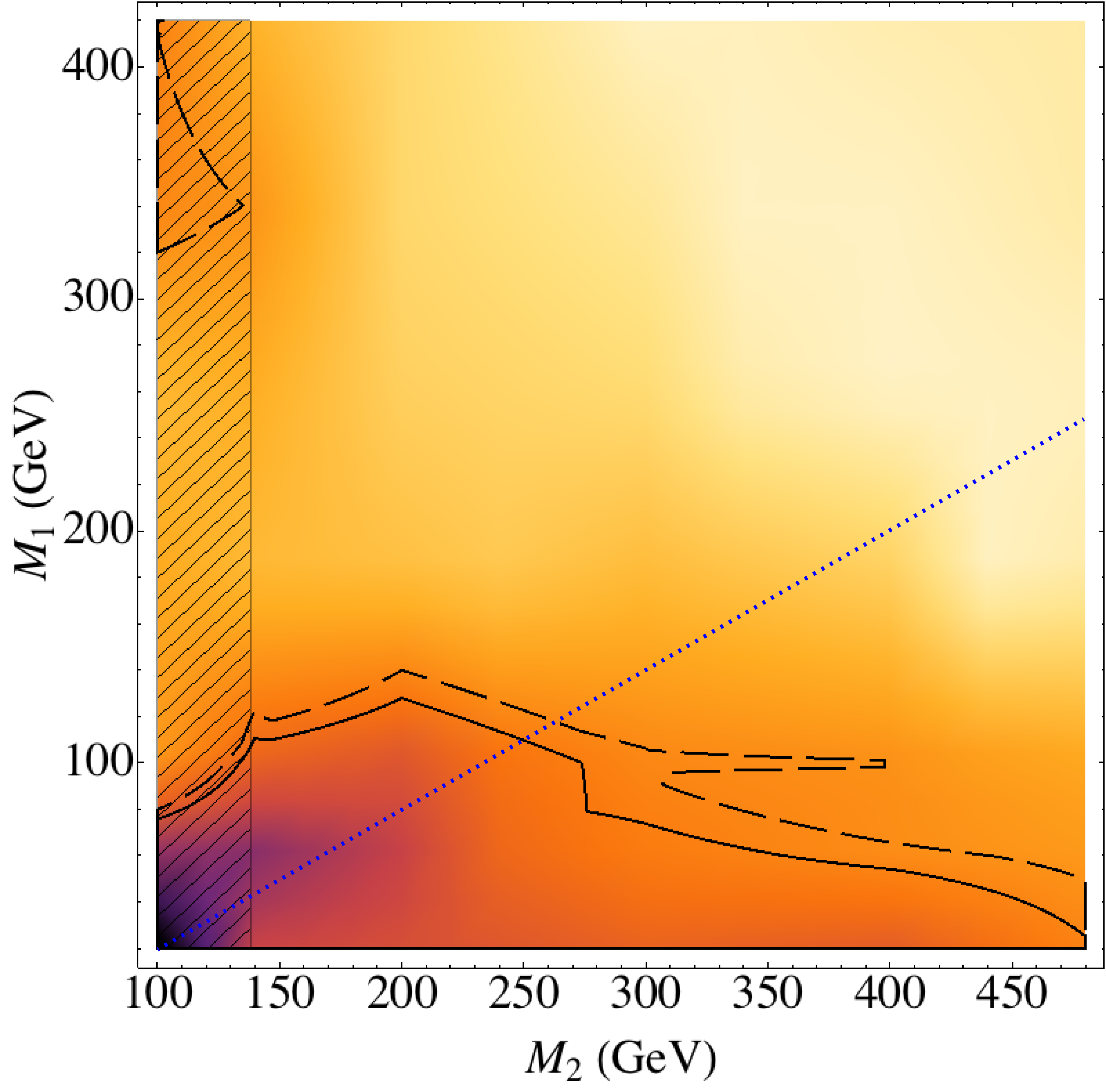}}
			\put(40,105){162 GeV}
			\put(0,105){$\mu=$}
		\end{picture}
		\label{}
	\end{subfigure}
	\begin{subfigure}{0.23\textwidth}
		\begin{picture}(100,100)
			\put(0,0){\includegraphics[width=\textwidth]{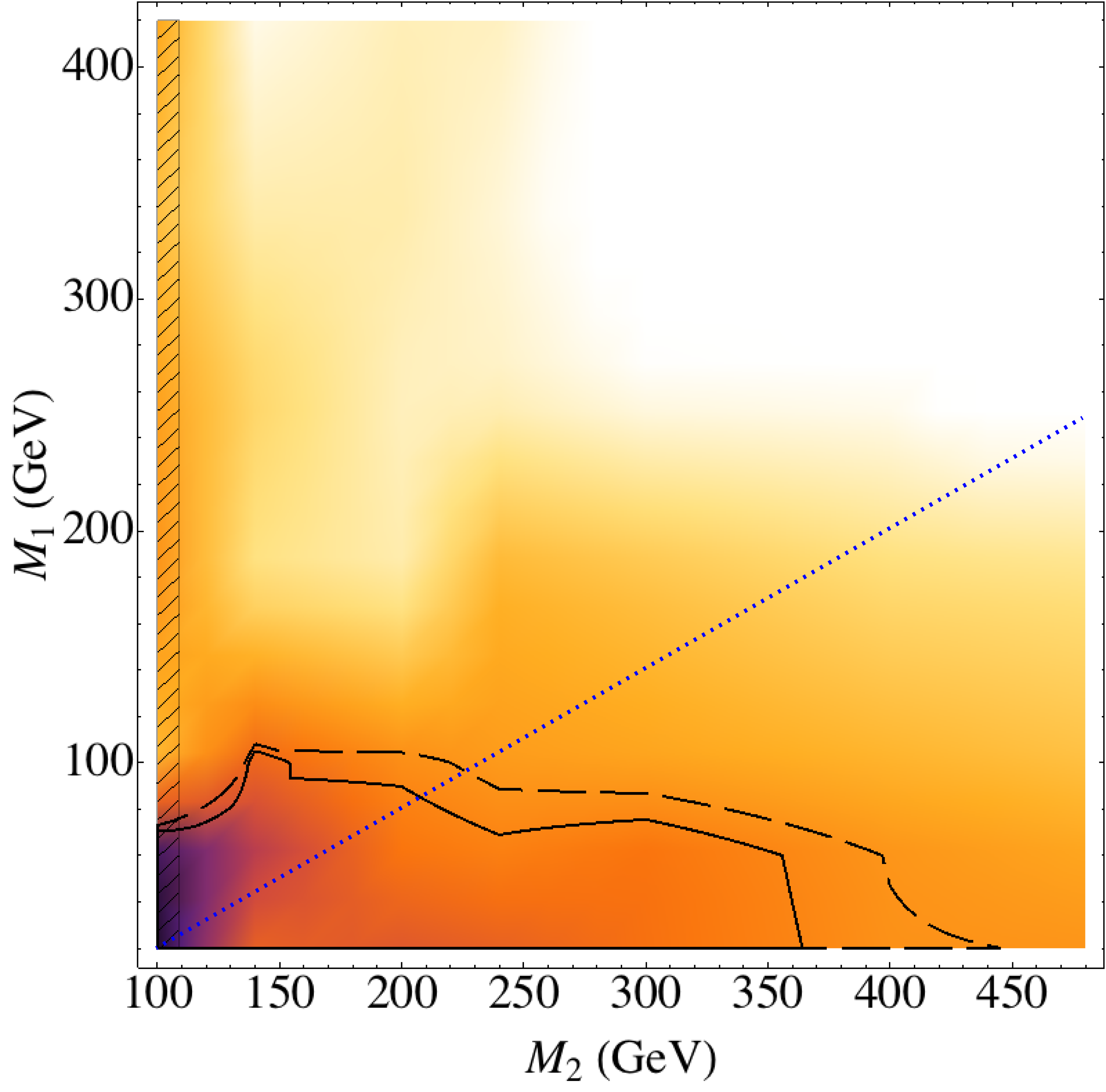}}
			\put(40,105){275 GeV}
		\end{picture}
		\label{}
	\end{subfigure}
	\begin{subfigure}{0.23\textwidth}
		\begin{picture}(100,100)
			\put(0,0){\includegraphics[width=\textwidth]{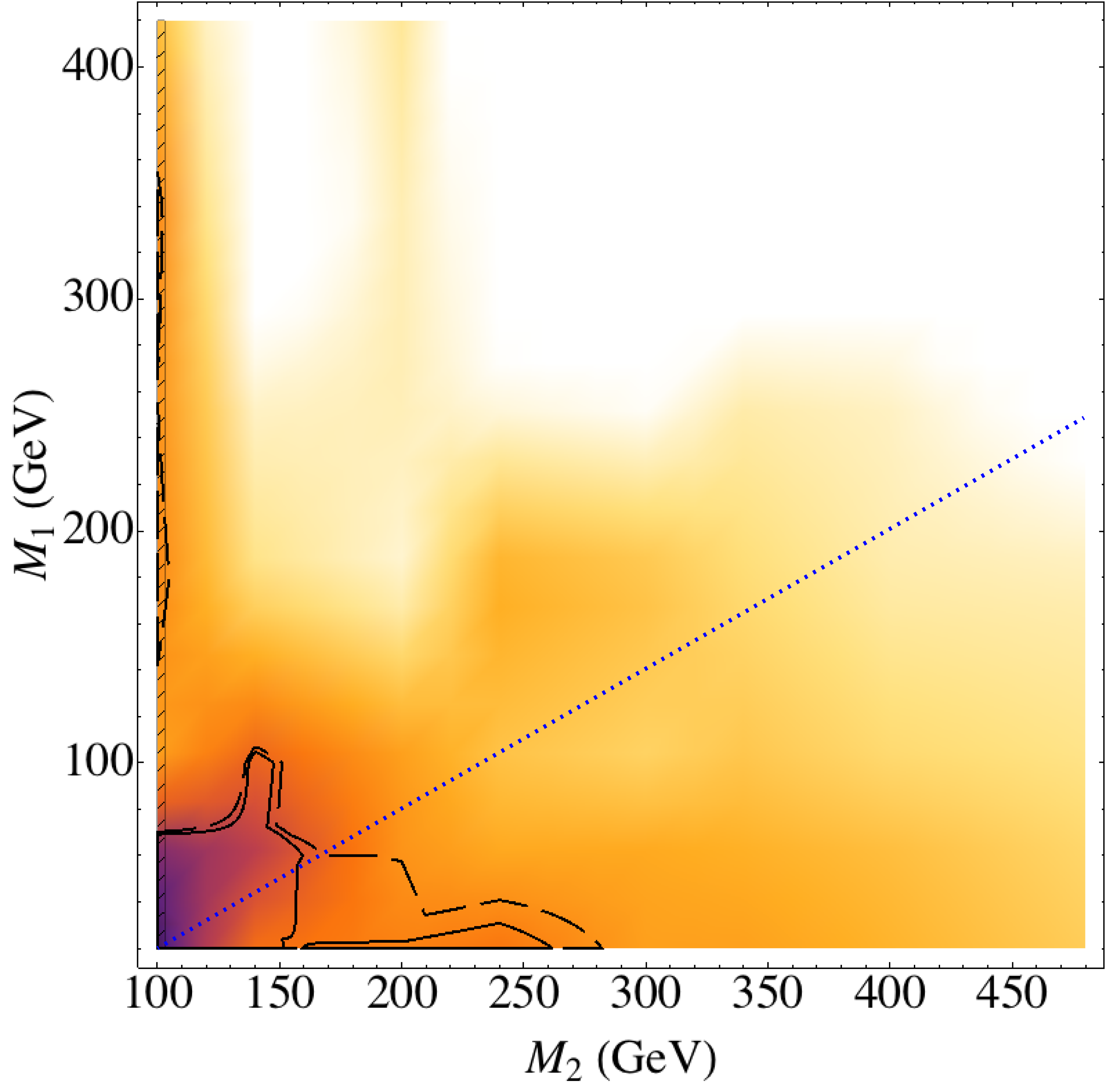}}
			\put(40,105){388 GeV}
		\end{picture}
		\label{}
	\end{subfigure}
	\begin{subfigure}{0.23\textwidth}
		\begin{picture}(100,100)
			\put(0,0){\includegraphics[width=\textwidth]{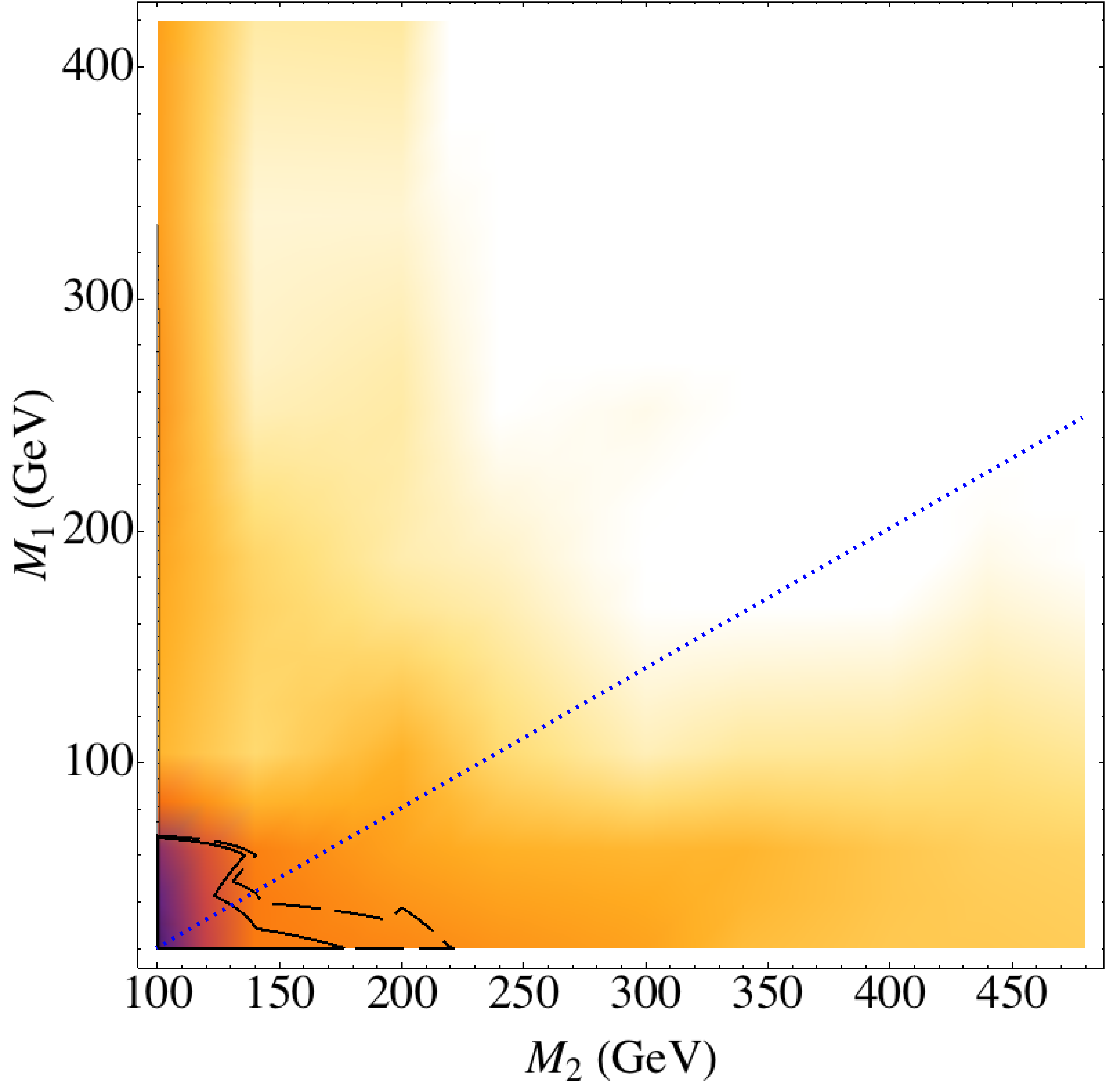}}
			\put(40,105){478 GeV}
		\end{picture}
		\label{}
	\end{subfigure}
	\begin{subfigure}{0.055\textwidth}
		\begin{picture}(100,100)
			\put(0,0){\includegraphics[width=\textwidth]{./Legend.pdf}}
		\end{picture}
		\label{}
	\end{subfigure}
	\vspace{-4mm}	
	\caption{Combined exclusions from the LHC analyses discussed in the text in the $M_2$--$M_1$ plane for several fixed values of $\mu$.  The boundaries of the 95\,\%~c.l. excluded regions are denoted by thick black solid lines (thick black dashed lines) assuming a $K$ factor of 1.0 (1.2).  Colour shading indicates the number of predicted signal events relative to the number excluded by the experimental analysis.  The hatched area shows the 95\,\%~c.l. exclusion from LEP. The dotted blue line indicates  where $M_2=2M_1$.} 
	\label{fig:totalmu}
\vspace{-0.0cm}
\end{figure}

  To investigate the $M_1$ dependence of these exclusions 
in more detail, we show in Fig.~\ref{fig:totalmu} the combined sensitivity 
of all LHC searches considered in the $M_2$--$M_1$
plane for $\mu = 162$, 275, 388, 478\,GeV.
In each of these plots we also indicate the gaugino universality
condition of $M_2\simeq 2M_1$ with a blue dotted line.
The excluded region only reaches to $M_1 \sim 100\,\gev$.  
For larger $M_1$ values (and accounting for the LEP limits on charginos), 
either the mass splittings $\nn-\n$ and $\x-\n$ become small 
or the non-LSP states become heavy.
Small mass splittings lead to a poor acceptance by the searches considered,
while heavier non-LSP states are produced less frequently.

  The excluded regions also shrink as $\mu$ becomes large.
In particular, the exclusion in the $\mu=478\,\gev$ panel 
of Fig.~\ref{fig:totalmu}, where the LSP is typically Bino-like 
and the $\x$ and $\nn$ states are Wino-like, is much weaker than
the exclusion than the exclusion quoted for a Bino-Wino simplified
model in Ref.~\cite{ATLAS:2013rla}.  In their analysis, they set
$\text{BR}(\nn\to Z^0\n) = 1$.  In contrast, we find that
in this limit the alternative decay mode $\nn\to h^0\n$ can become
very significant at large $\mu$.  Since the Higgs boson $h^0$ decays
only rarely produce more than a single lepton, this strongly suppresses
the trilepton signal.\footnote{We have also checked that our analysis
methods give exclusions similar to Ref.~\cite{ATLAS:2013rla} when the Higgs
decay mode is turned off.}  Decreasing $\mu$ increases the probability
of the $Z^0$ decay, and larger exclusions are found.

  Note that in this work we have not examined the detailed dependence 
of the excluded regions on $\tan\beta$, having fixed its value to $\tan\beta=10$.
However, as discussed previously, we find very similar production 
cross sections and decay branching fractions for $2<\tan\beta < 50$.
Thus, we expect qualitatively similar results for other values of
$\tan\beta$.

\section{Conclusions\label{sec:conc}}

  In this work we have investigated the sensitivity of current LHC searches
to the general chargino and neutralino sector of the MSSM in the limit
where all the other superpartners are heavy enough to be neglected.
This leaves a simple four-dimensional parameter space of
$\{M_1,\,M_2,\,\mu,\,\tan\beta\}$.  We have reinterpreted a diverse
set of studies by ATLAS and CMS to derive exclusions on this space.  

  The greatest LHC sensitivity to general electroweakinos comes
from searches requiring multiple leptons and missing energy.  
This helps to reduce the dominant background to vector diboson production.
However, distinguishing the signal from electroweakinos from this remaining
background is challenging, especially when the mass spectrum is compressed.
For this reason, we only find significant parameter exclusions for relatively
small values of $M_1 \lesssim 100\,\gev$ with a Bino-like LSP.  
In this case, signals come from the production of heavier Wino- or Higgsino-like
charginos and neutralinos followed by their decays to the LSP, 
primarily through weak vector bosons.  

  Despite the limited reach of existing LHC searches, our results show that
they have a reasonable acceptance for larger electroweak masses.  
For this reason, we expect that much larger exclusions will be possible 
using similar analysis techniques with improved data sets from 
upcoming LHC runs.  
Additional data should also allow for the investigation of scenarios with a
Wino- or Higgsino-like LSP. Further improvements may also be possible
with modified analysis techniques, such as those proposed in 
Refs.~\cite{Schwaller:2013baa,Gori:2013ala,Dreiner:2012gx,Mukhopadhyay:2014dsa}.

  Our results can be applied to test scenarios where the charginos 
and neutralinos play an important role.  One example is dark matter, 
where the relic density is very sensitive to the gauge content 
of the LSP~\cite{ArkaniHamed:2006mb,Giudice:2010wb,Cheung:2012qy}.  
A second case is supersymmetric electroweak baryogenesis, in which the charginos
and neutralinos are frequently the dominant new source of CP violation
required for the net creation of baryons~\cite{Cline:2000nw,Carena:2000id}.
In particular, our results suggest that the Bino-driven scenario
of Refs.~\cite{Li:2008ez,Cirigliano:2009yd,Kozaczuk:2011vr} 
is not significantly constrained by current LHC data.

\section*{Acknowledgements}

The authors would like to thank Anadi Canepa, Zoltan Gesce, Alejandro de la Puente, Isabel Trigger, and Peter Winslow for helpful discussions. This work was supported by the National Science and Engineering Research Council of Canada~(NSERC) and by the Perimeter Institute for Theoretical Physics.  Research at Perimeter Institute is supported by the Government of Canada through Industry Canada and by the Province of Ontario through the Ministry of Economic Development \& Innovation.

\newpage


\appendix

\section{Appendix: Mass Matrices and Mixings\label{sec:appa}}

  The charginos are mixtures of the charged components of the Winos and
Higgsinos.  Writing 
${\psi}^{+} = (-i\widetilde{W}^+,\widetilde{H}_u^+)^t$
and ${\psi}^{-} = (-i\widetilde{W}^-,\widetilde{H}_d^-)^t$ 
the corresponding mass term (in two-component notation) 
is~\cite{Haber:1984rc,Martin:1997ns}
\beq
-\mathscr{L}_{\pm} \supset 
({\psi}^{-})^tX{\psi}^{+} + (h.c.)
\eeq
with
\beq
X = 
\left(
\begin{array}{cc}
M_2&\sqrt{2}s_{\beta}m_W\\
\sqrt{2}c_{\beta}m_W&\mu
\end{array}
\right) \ .
\eeq
The matrix $X$ is not Hermitian in general, so there may not exist a unitary
matrix that diagonalizes it.  However, it can always be bi-diagonalized
with a pair of unitary matrices $U$ and $V$ such that
\beq
VX^{\dagger}XV^{\dagger} = U^*XX^{\dagger}U^t 
= \text{diag}(m_{\x},m_{\xx}) \ ,
\eeq
where $|m_{\x}| \leq |m_{\xx}|$.
In terms of $U$ and $V$, the mass and gauge eigenstates are related by
\beq
\chi_i^{+} = V_{ij}{\psi}^+_j \ , \hspace{3cm}
\chi_i^{-} = U_{ij}{\psi}^-_j \ .
\eeq
It is conventional to combine these into four-component Dirac fermions
with $\chi_i^+ = (\chi_i^+,(\chi_i^-)^{\dagger})$ (in an obvious abuse
of notation).

  For the neutralinos, the mass term in the basis
${\psi}^0 = 
(-i\widetilde{B}^0,-i\widetilde{W}^0,\widetilde{H}_d,\widetilde{H}_u)^t$
is~\cite{Haber:1984rc,Martin:1997ns}
\beq
-\mathscr{L} \supset \frac{1}{2}({\psi}^0)^tY{\psi}^0 + (h.c.)
\eeq
with
\beq
Y = \left(
\begin{array}{cccc}
M_1&0&-c_{\beta}s_Wm_Z&s_{\beta}s_Wm_Z\\
0&M_2&c_{\beta}c_Wm_Z&-s_{\beta}c_Wm_Z\\
-c_{\beta}s_Wm_Z&c_{\beta}c_Wm_Z&0&-\mu\\
s_{\beta}s_Wm_Z&-s_{\beta}c_Wm_Z&-\mu&0
\end{array}
\right) \ .
\eeq
This matrix is complex symmetric, and can be diagonalized by 
a unitary matrix $N$ such that
\beq
N^*YN^{\dagger} 
= \text{diag}(m_{\n},m_{\nn},m_{\nnn},m_{\nnnn}) \ ,
\eeq
where $|m_{\n}| \leq |m_{\nn}| \leq |m_{\nnn}| \leq |m_{\nnnn}|$.
The mass eigenstates are related to the gauge eigenstates via
\beq
\chi_i^0 = N_{ij}\psi^0_j \ .
\eeq
These two-component fermions can be combined into four-component
Majorana spinors $\chi_i^0 = (\chi_i^0,\,(\chi_i^{0})^{\dagger})^t$
(with another abuse of notation).

\section{Appendix: Couplings to the Standard Model\label{sec:appb}}

  For our purposes, we need the couplings of the charginos and neutralinos 
to the weak vector bosons and the light SM-like Higgs boson.  

\subsection{Vector Boson Couplings}

  These can be found in Refs.~\cite{Haber:1984rc,Chung:2003fi}.
We will write everything in four-component notation.

\paragraph{
$\mathbf{W^-\chi_i^0\chi_j^+}:$}
\beq
-\mathscr{L} \supset -g\,W^{-}_{\mu}\overline{\chi}_i^0\gamma^{\mu}(
\mathcal{O}^L_{ij}P_L+\mathcal{O}^R_{ij}P_R)\chi_j^+ + (h.c.) \ ,
\eeq
where
\beq
\mathcal{O}^L_{ij} &=& -\frac{1}{\sqrt{2}}N_{i4}V_{j2}^*+N_{i2}V_{j1}^* \ ,\\
\mathcal{O}^R_{ij} &=& \frac{1}{\sqrt{2}}N_{i3}^*U_{j2}+N_{i2}^*U_{j1} \ ,
\eeq
with $g$ the $SU(2)_L$ gauge coupling.  These terms derive from
the $SU(2)_L$-covariant derivatives of the Higgsinos (first terms) 
and the Wino (second terms).

\paragraph{$\mathbf{Z^0\chi_i^-\chi_j^+}:$}
\beq
-\mathscr{L} \supset -\bar{g}\,Z^{0}_{\mu}\overline{\chi}_i^+\gamma^{\mu}(
\mathcal{O'}^L_{ij}P_L+\mathcal{O'}^R_{ij}P_R)\chi_j^+ \ ,
\eeq
where
\beq
\mathcal{O'}^L_{ij} &=& -V_{i1}V_{j1}^*-\frac{1}{2}V_{i2}V_{j2}^* + \delta_{ij}s_W^2
\ , \\
\mathcal{O'}^R_{ij} &=& -U_{i1}^*U_{j1}-\frac{1}{2}U_{i2}^*U_{j2} + \delta_{ij}s_W^2
\ ,
\eeq
with $\bar{g} = g/c_W$.  These come from the $SU(2)_L\times U(1)_Y$
gauge couplings of the Winos (first terms) and Higgsinos (second terms).

\paragraph{
$\mathbf{Z^0\chi_i^0\chi_j^0}:$}
\beq
-\mathscr{L} \supset -\frac{1}{2}\bar{g}\,Z^{0}_{\mu}
\overline{\chi}_i^0\gamma^{\mu}
(\mathcal{O''}^L_{ij}P_L+\mathcal{O''}^R_{ij}P_R)\chi_j^0  \ ,
\eeq
where
\beq
\mathcal{O}^L_{ij} &=& -\frac{1}{2}N_{i3}N_{j3}^*+\frac{1}{2}N_{i4}N_{j4}^* \ ,\\
\mathcal{O}^R_{ij} &=& -(\mathcal{O''}^{L}_{ij})^* \ ,
\eeq
with $\bar{g} = g/c_W$.  Note that these couplings come only from
the Higgsinos.  The Bino has no gauge couplings at all, while the
$\widetilde{W}^0$ has $t^3=0=Y$ and therefore does not couple to the $Z^0$.
For $i=j$, this coupling is purely axial.  It also vanishes for $i=j$ in the
limit that $i$ corresponds to a pure Higgsino state.  

\paragraph{
$\mathbf{\gamma\chi_i^-\chi_j^+}:$}
\beq
-\mathscr{L} \supset e\,A^{\mu}\overline{\chi}_i^+\gamma^{\mu}\chi_i^+ \ ,
\eeq
which is purely diagonal and present only for the charginos due
to conservation of electric charge.  Off-diagonal couplings and
couplings to neutralinos can only occur by way of higher-dimensional
operators such as the electric and magnetic moment forms.

\subsection{(SM-like) Higgs Couplings}

  These are listed in Ref.~\cite{Gunion:1989we}.  We will focus
exclusively on the couplings to the SM-like Higgs $h^0$.
The corresponding mixing angles with the $H_u^0$ and $H_d^0$
gauge eigenstates are
\beq
\left(\begin{array}{c}
h^0\\H^0
\end{array}
\right)
= 
\left(\begin{array}{cc}
c_{\alpha}&-s_{\alpha}\\
s_{\alpha}&c_{\alpha}
\end{array}
\right)
\left(\begin{array}{c}
\sqrt{2}(ReH^0_u - v_u)\\
\sqrt{2}(ReH^0_d-v_d)
\end{array}
\right) \ .
\eeq
In the \emph{decoupling limit}, the couplings of the lighter $h^0$ state
to matter are identical to the SM.  In this limit, the mixing angle
reduces to $\alpha = \beta - \pi/2$ so that $c_{\alpha} = s_{\beta}$
and $s_{\alpha} = -c_{\beta}$.  

\paragraph{$\mathbf{h^0\chi_i^-\chi_j^+:}$}
\beq
-\mathscr{L} \supset g\,h^0\,\overline{\chi}_i^+\left[
(c_{\beta}Q_{ji}^*+s_{\beta}S_{ji}^*)P_L
+ (c_{\beta}Q_{ij}+s_{\beta}S_{ij})P_R
\right]
\chi_j^+
\eeq
with
\bea
Q_{ij} &=& \frac{1}{\sqrt{2}}V_{i1}U_{j2} \ ,\label{eq:qij}\\
S_{ij} &=& \frac{1}{\sqrt{2}}V_{i2}U_{j1} \ .\label{eq:sij}
\eea
These couplings involve one Higgsino component and one Wino component.
They come from the $-i\sqrt{2}g\widetilde{W}^aH_a^*t^a\widetilde{H}_a$
supersymmetrizations of the Higgs boson gauge couplings.

\paragraph{$\mathbf{h^0\chi_i^0\chi_j^0:}$}
\beq
-\mathscr{L} \supset g\,h^0\,\overline{\chi}_i^0\left[
(c_{\beta}{Q''}_{ji}^*-s_{\beta}{S''}_{ji}^*)P_L
+ (c_{\beta}{Q''}_{ij}-s_{\beta}{S''}_{ij})P_R
\right]
\chi_j^0
\eeq
with
\bea
{Q''}_{ij} &=& \frac{1}{{2}}\left[
N_{i3}(N_{j2}-t_WN_{j1}) + N_{j3}(N_{i2}-t_WN_{i1})
\right]\epsilon_i \ ,\\
{S''}_{ij} &=& \frac{1}{{2}}\left[
N_{i4}(N_{j2}-t_WN_{j1}) + N_{j4}(N_{i2}-t_WN_{i1})
\right]\epsilon_i \ ,
\eea
where $\epsilon_i$ is the sign of the $i$-th mass eigenvalues
(for real parameters).  As above, these couplings
involve one Higgsino component and one Wino or Bino component,
and they come from the supersymmetrizations of the Higgs boson gauge couplings.


\end{document}